\documentstyle[epsf,12pt]{article}

\catcode`\@=11
\long\def\@makefntext#1{
\protect\noindent \hbox to 3.2pt {\hskip-.9pt
$^{{\ninerm\@thefnmark}}$\hfil}#1\hfill}		

\def\@makefnmark{\hbox to 0pt{$^{\@thefnmark}$\hss}}  

\def\ps@myheadings{\let\@mkboth\@gobbletwo
\def\@oddhead{\hbox{}
\rightmark\hfil\ninerm\thepage}
\def\@oddfoot{}\def\@evenhead{\ninerm\thepage\hfil
\leftmark\hbox{}}\def\@evenfoot{}
\def\sectionmark##1{}\def\subsectionmark##1{}}

\renewcommand{\thefootnote}{\fnsymbol{footnote}}

\newcounter{sectionc}\newcounter{subsectionc}\newcounter{subsubsectionc}
\renewcommand{\section}[1] {\vspace*{0.6cm}\addtocounter{sectionc}{1}
\setcounter{subsectionc}{0}\setcounter{subsubsectionc}{0}\noindent
	{\normalsize\bf\thesectionc. #1}\par\vspace*{0.4cm}}
\renewcommand{\subsection}[1] {\vspace*{0.6cm}\addtocounter{subsectionc}{1}

	\setcounter{subsubsectionc}{0}\noindent
	{\normalsize\it\thesectionc.\thesubsectionc. #1}\par\vspace*{0.4cm}}
\renewcommand{\subsubsection}[1]
{\vspace*{0.6cm}\addtocounter{subsubsectionc}{1}
	\noindent {{\begin{center}\normalsize\it
	\thesectionc.\thesubsectionc.\thesubsubsectionc.
	#1\end{center}}}\par\vspace*{0.4cm}}

\newcounter{appendixc}
\newcounter{subappendixc}[appendixc]
\newcounter{subsubappendixc}[subappendixc]

\renewcommand{\appendix}[1] {\vspace*{0.6cm}
        \refstepcounter{appendixc}
        \setcounter{figure}{0}
        \setcounter{table}{0}
        \setcounter{equation}{0}
        \renewcommand{\thefigure}{\Alph{appendixc}.\arabic{figure}}
        \renewcommand{\thetable}{\Alph{appendixc}.\arabic{table}}
        \renewcommand{\theappendixc}{\Alph{appendixc}}
        \renewcommand{\theequation}{\Alph{appendixc}.\arabic{equation}}
        \noindent{\bf Appendix \theappendixc #1}\par\vspace*{0.4cm}}

\def\abstracts#1{{

\centering{\begin{minipage}{12.2truecm}\footnotesize\baselineskip=12pt\noindent
	\centerline{\footnotesize ABSTRACT}\vspace*{0.3cm}
	\parindent=0pt #1
	\end{minipage}}\par}}

\renewenvironment{thebibliography}[1]
	{\begin{list}{\arabic{enumi}.}
	{\usecounter{enumi}\setlength{\parsep}{0pt}
\setlength{\leftmargin 1.25cm}{\rightmargin 0pt}
	 \setlength{\itemsep}{0pt} \settowidth
	{\labelwidth}{#1.}\sloppy}}{\end{list}}

\newcounter{itemlistc}
\newcounter{romanlistc}
\newcounter{alphlistc}
\newcounter{arabiclistc}

\newcommand{\fcaption}[1]{
        \refstepcounter{figure}
        \setbox\@tempboxa = \hbox{\footnotesize Fig.~\thefigure. #1}
        \ifdim \wd\@tempboxa > 6in
           {\begin{center}
        \parbox{6in}{\footnotesize\baselineskip=12pt Fig.~\thefigure. #1}
            \end{center}}
        \else
             {\begin{center}
             {\footnotesize Fig.~\thefigure. #1}
              \end{center}}
        \fi}

\newcommand{\tcaption}[1]{
        \refstepcounter{table}
        \setbox\@tempboxa = \hbox{\footnotesize Table~\thetable. #1}
        \ifdim \wd\@tempboxa > 6in
           {\begin{center}
        \parbox{6in}{\footnotesize\baselineskip=12pt Table~\thetable. #1}
            \end{center}}
        \else
             {\begin{center}
             {\footnotesize Table~\thetable. #1}
              \end{center}}
        \fi}

\def\@citex[#1]#2{\if@filesw\immediate\write\@auxout
	{\string\citation{#2}}\fi
\def\@citea{}\@cite{\@for\@citeb:=#2\do
	{\@citea\def\@citea{,}\@ifundefined
	{b@\@citeb}{{\bf ?}\@warning
	{Citation `\@citeb' on page \thepage \space undefined}}
	{\csname b@\@citeb\endcsname}}}{#1}}

\newif\if@cghi
\def\cite{\@cghitrue\@ifnextchar [{\@tempswatrue
	\@citex}{\@tempswafalse\@citex[]}}
\def\citelow{\@cghifalse\@ifnextchar [{\@tempswatrue
	\@citex}{\@tempswafalse\@citex[]}}
\def\@cite#1#2{{\if@cghi$\null^{#1}$\else#1\fi\if@tempswa\typeout
	{IJCGA warning: optional citation argument
	ignored: `#2'} \fi}}

 1
 1
 1

\font\ninerm=cmr9

\newcommand{\gsim}{\mathrel{\mathpalette\@versim>}}
\def\@versim#1#2{\vcenter{\offinterlineskip\ialign{$\m@th#1\hfil##\hfil
		$\crcr#2\crcr\sim\crcr}}}

\begin{document}
\centerline{\normalsize\bf THE CHIRAL PHASE TRANSITION IN QCD:}
\baselineskip=20pt
\centerline{\normalsize\bf CRITICAL PHENOMENA}
\baselineskip=20pt
\centerline{\normalsize\bf AND LONG WAVELENGTH PION
OSCILLATIONS}

\centerline{\footnotesize KRISHNA RAJAGOPAL}
\baselineskip=13pt
\centerline{\footnotesize\it Lyman Laboratory of Physics, Harvard University,}
\baselineskip=12pt
\centerline{\footnotesize\it Cambridge, MA, USA. 02138}
\centerline{\footnotesize E-mail: rajagopal@huhepl.harvard.edu}

\vspace*{0.8cm}
\abstracts{In QCD with two flavours of massless quarks, the
chiral phase transition is plausibly in the same universality
class as the classical four component Heisenberg magnet.
Therefore, renormalization group techniques developed
in the study of phase transitions can be applied to calculate
the critical exponents which characterize the scaling behaviour
of universal quantities near the critical point.
As a consequence of this observation, a quantitative
description of the universal physics of the chiral
phase transition in circumstances where the plasma is
close to thermal equilibrium as it passes through
the critical temperature can be obtained.  This
approach to the QCD phase transition has implications
both for lattice gauge theory and for heavy ion collisions.
Future lattice simulations with longer correlation lengths
will provide quantitative measurements of the various exponents and
the equation of state for the order parameter as a
function of temperature and quark masses.  Present lattice
simulations already allow many qualitative tests.
In a heavy ion collision, the consequence of a long correlation
length would be large fluctuations in the number ratio
of neutral to charged pions.  Unfortunately, we will
see that because of the explicit chiral symmetry breaking
introduced by the masses of the up and down quarks, no equilibrium
correlation length gets long enough for this phenomenon to occur.
In the third chapter of this article, I discuss attempts
to model the dynamics of the chiral order parameter
in a far from equilibrium QCD phase transition
by considering quenching in the O(4) linear sigma
model.  I argue, and present numerical evidence, that
in the period immediately following the quench
long wavelength modes of the pion field are amplified.
This could have dramatic phenomenological consequences
in heavy ion collisions.  I review various recent theoretical
advances --- attempts to include expansion and relax the
quench assumption; attempts to include quantum mechanical
effects.  Long wavelength pion oscillations have
now been seen in a number of simulations in which
different assumptions are made.  However, all of
the theoretical treatments involve idealizations and
assumptions, and should only be regarded as qualitative
guides to what can happen.
It is up to experimentalists
to determine whether or not such phenomena occur;
detection in a heavy ion collision
would imply an out of equilibrium chiral transition.}
\vfill
\centerline{\normalsize To appear in Quark-Gluon Plasma 2, edited by
R. Hwa, World Scientific, 1995.}
\vfill
\centerline{\normalsize April, 1995. \hfill HUTP-95-A013}
\vfill
\pagebreak

\normalsize\baselineskip=15pt
\setcounter{footnote}{0}
\renewcommand{\thefootnote}{\alph{footnote}}

\section{Introduction}

Shortly after the discovery that QCD is asymptotically free\cite{asympfree}
and that
therefore quarks are almost noninteracting at short distances,
Collins and Perry\cite{collinsperry}
noted that this means that QCD at high temperature
or high density is qualitatively different than at low temperature
and density.
At high temperatures ($T\gg \Lambda_{{\rm QCD }}$)
or high baryon number density ($n\gg \Lambda_{{\rm QCD }}^{-3}$)
QCD describes a world of weakly interacting quarks and gluons very
different from the hadronic world in which we live.
This raises the possibility of phase transitions as the
temperature or density is increased.  While the subject
of phase transitions as a function of increasing density has interesting
applications in the physics of neutron stars and heavy ion collisions,
in this article I will limit myself to discussing the physics of
a QCD plasma with zero baryon number density.  There are at least
two qualitative differences
between such a plasma at $T\gg \Lambda_{{\rm QCD }}$
and at $T\sim 0$.  First, as Collins and Perry discussed, at low temperatures
one has a plasma of quarks and gluons
confined in mesons and baryons, while
at high temperatures the quarks and gluons are deconfined.  Second,
at low temperatures a $q \bar q$ condensate spontaneously breaks
chiral symmetry while at high temperatures the interactions
among the quarks and anti-quarks are weak, no such condensate exists,
and chiral symmetry is manifest.  Thus, in studying the phase structure
of QCD as a function of increasing temperature, there are two
possible phase transitions to consider.

Studying the QCD phase transition(s) has a certain intrinsic appeal.
The question involved is simple enough that it can be stated
in a sentence:  What happens if you take a chunk of any substance
around you and heat it up until it is one hundred thousand times hotter
than the centre of the sun?  While the question is simple,
if one tries to answer it experimentally by colliding
heavy ions at ultra-relativistic energies, the phenomena
seem at first glance complicated and intractable.  And yet,
using techniques from different areas of theoretical physics ---
quantum field theory, particle physics,
nuclear physics, lattice gauge theory,
condensed matter physics, and cosmology --- much progress is possible.
Quantitative predictions for processes in which
a plasma cools through the QCD phase transition in thermal
equilibrium can be extracted, and
striking phenomena with qualitative signatures
may occur if the plasma is far from equilibrium
as it cools through the transition.

In addition to having intrinsic appeal, the QCD phase transition
is of interest from several different points of view.
First, there can be no doubt that it occurred throughout the universe
about a microsecond after the big bang.
Second, it is reasonable to hope that in a heavy ion collision of
sufficiently high energy, a small region of the high temperature phase
is created which then cools through the phase transition.
Third, lattice gauge theory is well suited to calculating the equilibrium
properties of QCD at high temperatures.
{}From all these perspectives, it is important to learn as much as
can be learned analytically about the phase transition, relying as much
as possible only on fundamental symmetries and universality arguments and
as little as possible on specific assumptions and models.
In Chapter 2, following original work done with Wilczek,\cite{us}
we will see that it is possible
to use this approach to learn much about the equilibrium QCD phase transition.
In studying heavy ion collisions, however, it is worth considering
the possibility that the transition process may be far from thermal
equilibrium.  Without thermal equilibrium, it is much harder
to be quantitative both because the phenomena are more varied and
because there is no notion of universality.  We will nevertheless
find that the distinctive qualitative signature of the
QCD phase transition in heavy ion collisions discussed in Chapter 3 ---
the amplification of long wavelength oscillations of the
pion field --- can only occur if the plasma is
far from  thermal equilibrium as it experiences the
QCD phase transition.\cite{quenchpaper}

The physics of QCD at high temperatures is an enormous subject
with many facets, which I will not attempt to review here.
The reader interested in entering the voluminous literature
might consider starting with the review by Gross,
Pisarski, and Yaffe,\cite{gpy}
the review by McLerran,\cite{mclerran}  the book by Shuryak,\cite{shuryak}
the previous collection of articles edited by Hwa,\cite{hwa}
recent reviews by M\"uller,\cite{muller}
recent conference proceedings in the
Quark Matter series,\cite{qmi,qmii,qmiii,qmiv}
or other articles in this volume.

The remainder of this introduction has two goals.  First,
a brief discussion of the confinement/deconfinement
phase transition seems in order,
since the bulk of this article is concerned with the chiral
phase transition.
Second,
I will give a very brief introduction
to some of the physics relevant for a discussion of the QCD
phase transition in
heavy ion collisons,  in order to set the
stage for the third chapter.

Let us recall the
arguments\cite{svetitsky} which lead to the conclusion that the
confinement/deconfinement phase transition is first order
in pure $SU(3)$ gauge theory and is smoothed into a crossover when
light quarks are present.
In studying $SU(N)$ gauge theory in thermal equilibrium at a
temperature $T$, the partition function is given by
a Euclidian path integral
\begin{equation}
Z= \int_{\rm PBC} {\cal D} A_\mu \exp (-S_E)
\label{inta}
\end{equation}
over a slab of spacetime which is infinite in the three spatial
directions and is $\beta \equiv 1/T$ thick
in the $t$ direction.  $A_\mu \equiv A_\mu^a \lambda^a$
where the $\lambda^a$ are representation matrices for
the fundamental representation of $SU(N)$.
The path integral
is over gauge field configurations which satisfy periodic boundary
conditions $A_\mu(\vec x,0) = A_\mu(\vec x,\beta)$.
The Euclidian action is given by
\begin{equation}
S_E = {1 \over 4g^2} \int_0^\beta {\rm d}t \int {\rm d}^3\vec x~
{\rm tr} F_{\mu \nu}F_{\mu \nu}  ~~.
\label{intb}
\end{equation}
The expectation value of operators is given by
\begin{equation}
\langle {\cal O}(A_\mu) \rangle = Z^{-1} \int_{\rm PBC} {\cal D}A_\mu
\exp (-S_E) {\cal O}(A_\mu) ~~.
\label{intc}
\end{equation}
We will be particularly interested in the Wilson loop operator
\begin{equation}
L(\vec x) = {\rm tr} ~\Omega(\vec x) ~;~~~
\Omega(\vec x) = P \exp \Bigl[ i \int_0^\beta {\rm d}t A_0(\vec x ,t) \Bigr]
\label{intd}
\end{equation}
for a
loop at fixed $\vec x$ which loops around the periodic time
direction once.  $P$ denotes path ordering.
This operator, conventionally called the Polyakov loop,
is of interest because the free energy of a single static
quark is
\begin{equation}
\beta (F_q - F_0) = - \log \langle L(\vec x) \rangle
\label{inte}
\end{equation}
where $F_0 = - T \log Z$ is the free energy of the vacuum.
If $\langle L \rangle=0$,
inserting a quark into the ensemble costs an infinite free energy.
This implies that the
flux cannot be screened --- the state of the
$A_\mu$ fields is changed all the way out to spatial infinity
and costs an infinite free energy.
Therefore, $\langle L \rangle =0$ indicates that the theory is
in the confining phase.
If $\langle L \rangle \neq 0$, then $F_q$ is finite and the
theory is in a deconfined phase characterized by Debye screening.

Under a gauge transformation
\begin{equation}
A_\mu \rightarrow U A_\mu U^{-1} + i U \partial_\mu U^{-1}~,~~~
U(\vec x,t) \in SU(N)
\label{intg}
\end{equation}
$L$ transforms as
\begin{equation}
L(\vec x) \rightarrow {\rm tr} U(\vec x,0) \Omega(\vec x) U(\vec x,\beta) ~,
\label{inth}
\end{equation}
so $L$ is invariant when $U$ is periodic.
Local observables are invariant under those
aperiodic gauge transformations
\begin{equation}
U(\vec x,t+\beta)=zU(\vec x,t)
\label{inti}
\end{equation}
where $z=\exp(2\pi i n/N)$ is an element of the $Z(N)$
centre of the gauge group.  This is not true of the
Polyakov loop, which transforms under (\ref{inti}) according
to
\begin{equation}
L\rightarrow zL ~~.
\end{equation}
Therefore, $\langle L \rangle$ is an order parameter.
Going from $\langle L \rangle =0$ to $\langle L \rangle \neq 0$
requires the spontaneous breaking of a symmetry --- in this
case, spontaneous breaking of $Z(N)$.
Hence, the realization of the $Z(N)$ symmetry directly
determines whether or not the $SU(N)$ gauge theory is
in a confining phase.

The confinement/deconfinement phase transition as a function
of temperature in $SU(N)$ gauge theory is a transition
in which a global $Z(N)$ symmetry is spontaneously broken.
If the order parameter $\langle L \rangle$ varies continuously
(but non-analytically) from $\langle L \rangle =0 $ for
$T<T_c$ to $\langle L \rangle \neq 0$ for $T>T_c$, then
the transition is second order.  If, on the other
hand, $\langle L \rangle$ is discontinuous at $T_c$,
the transition is first order.  Following Landau,\cite{landau}
we write an effective Lagrangian for $L$ at long length scales whose
terms are determined by the symmetry of the order parameter
\begin{equation}
S_{\rm eff} = \int {\rm d}^3\vec x \bigl[ (\partial_i L)^2 + V(L) \bigr]
\label{intk}
\end{equation}
where the effective potential is
\begin{equation}
V(L) = f\bigl( |L|^2\bigr) + g\bigl(|L|^2, {\rm Re}(L^N) \bigr)~~.
\label{intl}
\end{equation}
In general, $f$ and $g$ are arbitrary functions.
To look for a second order phase transition, we expand
them about the origin giving
\begin{equation}
V(L) = a |L|^2 + b |L|^4 + c {\rm Re}(L^3) + \cdots
\label{intm}
\end{equation}
where $a$, $b$, and $c$ are functions of temperature
and we have specialized to $SU(3)$.
Taking $b>0$, we see that if $c$ were zero,
the minimum of the potential would increase continuously
from $L=0$ to $L\neq 0$ as $a$ went through zero.
However, with the $c$ term present $|L|$ will {\it always}
jump discontinuously away from zero as $a$ is lowered.
In fact, no cases are known in which there is a second
order phase transition involving an order parameter whose
symmetries allow a cubic invariant in the free energy.
The cubic term which is present because the symmetry is $Z(3)$
forces the phase transition to be first order.

The QCD phase transition is an inherently non-perturbative
phenomenon.   Perhaps the best long term strategy
for studying it in full (non-universal) detail
from first principles is to use lattice gauge theory.
Following Wilson,\cite{wilson} the theory can
be formulated on a lattice and the
Euclidian path
integral (\ref{inta}) can be evaluated using Monte Carlo
techniques.\cite{montecarlo}
This involves evaluating expectation values of operators by
importance sampling --- one chooses configurations
weighted by a known probability measure --- and is therefore
limited to studying equilibrium situations in which no
time evolution occurs.
The prediction of
Svetitsky and Yaffe described above that the pure $SU(3)$
phase transition is first order has been confirmed in lattice
gauge theory simulations.\cite{kogut}

Including dynamical quarks in the partition function
changes the situation entirely.
$\langle L \rangle$ is nonzero even in the confined
phase.  This can be understood heuristically by noting
that a static quark source can always be screened by virtual
$q \bar q$ pairs in the ensemble.  It can also be understood
more formally\cite{svetitsky} by showing that after
integrating out the fermions, the effective action
for the gauge fields includes a term which breaks the $Z(N)$
symmetry.  No order parameter is known for the confinement/deconfinement
transition in $SU(N)$ gauge theory when dynamical fermions are present.
This would seem to preclude the existence of a second order transition.
A first order transition is logically possible, and
presumably is obtained for heavy quarks.  For
two species of light quarks,
on the other hand, lattice gauge theory
simulations\cite{oldgott,bernard,columbia,fukugita,gottlieb,zhu,iwasaki,b2,latreviews}
show no signs of
a first order deconfinement transition.  $L$ increases smoothly as
a function of increasing temperature
and no discontinuities are apparent.  As in the ionization of
a gas, there is no sharply defined transition, but rather
a smooth change from one regime to another.

Arguing only from fundamental symmetries, we have
found that the phase transition is first order in pure gluon QCD.
We found an order parameter, but its symmetry precluded
a second order transition.  In the presence of dynamical
quarks, we were unable even to find an order parameter.
Had we found a second order transition, we could do much more.
Powerful renormalization group techniques could be brought to bear
on the problem, superceding the mean field analysis above.
When there
are massless quarks present, while there is no order parameter
for the confinement/deconfinement transition, there is
an order parameter for
the chiral phase transition.  In Chapter 2,
following arguments first advanced by Pisarski
and Wilczek\cite{piswil}
which are similar in philosophy
to those above, we will discover that for two species of massless
quarks the chiral phase transition {\it can} be second order.
This will have many implications and quantitative consequences.

All the results of Chapter 2
describe quantities
observable in thermal equilibrium where there is no
time evolution.  As we have discussed, this makes them ideal
for comparison with lattice ``experiments''.   Working on the
lattice, one has more freedom to adjust the parameters
of the theory than in the real world.  ``Experimentalists''
can choose what quark masses to use, while experimentalists cannot.
We will see that this is unfortunate, because in order to
observe the critical phenomena discussed in
Chapter 2 on the lattice or in the real world, it will be
desirable to use smaller up and down quark masses than
nature does.  This is possible on the lattice,
although difficult in practice in current simulations.  The future
for ``experimentalists'' is bright.

What about experimentalists?  Experimentalists hope to study
the physics of the QCD phase transition in relativistic heavy
ion collisions.  We will argue in Chapter 2 that if
the plasma in these events stays in equilibrium, it is unlikely
that the chiral transition will have a dramatic signature.
Of course, it is still possible to hope to detect the
difference between $T\gg T_c$ and $T\ll T_c$.  It is just
that the physics at $T\sim T_c$ will be smooth.
In studying heavy ion collisions,  however, we really should
be open to the possibility that the plasma cools through
the transition without staying in thermal equilibrium.  This will
be our concern in Chapter 3.  We will find that it is possible
that long wavelength oscillations of the pion field are
excited, leading to the production of clusters of pions
within which the number ratio of neutral pions is
far from 1/3.  This is a potentially dramatic signature
of the chiral phase transition in relativistic heavy ion
collisions in which thermal equilibrium is not maintained.

As previously mentioned,
the second goal of this introduction is to sketch some standard ideas
about what may happen in ultra-relativistic
heavy ion collisions.
We will not review this enormous field comprehensively.
References to reviews wherein the interested reader
can find both discussions of greater depth and further
references will be given. The reader should also
consult other articles in this volume.

Let us recall
a few qualitative
facts which are known empirically from studying high energy proton-proton
collisions with multiple production of secondary
particles,\cite{feynman,perkins} and which are presumed to apply to
high enough energy collisions
between nuclei also.\cite{bjorken,salmeron}
The momentum distribution of the secondary particles which
could in general be a function of the longitudinal and transverse
momenta of the particles ($p_L$,$p_T$) and the centre of
mass energy of the incident protons ($\sqrt{s}$)
is in fact observed to be independent of $\sqrt{s}$ for $\sqrt{s}$ larger
than about $10~{\rm GeV}$
and the invariant cross-section factorizes according to
\begin{equation}
\frac{E{\rm d}^3 \sigma}{ {\rm d}^3p}
= \frac{E{\rm d}^3 \sigma }{ {\rm d}^2 (p_T) ~{\rm d}p_L}
= F_1(x) F_2(p_T)~~.
\label{intn}
\end{equation}
$E$, $\vec p$, and $m$ are the energy, momentum and mass of a particular
secondary particle.
The Feynman $x$ variable is given by
\begin{equation}
x = \frac{p_L }{ p_{L~{\rm max}}} \simeq \frac{2 p_L }{ \sqrt{s}}\ .
\label{intp}
\end{equation}
The transverse momentum distribution $F_2$ falls rapidly
with increasing $p_T$ and the mean transverse momentum
is $p_T \sim 0.35~ {\rm GeV}$.\cite{perkins}  (Events
in which hard parton-parton scatterings give rise to
jets with large $p_T$ occur about once per million
collisions.\cite{perkins}  Most high energy proton-proton
collisions yield secondary particles with small transverse momenta.)
To discuss the longitudinal momentum distribution it
is useful to define a variable $y$, called rapidity, where
\begin{equation}
y \equiv \frac{1}{2} \log \Biggl( \frac{E + p_L }{ E - p_L} \Biggr)
= \log \Biggl( \frac{E + p_L }{ \sqrt{p_T^2 + m^2} } \Biggr) ~.
\label{intq}
\end{equation}
One reason why $y$ is convenient is that boosting in the
beam direction corresponds simply to adding a constant to $y$.
Another reason is that for $p\gg m$,
\begin{equation}
y\simeq \eta \equiv - \log \tan \theta /2
\label{intqaa}
\end{equation}
where the pseudorapidity $\eta$ is easily measured since it
depends only on the polar angle $\theta$.
A third useful property of $y$ is that
\begin{equation}
{\rm d}y = \frac{{\rm d}p_L }{ E}~.
\label{intqa}
\end{equation}
We will work in the centre of mass frame,
where $p_L = x = 0$ corresponds to $y=0$ and $p_L = \pm p_{L~{\rm max}}$
corresponds to $y = \pm y_{{\rm max}}$ where
\begin{equation}
y_{{\rm max}} = \frac{1}{2} \log \Biggl( \frac{s}{p_T^2 + m^2} \Biggr)~.
\label{intr}
\end{equation}
Feynman postulated\cite{feynman}
that at large $s$ there is a central rapidity region
($|y|\ll y_{\rm max}$, or, equivalently,
$|p_L|\ll p_{L~{\rm max}}$) where $F_1(x)$ is a constant.
In this region,
\begin{equation}
E{\rm d}^3 \sigma = F_1 ~F_2(p_T)~ {\rm d}^2 (p_T) {\rm d}p_L
\label{ints}
\end{equation}
and, integrating over $p_T$ and using (\ref{intqa}),
\begin{equation}
\frac{{\rm d}\sigma }{ {\rm d}y} = {\rm constant} ~.
\label{intt}
\end{equation}
Thus, there is a central rapidity region in which the multiplicity per
unit rapidity is approximately constant. This result has indeed been
seen in proton-(anti)proton collisions.\cite{perkins}
The qualitative features
found in high energy hadron-hadron collisions --- low $p_T$ and
a central rapidity plateau --- are expected\cite{bjorken} to
characterize ultra-relativistic nucleus-nucleus collisions also.

In a relativistic heavy ion collision of sufficiently high
energy, the decay products divide naturally into two regions.
In the centre of mass reference frame,
the fragments of the incident nuclei are found at high (positive
and negative) rapidity, while the central rapidity region
is characterized by approximately constant multiplicity per
unit rapidity and by
low baryon number density.\cite{bjorken}
At early times, this central rapidity region
experiences boost invariant longitudinal expansion\cite{bjorken} in
which the momentum and position
of particles are related in such a way that the rapidity is
given by
\begin{equation}
y=\frac{1}{2}\log\left(\frac{t+z}{t-z}\right)~,
\label{posrap}
\end{equation}
where $t$ is the time and $z$ is the position coordinate
in the beam direction.
Mean physical quantities depend only on the proper time
\begin{equation}
\tau\equiv\sqrt{t^2-z^2}
\label{taudef}
\end{equation}
and transverse position, and are independent of $y$.

Describing the
fragmentation regions requires consideration of QCD at nonzero
temperature and baryon number density, and is not discussed
in this article.  We limit ourselves to addressing the physics
of the ``hot vacuum'' in the central rapidity region.
How high
an energy is sufficient for there to be a reasonably clean
separation between distinct
fragmentation and central rapidity regions?
The parameter of interest is $\delta y$, the
extent in rapidity of one of the fragmentation regions.
$\delta y$ is not well known --- estimates
range from 4 to 8.\cite{satz}
Even if $\delta y \sim 4$,
present fixed target heavy ion collision experiments at the
AGS and SPS do not access sufficient rapidities for there to
be a baryon poor central rapidity region.  At the AGS, $y_{\rm max} \sim 2$
and at the SPS $y_{{\rm max}}\sim 3$.  In the coming  generation
of colliding beam heavy ion accelerators, the situation will be much
improved.  At the Relativistic Heavy Ion Collider currently
under construction at Brookhaven National Laboratory,
it is expected that Au-Au collisions with $\sqrt{s}=200~{\rm GeV/A}$
with multiplicities of at least 1400 per unit
rapidity will be achieved.\cite{satz}
When the LHC is built at CERN and is run as a Pb-Pb collider,
collisions with $\sqrt{s}=6300~{\rm GeV/A}$ and with multiplicities
of at least 2600 per unit rapidity
will be possible.\cite{satz}  These multiplicity estimates
are qualitative --- they are based
on extrapolating from data from proton-nucleus collisions, and
results from event generators suggest that they could
be too low.\cite{satz}
At RHIC ($y_{{\rm max}}\sim 5$) the central rapidity
region will have essentially zero baryon number density if $\delta y < 5$
and at the LHC ($y_{{\rm max}}\sim 9$) there will be such a
region for $\delta y < 9$.\cite{satz}  Even if there
are baryons in the central region at RHIC, the number ratio
of mesons to baryons is estimated to be $\sim 10^2$.\cite{satz}
(At the
LHC, this ratio is expected to be $\gg 10^5$.\cite{satz})
Thus, analyzing the physics of QCD at finite temperature and
zero baryon number density will be of relevance in describing
the physics of the central rapidity regions of heavy ion collisions
at RHIC and the LHC.

What happens in the central rapidity region?  The standard
description involves three phases --- a pre-equilibrium phase,
a hydrodynamic evolution phase during which thermal
equilibrium is assumed, and a hadronization phase.
There have been a number of different approaches to treating
the earliest phase of a heavy ion collision.
For example,
in the string model, one imagines that the incident nuclei
pass through each other and colour flux tubes, or strings, are
drawn out between the receding fragmented nuclei.  These
strings then break apart by quark--anti-quark pair production,
and realizations of the scenario are based on Monte Carlo
simulation of this process.
This scenario has been reviewed by Bia\l as and Czy\'z.\cite{bialas}
It runs into difficulties at energies high enough that
the string density becomes large.  At high enough energies,
it should be possible to describe
the earliest phase of the evolution
in terms of parton interactions by perturbative QCD.
This approach, called the parton cascade
model,\cite{geiger,biro,muller} has begun to
yield results.
One starts with distribution functions for the partons
in the nucleons of the incident nuclei,
and follows the quark-gluon interactions in the framework
of renormalization group improved perturbative QCD by solving
transport equations for the parton distributions in phase-space
with Monte Carlo
methods.  The results of this work give support at the parton
level to the idea that a thermalized state is formed.
Quarks thermalize more slowly than gluons because
the relevant QCD cross-sections are larger by a factor of two or
three.\cite{twostage}
Also, for both quarks and gluons, a thermal
distribution of momenta is achieved more quickly than
chemical equilibrium (equilibrium number density) is achieved.
(It is not clear whether there will be enough
time in a RHIC collision for quark chemical equilibrium
to be established.\cite{twostage,biro})
It is of interest that at the latest times considered
by Geiger\cite{geiger} ($\sim 3 ~{\rm fm}/c$), the baryon number is
in fragmentation regions of high rapidity, and the central
rapidity region contains partons with thermalized momenta which are
expanding longitudinally with little
transverse motion.
While these calculations are far from the end of the story
of the pre-equilibration physics, they do provide justification
for supposing that in an energetic enough collision (perhaps at
RHIC) there is a low baryon number central rapidity region
in which particle momenta are in local thermal equilibrium
at $T>T_c$.

If equilibrium is achieved, and while it is maintained,
the evolution of the plasma can be described hydrodynamically.
The matter is treated as a relativistic fluid described by
smoothly varying four-velocity, pressure, temperature,
and energy density without reference to the microscopic
description in terms of partons.
This approach has a long history, but perhaps the first
quantitative treatment was that of Bjorken.\cite{bjorken}
The review by Blaizot and Ollitrault\cite{hydroreview}
is a useful introduction to the subject.
Bjorken\cite{bjorken} used the observation that the central rapidity
region is approximately boost invariant to find solutions
to the hydrodynamic equations describing the longitudinal
expansion and consequent cooling of the plasma.
Blaizot and Ollitrault\cite{hydroreview} review various
numerical simulations
of the hydrodynamic equations which include transverse expansion.

In using the hydrodynamical approach one {\it assumes}
that thermal equilibrium is valid throughout the transition
and until the hadrons no longer interact.   ``This is
an assumption which is hard to justify in any satisfying
way.''\cite{hardtojustify}
It is particularly difficult to see how long wavelength degrees
of freedom can stay in equilibrium as the plasma expands
and cools through the phase transition.
Given the uncertainties, in considering the
physics of the chiral phase transition in relativistic
heavy ion collisions we will discuss both the phenomena
that would occur if the plasma stays in equilibrium and those
that could characterize a far from equilibrium transition.

In the final phase of a heavy ion collision, regardless of
what has happened earlier, hadrons, photons,  and leptons
impinge upon an experimentalist's detector.  How is the
experimentalist supposed to unravel what has gone on before?
The zeroth order question is how can she determine whether
for a brief instant a small region of quark-gluon plasma
was created?
Many signatures have been proposed for the production of
a region of quark-gluon plasma in a heavy ion collision
and its consequent passage through
the QCD phase transition.
All have problems to overcome, arising either from background,
or from insufficient theoretical understanding, or from
the possibility that the signature can be mimicked by hadronic
processes.  This suggests that
the QCD phase transition will be detected by a process where
various
pieces of
evidence become more and more convincing rather than by
the observation of a ``smoking gun'' signature.
If the speculations of Chapter 3 are correct, however,
the situation could be improved.  We will see
that if the chiral phase transition occurs by a process that
is far enough from thermal equilibrium that it can be
idealized as a quench, then coherent long wavelength
oscillations of the pion field may emerge.  These would
manifest themselves as clusters of pions
with similar rapidity in which the number ratio of neutral pions
is dramatically different from 1/3.
If seen, this phenomenon would be a distinctive, qualitative
signature that a far from equilibrium chiral transition
had occurred.

\section{Critical Phenomena at a
Second Order QCD Phase Transition}

The QCD phase transition has been studied from many different perspectives
using many different models and approximations.  Here, however,
we will attempt to learn as much as
can be learned analytically about the phase transition, relying as much
as possible only on fundamental symmetries and universality arguments and
as little as possible on specific assumptions and models.
This philosophy yields little fruit when applied to the
confinement/deconfinement transition.  We saw in the previous chapter
that in QCD with light quarks, this transition is
neither first order nor second order --- it is a smooth crossover
like the ionization of a gas.
The chiral phase transition is another story.
Wilczek\cite{wilczek} has
emphasized that in the chiral limit
where there are two species of quarks with zero current algebra mass, the
order parameter for the chiral phase transition has the same symmetry as
the magnetization of a four component Heisenberg magnet, which
has a second order phase transition.
At an equilibrium second order phase transition, thermodynamic quantities
have non-analytic behaviour determined by the physics of the
modes which develop long correlation lengths at the critical point.
Thus, by studying these universal characteristics in a much simpler
model --- the $N=4$ Heisenberg magnet --- it is possible to learn
much about the chiral phase transition in QCD.

In this chapter, which to a large extent follows
work done with Wilczek,\cite{us}
we explore the consequences of this approach
to the QCD phase transition in thermal equilibrium.
In the following section, we use renormalization group arguments
to
establish a dictionary between QCD and
the magnetic system.  Using this dictionary, we apply results
for the static critical exponents obtained in the magnet model to QCD.
In section 2.2, we discuss
the behaviour of the pion and sigma masses at the transition.
In section 2.3, we review recent work in lattice gauge theory
which has begun to test these ideas.
In section 2.4, we consider how the
strange
quark affects the phase transition, and again compare
with lattice results.
In the concluding
section of this chapter, we discuss the
implications of all this for cosmology and lattice gauge theory,
and for heavy ion collisions under the assumption that the
chiral order parameter
remains close to thermal equilibrium through the phase transition.

\subsection{QCD and the $O(4)$ Magnet}

The physics of the QCD phase transition is
qualitatively different in the cases of zero, one, two, or three or more
flavours of quarks.\cite{wilczek}
In this section we consider QCD with two species
of quarks.  Pisarski and Wilczek\cite{piswil} used an analysis
similar to the one
which follows to show
that for three or more flavours of massless quarks, the
chiral phase transition is first order.  In section 2.4 we will
discuss the effects of a third quark species, massless or massive, and
for now, therefore, we postpone our discussion of the effects of
the strange quark.
If there are two flavours of massless quarks, the
QCD lagrangian is symmetric under
global chiral transformations in the group $SU(2)_L \times SU(2)_R \times
U(1)_{L+R}$ of independent special unitary transformations of the left and
right handed quark fields, and a vector $U(1)$ transformation which
corresponds to baryon number symmetry.  (The axial $U(1)$ which would make
the symmetry group into $U(2) \times U(2)$ is a symmetry of the classical
theory, but not of the quantum theory.\cite{thooft})
This chiral symmetry
breaks spontaneously down to $SU(2)_{L+R} \times U(1)_{L+R}$
at low temperatures, and is restored at sufficiently high temperatures.
The order parameter for this phase transition is the expectation value
of the quark bilinear
\begin{equation}
{\cal M}^i_j ~\equiv ~ \langle  {\bar q_L}^i {q_R}_j \rangle
\label{ab}
\end{equation}
which breaks the symmetry when it acquires a nonzero value below some
critical temperature $T_c$.

At a second order phase transition, the system is at
an infrared fixed point of the renormalization group.
Renormalization group transformations correspond to
coarsening one's view of the world.  Thus, at an infrared
fixed point, physics is scale invariant.  The order parameter
fluctuates
on all length scales and, in particular, on arbitrarily long
length scales.  The correlation length associated with
these scale invariant fluctuations is necessarily infinite,
and the correlation function which describes them is a power law.
Our goal is not to describe all of the physics occurring
at the transition.  Rather, we limit our attention
to the nonanalyticities of thermodynamic quantities.
The partition function $Z={\rm Tr}e^{-H/T}$ is the sum
of analytic terms, and so in a system with finitely many
degrees of freedom
must be an analytic
function of the temperature.
Nonanalyticity can only arise in the infinite volume limit,
and so by limiting the scope of our treatment in this way
we concentrate on the physics of the long wavelength
degrees of freedom.
At an infrared fixed point,
renormalization ({\it i.e.} coarsening one's view)
does not change the physics of
the long wavelength fluctuations of the order parameter,
which is therefore independent of all short wavelength physics.
The nonanalytic critical phenomena are universal: they
only depend on the modes with long wavelength fluctuations
(and these are determined by the symmetries of the order parameter)
and on the scaling behaviour near the fixed point of low dimension operators
constructed from these modes.

In order to describe a second-order transition
quantitatively, we must find a tractable model in the
same universality class.  For the chiral order parameter
(\ref{ab}) the relevant symmetries are independent unitary transformations
of the left- and right-handed quark fields, under which
\begin{equation}
{\cal M}~\rightarrow~ U^\dagger {\cal M} V ~.
\label{ba}
\end{equation}
These transformations generate an
$U(2)_L\times U(2)_R $ symmetry, which is not quite what is needed,
since it includes the axial baryon number symmetry which is not present
in QCD.  This problem is solved\cite{wilczek} by restricting ${\cal M}$ to
multiples of
unitary matrices with positive determinant, instead of general complex
matrices.  Matrices ${\cal M}$ in this restricted class remain in this
restricted class under the transformation (\ref{ba}) only if $U$ and $V$ have
equal phases.  Hence, the axial $U(1)$ has indeed been removed.  The
$2 \times 2$ matrices ${\cal M}$ can conveniently be parametrized
using four real parameters $(\sigma ,\vec \pi)$ and the Pauli matrices as
\begin{equation}
{\cal M}~=~ \sigma + i\vec\pi \cdot \vec\tau~.
\label{bc}
\end{equation}
In fact the order parameter can be written as
a four-component vector
\begin{equation}
\phi^\alpha  \equiv (\sigma, \vec \pi)
\label{bca}
\end{equation}
and the transformations (\ref{ba}) are simply $O(4)$ rotations
in internal space.  Hence, the order parameter appropriate for the chiral
phase transition in QCD with two flavours of massless quarks has the
symmetries of the standard $O(4)$ invariant
$N=4$ Heisenberg magnet.  For smaller number of components,
this sort of model is a much-studied model for the critical
behaviour of magnets, with the order parameter representing the
magnetization of a ferromagnet or the staggered magnetization
of an antiferromagnet.

In order to discuss the universal critical phenomena,
it is sufficient to retain those degrees of freedom
which develop large correlation lengths at the critical point.
These are just long wavelength
fluctuations of the order parameter, which is small in magnitude near
the critical point and therefore fluctuates at little cost in energy.
Thus, the most plausible
starting point for analyzing the critical behaviour of a
second-order chiral
phase transition in QCD is the Landau-Ginzburg free energy
\begin{equation}
F = \int d^3 x \Biggl\lbrace \frac{1}{ 2}~\partial^i \phi^\alpha
     \partial_i \phi_\alpha
    ~+~ \frac{\mu^2}{2} ~\phi^\alpha \phi_\alpha ~+~
    \frac{\lambda}{4}(\phi^\alpha \phi_\alpha)^2
    ~\Biggr\rbrace ~.
\label{bb}
\end{equation}
Here $\mu^2$ is the temperature dependent
renormalized mass squared.  $T_c$
is the temperature at which $\mu^2$ vanishes.  $\mu^2$
is positive above the critical temperature and negative below it.
The quartic coupling $\lambda$ depends smoothly on the temperature.
In this section we consider the case in which $\lambda$ is positive;
we will discuss what happens if $\lambda$ is not positive at $T_c$
in section 2.4.
We neglect higher dimension operators, because they are irrelevant
at an infrared fixed point --- their coefficients go to zero upon
renormalization toward the infrared.\footnote{~$\phi^6$ is not
irrelevant in 3-dimensions, but as long as $\lambda$ is positive
at $T_c$, $\phi^6$ can be neglected relative to $\phi^4$
since $|\phi|$ is small near the transition.
The $\phi^6$
term will be important in section 2.4.}~~The
symmetry breaking pattern we want is $\langle {\cal M} \rangle
\propto {\bf 1}$ (equivalently,
$\langle \sigma \rangle \neq 0; \langle \vec \pi \rangle = 0$)
below the
transition which is indeed what we find at the minimum of the
potential for positive $\lambda$.  This model has been studied in depth
for arbitrary $N$ and spatial dimension $d$, and the existence of
an infrared stable fixed point of the renormalization group has been
established.\cite{amitma}  Hence, it is a model for a second order QCD chiral
phase transition for two massless quarks.

Before proceeding, let us briefly outline the
the physics of the theory given by (\ref{bb})
in thermal equilibrium
above, below, and at the critical temperature at which
the second order phase transition occurs.
At temperatures above $T_c$, the direction in internal space
in which $\phi$
is oriented is correlated only over
spatial distances of order a finite correlation length;
over longer distances all orientations of $\phi$
are equally probable. Thus, $\phi$ is disordered and $\langle \phi \rangle=0$.
At temperatures below $T_c$, $\phi$ is ordered: $\langle \phi \rangle$
is nonzero and selects a particular direction (defined
to be the $\sigma$ direction) in internal space.
Whereas the theory has an explicit $O(4)$ symmetry,
the equilibrium configuration is not $O(4)$ symmetric ---
the symmetry is spontaneously broken.
Fluctuations of $\phi$ about $\langle \phi \rangle$
in the $\sigma$ and $\vec \pi$ directions are described by
differing correlation functions.  The explicit $O(4)$ symmetry implies
that $\langle \phi \rangle$ could choose any direction in
internal space.  Below $T_c$, therefore, fluctuations of $\phi$
in the $\vec \pi$ directions
(which correspond to fluctuations of the orientation of
$\langle \phi \rangle$) are massless
Goldstone modes.
At $T=T_c$, $\langle \phi \rangle =0$,
the correlation length is infinite, and the
fluctuations of $\phi$ are scale invariant.

We have established
that there is an infrared fixed point of the renormalization group
in the same universality class
as QCD with two massless quarks.
The theories in this universality class live in an infinite dimensional
parameter space spanned by $\mu$, $\lambda$, and
the coefficients of an infinite number
of irrelevant higher dimension operators.  The second order fixed point
has some basin of attraction in this space of theories.  For the rest of this
chapter, we will explore the hypothesis that QCD is indeed in the
basin of attraction of the infrared fixed point.
We will
calculate much about the QCD phase transition based
on this assumption.  We will also be show that it is
consistent with many phenomena found
to date in lattice gauge theory simulations.
We can never be sure, however, that upon renormalization toward
the infrared QCD is in fact driven to this fixed point.
It is
always a logical possibility that, say, QCD may live in a region
of theory space from which infrared renormalization drives
the theory into a region of negative $\lambda$, making the phase transition
first order.
Hence, the strongest statement that it is possible to make is that it
is plausible that in QCD with two species of massless quarks
the chiral phase transition is second order.
Our strategy, then, is to establish quantitative
consequences of this hypothesis and test them against lattice
simulations of QCD.

When the free energy (\ref{bb}) is written in terms of $\sigma$ and $\vec \pi$
it looks much like the original model of Gell-Mann and L\'evy\cite{gm}
with two
changes: there are no nucleon fields and only three (spatial) dimensions.
These two changes reflect an important distinction.\cite{ginsparg}
We are only proposing
(\ref{bb}) as appropriate near the second order phase transition point.  This
is because it is only there that we can appeal to universality --- the
long-wavelength
behaviour of the $\sigma$ and $\vec \pi$ fields is determined by
the infrared fixed point of the renormalization group, and microscopic
considerations are irrelevant to it.
In Euclidean field theory at finite temperature, the integral over $\omega$
of zero temperature field theory is replaced by a sum over Matsubara
frequencies $\omega_n$ given by $2n\pi T$ for bosons and $(2n+1)\pi T$ for
fermions with $n$ an integer.  Hence, one is left with a Euclidean theory
in three spatial dimensions with massless fields from the $n=0$ terms
in the boson sums and massive fields from the rest of the boson sums and
the fermion sums.   Hence, to discuss the massless modes of interest at
the critical point, (\ref{bb}) is sufficient and
we do not need to
introduce nucleon fields or constituent quark
fields.  In arriving at an effective three dimensional
theory of the long wavelength fluctuations of the order parameter
near $T_c$, all other bosonic degrees of freedom and all
fermionic degrees of freedom have been integrated out.

We have motivated a very definite hypothesis for the nature of the
phase transition for QCD with two species of massless quarks, namely that
it is in the universality class of the $N=4$ Heisenberg magnet.
This means that under renormalization toward the
infrared, QCD with two massless quarks is driven to
the infrared fixed point of (\ref{bb}) and that
therefore the nonanalytic behaviour of thermodynamic
quantities near $T_c$ is the same in the two theories.
This hypothesis has numerous consequences which are the subject
of the rest of this chapter.
In the
remainder of this section, we review the
resulting predictions for the
critical exponents.\cite{wilczek}

First, we define
the reduced temperature $t~=~(T-T_c)/T_c$.
The exponents $\alpha$,  $\beta$, $\gamma$, $\eta$, and $\nu$
describe the singular
behaviour of the theory with strictly zero quark masses as
$t \rightarrow 0$.
For the specific heat one finds
\begin{equation}
C(T) \sim |t|^{-\alpha} + {\rm less~ singular.}
\label{da}
\end{equation}
The behaviour of the order parameter defines $\beta$.
\begin{equation}
\langle | \phi |\rangle \sim  | t |^{\beta}~~ {\rm for}~ t<0~.
\label{dab}
\end{equation}
$\eta$ and  $\nu$ describe the behaviour of the correlation length $\xi$
where
\begin{equation}
 G_{\alpha \beta}(x)~ \equiv~
 \langle \phi(x)_\alpha \phi(0)_\beta \rangle
 -\langle \phi_\alpha \rangle \langle \phi_\beta \rangle
 ~\rightarrow~
 \delta_{\alpha \beta} \frac{A}{|x|^{(d-2+\eta )}}
\exp (-|x|/\xi) ~~{\rm for}~|x|\rightarrow \infty~.
\label{df}
\end{equation}
$A$  is independent of $|x|$, but may depend on $t$.
The correlation length diverges as $|t|\rightarrow 0$, and
the correlation length
exponent $\nu$ is defined by
\begin{equation}
\xi \sim |t|^{-\nu }~.
\label{de}
\end{equation}
Above $T_c$, where the correlation lengths are equal in the sigma and pion
channels,
the susceptibility exponent $\gamma$ is defined by
\begin{equation}
\int d^3 x~G_{\alpha \beta}(x) \sim t^{-\gamma}.
\label{deb}
\end{equation}
We will discuss the behaviour of the susceptibility below the transition
in the following section. At the critical point, the correlation
length is infinite and the correlation function is a power law.
The exponent $\eta$ is defined through the
behaviour of the Fourier transform of the correlation function at $t=0$:
\begin{equation}
G_{\alpha \beta}(k\rightarrow 0) \sim k^{-2+\eta} ~.
\label{dea}
\end{equation}

The last exponent, $\delta$, is related to the behaviour
of the system in a small magnetic field $H$ which explicitly
breaks the $O(4)$ symmetry.  Let us first show that in a QCD context,
$H$ is proportional to a common quark mass $m_u = m_d \equiv m_q$.
This common mass term may be represented by a $2\times 2$ matrix ${\cal D}$
given by $m_q$ times the identity matrix.
We are now allowed to construct the free energy from invariants involving
both ${\cal D}$ and ${\cal M}$.  The
lowest dimension term linear in ${\cal D}$
is just ${\rm tr} {\cal M}^\dagger {\cal D} = m_q \sigma$,
which in magnet language
is simply the coupling of the magnetization to an external
magnetic field $H \propto
m_q$.  In the presence of an external field, the order parameter is
not zero at $T_c$.  In fact,
\begin{equation}
\langle | \phi |\rangle (t=0,H\rightarrow 0) \sim H^{1/\delta}~.
\label{dec}
\end{equation}
Thus when the $O(4)$ symmetry is explicitly broken by a small
external magnetic field or
equivalently by small quark masses, the phase transition
is a smooth crossover.  It is, however, a very special crossover about
which much can be said, because as $H\rightarrow 0$ one approaches
the second order fixed point.

The six critical exponents defined above are related by four scaling
relations.\cite{amitma}  These are
\begin{eqnarray}
\alpha~&=&~2-d\nu \nonumber \\
\beta~&=&~{\nu \over 2} ( d - 2 + \eta ) \nonumber \\
\gamma~&=&~(2-\eta )\nu \nonumber \\
\delta~&=&~\frac{d + 2 - \eta}{d - 2 +\eta}~.
\label{db}
\end{eqnarray}
We therefore need values for $\eta$ and $\nu$ for the four component
magnet in $d=3$. These were obtained in the remarkable work of
Baker, Meiron and Nickel,\cite{bmn} who carried the perturbative
expansion in $\lambda$ for the theory (\ref{bb}) to
seven-loop order, and used information about the behaviour of
asymptotically large orders, and conformal mapping and Pad\'e
approximant techniques to obtain
\begin{eqnarray}
\eta ~&=&~~ .03 \pm .01 \nonumber \\
\nu  ~&=&~~ .73 \pm .02 ~.
\label{dc}
\end{eqnarray}
Using (\ref{db}), the remaining exponents are
\begin{eqnarray}
\alpha ~&=&~~ -0.19 \pm .06 \nonumber \\
\beta ~&=&~~ 0.38 \pm .01 \nonumber \\
\gamma ~&=&~~ 1.44 \pm .04\nonumber \\
\delta ~&=&~~ 4.82 \pm .05 ~.
\label{dba}
\end{eqnarray}
Since $\alpha$
is negative there is a cusp in the specific heat at $T_c$, rather than a
divergence.  There are other ways in which these exponents could
be calculated.  In $d=4$, mean field theory suffices
and one has $\eta=0$ and $\nu=1/2$.  We are interested in
$d=3$, however.  A standard approach\cite{wilsonfisher}
is to work in $4-\epsilon$
dimensions, evaluate the exponents order by order in
$\epsilon$, and then set $\epsilon=1$. The $\epsilon$-expansion
has not been pushed to high enough order to compete in accuracy
with the work of Baker, Meiron, and Nickel.
Another approach
is to do a finite temperature lattice simulation of the theory
(\ref{bb}) and measure the critical exponents.
This has recently been done to high accuracy by
Kanaya and Kaya,\cite{kanayakaya} who find
$\nu=0.748\pm0.009$ and $\gamma=1.48\pm.02$,
in agreement with the results of Ref. [\citelow{bmn}].

To reiterate, these exponents and the other critical phenomena
we discuss in subsequent sections are universal.  Whether
the QCD phase transition is studied
experimentally, or on the lattice, or by
instanton liquid techniques, or by using the Nambu--Jona-Lasinio model,
or by using the (four dimensional) linear or nonlinear sigma models,
or by any other means,
if the hypothesis that the chiral transition is a second order
transition in the $O(4)$ universality class is
correct the exponents should turn out to have the values we have described.
Of course, using any technique except appealing to the simplest
model in the same universality class as we have done would
make it much harder to study the theory near the critical point,
and hence much harder to obtain the critical exponents.

\subsection{The Equation of State and the Pion and Sigma Masses}

The expressions which define $\beta$, $\gamma$ and $\delta$
are actually special cases of a more general relationship between
the magnetization and the magnetic field called the critical
equation of state.  The equation of state has been calculated to
order $\epsilon^2$ by Br\'ezin, Wallace and Wilson.\cite{bww}
In this section, we review the
use of the equation of state to determine the behaviour
of the pion and sigma masses near the critical point.

First, we must define what we mean by the ``mass'' of the pion
and sigma.   We could choose either to define the mass as an inverse
correlation length or as an inverse susceptibility.  We choose the
latter, which is conventional in the condensed matter literature.
Specifically, we define
\begin{equation}
m_{\sigma}^{-2} = \int d^3 x G_{0 0}
\label{ea}
\end{equation}
and
\begin{equation}
m_{\pi}^{-2} \delta_{ij} = \int d^3 x G_{ij}
\label{eb}
\end{equation}
where $\phi_0 = \sigma$ and $\phi_i = \pi_i, i=1,2,3$.
Defining ``masses'' as inverse correlation lengths
would give different
scaling behaviours for masses as functions of $t$ and $H$.
(For $\eta=0$, however, both choices would yield the same
scaling behaviour, and in the theory of interest $\eta$
is small.)
We shall see that with the conventional choices (\ref{ea}) and (\ref{eb}),
the masses can be extracted conveniently from the equation of state.
It is worth noting that the masses we have defined are related only
to the behaviour of spatial correlation functions in the static
(equilibrium) theory.  They carry no dynamical information, and
should not be confused with, say, pole masses in a 4-dimensional
theory.
Also,
we will only be able to make universal statements about how the masses
scale at the transition.
Normalizing the magnitudes of the masses will require using some
specific model, and hence will not be universal.

The equation of state gives the magnetization as a function of $t$ and
$H$.  For the rest of this chapter, we will write the order parameter
as $M$, for magnetization, keeping in mind that $M = \langle \sigma \rangle
= \langle |\phi| \rangle $.  Also, recall that $H$, the external
field, is proportional to the quark mass $m_q$.
In order to define the equation of
state, we first define a shifted field $\tilde \sigma = \sigma - \langle
\sigma \rangle = \sigma - M$.  Then the equation of state is simply
the relation
\begin{equation}
\langle \tilde \sigma \rangle = 0\ .
\label{ec}
\end{equation}
This relation has been expanded to order $\epsilon^2$ by Br\'ezin, Wallace
and Wilson.\cite{bww}  The result
can be expressed conveniently in terms of
the variables
\begin{equation}
y \equiv H/M^\delta ~~{\rm and}~~ x\equiv t/M^{1/\beta}
\label{eca}
\end{equation}
as
\begin{equation}
y = f(x)~ .
\label{ed}
\end{equation}
\begin{figure}
\centerline{
\epsfysize=100mm
\epsfbox[72 216 540 576]{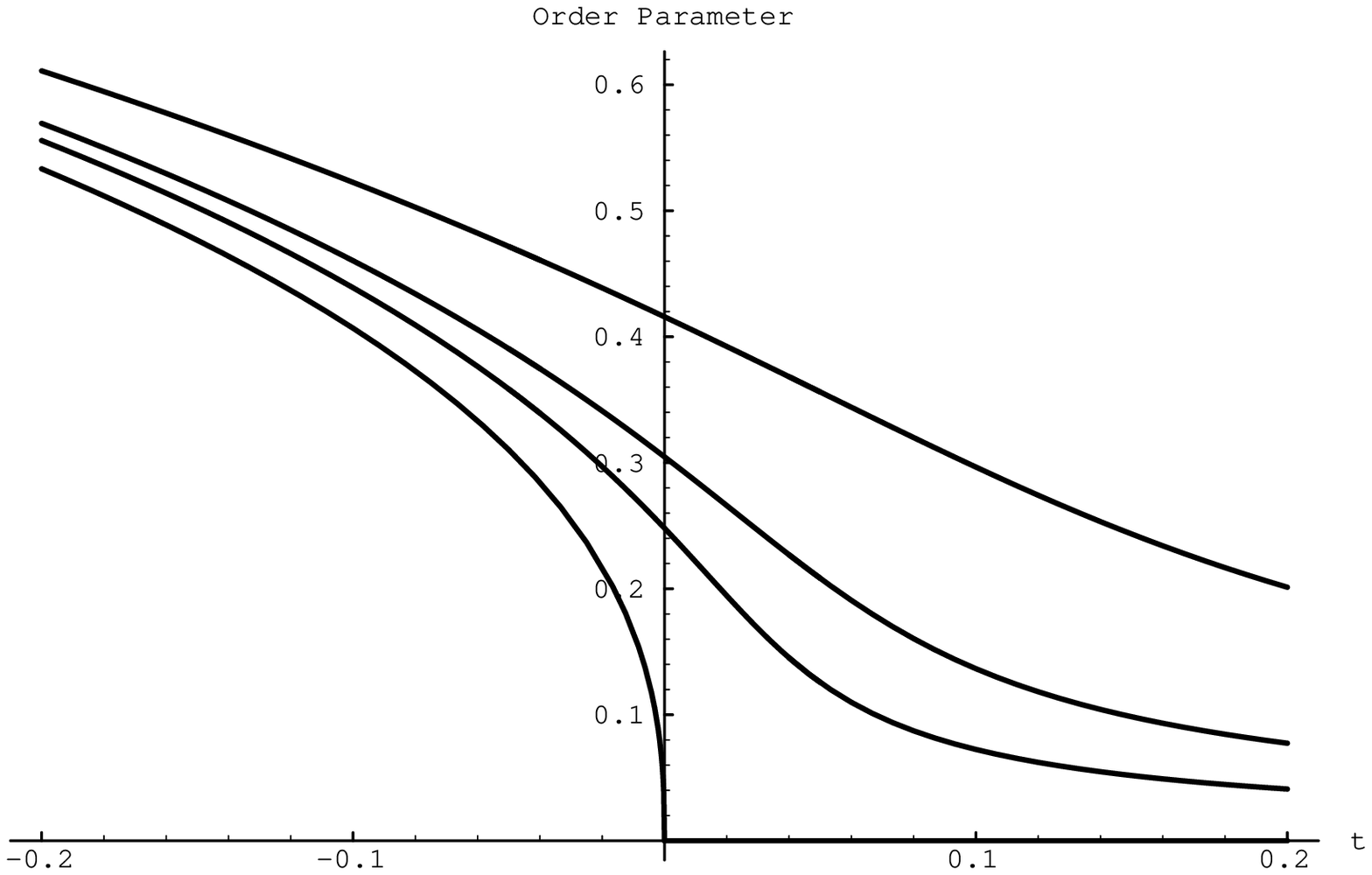}
}
\fcaption{From Ref. [\citelow{us}].
The order parameter $M$ as a function of reduced
temperature $t\equiv (T-T_c)/T_c$
for magnetic fields $H=$ 0, 0.002, 0.005, and 0.02.
Like $t$, both $M$ and $H$ are dimensionless. They are obtained from their
dimensionful counterparts by dividing by non-universal dimensionful constants
defined in such a way that
for $t<0$ and $H=0$ the order parameter is given by $M=(-t)^\beta$, and
for $t=0$ it satisfies $M=H^{1/\delta}$.}
\label{fig:orderparam}
\end{figure}
The function $f(x)$ is universal and
was calculated to order $\epsilon^2$
by Br\'ezin, Wallace, and Wilson.\cite{bww}
Their result is also quoted in Ref. [3].
The units in which $H$ and $M$ are
measured are chosen so that $f(0) = 1$ corresponds to $t=0$ and
$f(-1)=0$ corresponds to the $t<0,~H=0$ coexistence curve.
Knowing $f(x)$, we can calculate the value of the order parameter $M$
for a given $H$ and $t$ using (\ref{ed}).
The behaviour of the order parameter is illustrated in Fig. 1.
This figure and the other ones
in this section should be viewed as illustrations
of qualitative behaviour rather than quantitative predictions
because they are based on setting $\epsilon =1$
in the $O(\epsilon ^2)$ expression for $f(x)$. The values for the
critical exponents themselves which we quoted in the previous section are
quantitative predictions, complete with error estimates, because they are
based on the much more elaborate analysis of Baker, Meiron and
Nickel.\cite{bmn}

{}From the equation of state, we can deduce the behaviour of $m_\pi$
and $m_\sigma$ at nonzero (but small) $t$ and $H$.
The masses are given by
\begin{equation}
m_\sigma^2 = \frac{ \partial H }{ \partial M}
\label{ee}
\end{equation}
and
\begin{equation}
m_\pi^2 = \frac{H}{M}.
\label{ef}
\end{equation}
The first relation follows directly from the definition (\ref{ea}), and
the second follows from (\ref{eb}) and
from assuming that $\vec M \parallel \vec H$, so that a small change $\delta H
\perp H$ gives a small change $\delta M \perp M$ with $\delta M / \delta
H = M/H$.
Using the equation of state, we can rewrite (\ref{ee}) and (\ref{ef}) as
\begin{equation}
m_\pi^2 = M^{\delta - 1} f(x)
\label{efb}
\end{equation}
and
\begin{equation}
m_\sigma^2 = M^{\delta - 1} \left( \delta f(x) - \frac{x }{ \beta} f'(x)
\right)~.
\label{efc}
\end{equation}
Hence, $\delta$ can be determined by measuring the
ratio $m_\sigma^2 / m_\pi^2 $ at $t=0$.  In general, from $f(x)$ we can find
the pion and sigma masses for any $t$ and $H$.

There are two interesting limits which we will consider
explicitly.  First, for $t>0$ and $H\rightarrow 0$ which corresponds
to $x\rightarrow \infty $, we should find the full $O(4)$ symmetry,
and hence should find that the pion and sigma masses are identical.
For $x\rightarrow \infty $, the function $f(x)$
behaves as $f(x)=c x^\gamma $.  The constant $c$ is
given to $O(\epsilon)$ in Ref. [3].
Applying
(\ref{efb}) and (\ref{efc}) , we find that
\begin{equation}
m_\sigma^2 = m_\pi^2 = c t^\gamma~~{\rm for~}x\rightarrow \infty ~,
\label{eg}
\end{equation}
consistent with the symmetry.

We can also consider the limiting case of approaching the coexistence curve.
This means taking $t<0$ and $H\rightarrow 0$, which implies
$x\rightarrow -1$.  In this limit, $M$ tends to a nonzero constant,
and so from (\ref{ef}) , we obtain $m_\pi^2 \propto H$, a familiar result for
Goldstone bosons.  The behaviour of the pion mass is illustrated in
Fig. 2.

The result (\ref{ef})
may look peculiar to a particle physicist who is more familiar
with the zero temperature result
\begin{equation}
m_\pi^2 = {2 m_q \langle \bar q q \rangle \over f_\pi^2}~.
\label{ega}
\end{equation}
Before considering the sigma mass, we therefore pause here to explain how
(\ref{ega}) and (\ref{ef}) are related.
We have seen that $m_q \sim H$ and that
the order parameter $\langle \bar q q \rangle \sim \langle \sigma \rangle
\sim M$.
At zero temperature,
$f_\pi$ is defined in terms of the axial current by the relation
\begin{equation}
\langle 0 \mid A_\mu^\alpha (0) \mid \pi^\beta (q) \rangle =  i f_\pi q_\mu
\delta^{\alpha\beta}~.
\label{egb}
\end{equation}
In the zero temperature linear sigma model, the axial current is given by
\begin{equation}
A_\mu^\alpha (x) = \sigma (x) \partial_\mu \pi^\alpha (x) ~-~ \pi^\alpha (x)
\partial_\mu \sigma (x)~,
\label{egc}
\end{equation}
which means that $f_\pi$ defined in (\ref{ega}) is simply
\begin{equation}
f_\pi = \langle 0 \mid \sigma \mid 0 \rangle ~.
\label{egd}
\end{equation}
This result suggests that we make the identification $f_\pi \sim M$, which
does indeed make (\ref{ef}) and (\ref{ega}) equivalent.
\begin{figure}
\centerline{
\epsfysize=100mm
\epsfbox[72 216 540 576]{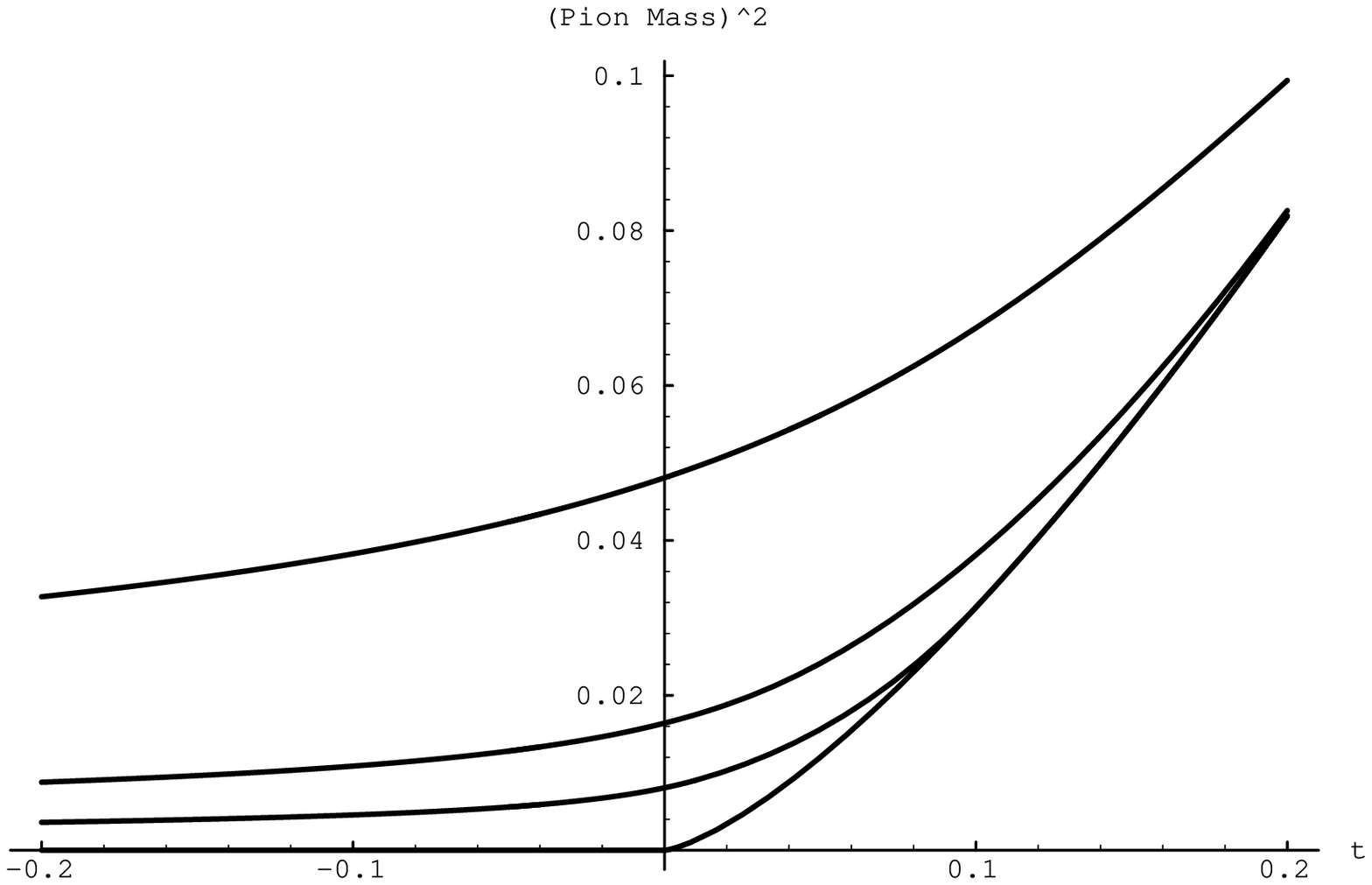}
}
\fcaption{From Ref. [\citelow{us}].
$m_\pi^2$ as a function of $t$ for
$H=$ 0, 0.002, 0.005, and 0.02.  Since $M$ and $H$ are in dimensionless
scaled units, so is $m_\pi^2$.
For $t=0$, $m_\pi^2 = H^{(\delta -1)/\delta}$.
For $H=0$ and $t>0$ (and for large enough $t$ for any $H$)
$m_\pi^2 \sim t^\gamma$.
For $t<0$ and $H\rightarrow 0$, $m_\pi^2 \sim H$.}
\label{fig:pionmass}
\end{figure}
However, it is important
to remember that
using the linear sigma model at zero temperature cannot be justified by
a universality argument in the way that using it near $T=T_c$ can.
Hence the argument of this paragraph is {\it not} a derivation of (\ref{ef})
from the zero temperature result (\ref{ega}) .
(\ref{ef}) is valid near $T=T_c$ while
(\ref{ega}) is valid at $T=0$.  Also, $m_\pi$ in (\ref{ega})
is a mass in a $3+1$
dimensional Lorentz invariant theory,
while $m_\pi^2$ in (\ref{ef}) is an inverse susceptibility in a 3 dimensional
theory.
We have simply shown that a reader familiar
with one expression should not be surprised by the other.

The behaviour of the sigma mass at the coexistence
curve is trickier to obtain than that of the pion mass.
First, we note that in mean field theory ($\epsilon = 0$) the equation
of state is simply $y = f(x) = 1+x$, and $m_\sigma$
is easily evaluated using (\ref{efc}) .  For $H\rightarrow 0$ at fixed $t<0$
the result is
\begin{equation}
m_\sigma^2 = \Bigl( \frac{\delta}{ |t|^\beta } \Bigr) H
+ \frac{|t|^{\beta (\delta -1 )} }{\beta } ~.
\label{ege}
\end{equation}
Hence, in mean field theory $m_\sigma^2$ decreases with $H$ to a nonzero
value at $H=0$.
However, for $d<4$ when
fluctuations are important, the result is quite different.  In words,
fluctuations of the massless pions produce new infrared singularities
in the longitudinal susceptibility, or, equivalently,
make the sigma massless. Now, let us see how this result can
be obtained from the equation of state.\cite{bw,wz}
In the limit $H\rightarrow 0$, $f(x) \sim H$ while $f'(x)$, we will see,
tends to zero more slowly.  Hence, the second term in (\ref{efc})
is dominant and
gives
\begin{equation}
\frac{ \beta m_\sigma^2 }{ M^{\delta -1}} \rightarrow f'(x)
{}~{\rm for}~x\rightarrow -1~.
\label{eh}
\end{equation}
\begin{figure}
\centerline{
\epsfysize=100mm
\epsfbox[72 216 540 576]{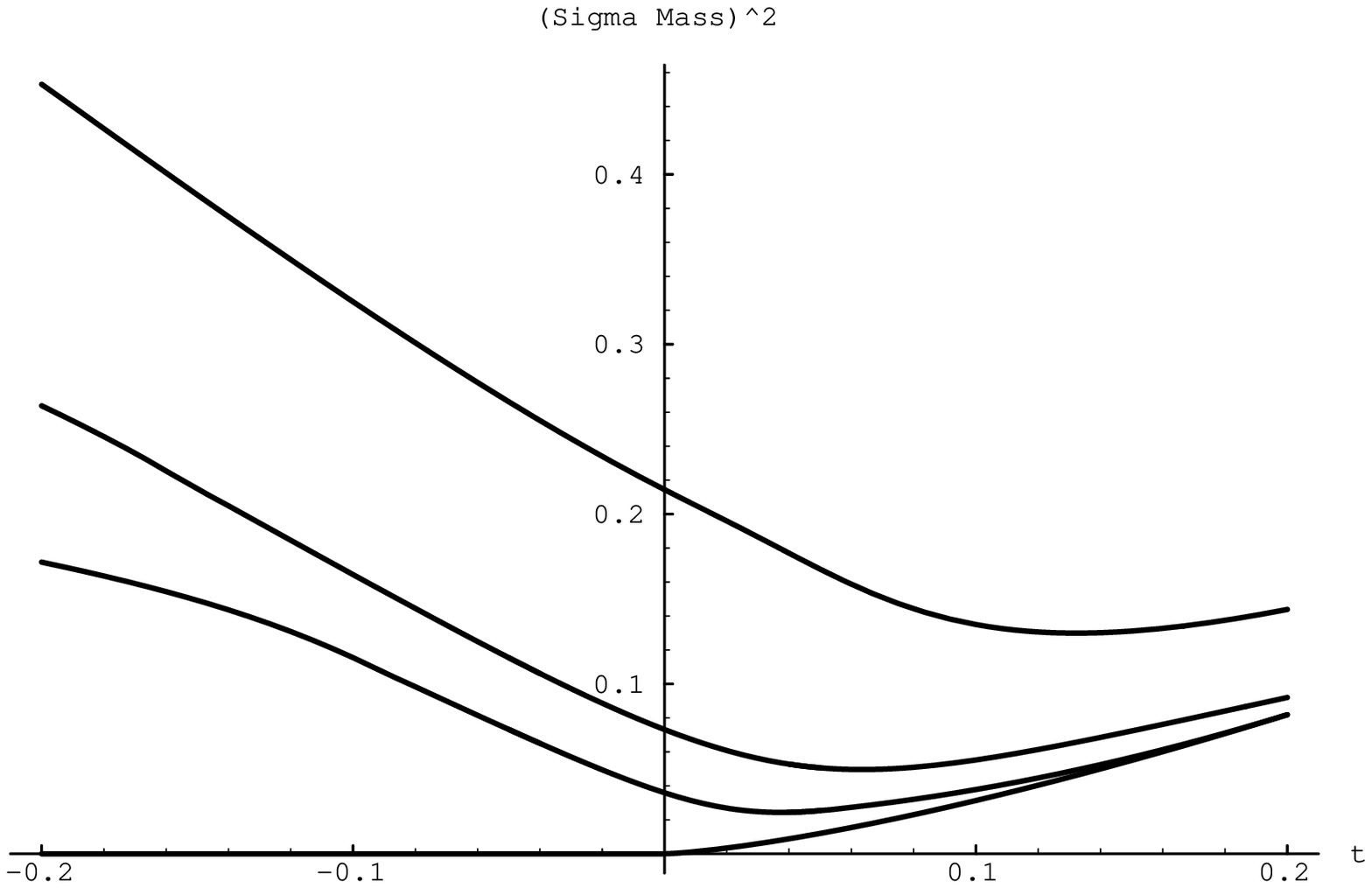}
}
\fcaption{From Ref. [\citelow{us}].
$m_\sigma^2$
as a function of $t$ for
$H=$ 0, 0.002, 0.005, and 0.02.  Since $M$ and $H$ are in dimensionless
scaled units, so is $m_\sigma^2$.
For $t=0$, $m_\sigma^2 = \delta m_\pi^2$.
The scale on the vertical axis is different than in Fig. 2.
For $H=0$ and $t>0$ (and for large enough $t$ for any $H$) $m_\sigma^2 =
m_\pi^2$, and both increase like $
t^\gamma$.  For $t<0$ and $H\rightarrow 0$,
the sigma mass decreases to zero as shown in Fig. 4.}
\label{fig:sigmass}
\end{figure}
The difficulty is that
for
$x\rightarrow -1$,
$f'(x)$ contains divergent terms like $\epsilon \log(x+1)$,
 $\epsilon^2 \log ^2(x+1)$ and $\epsilon^2 \log(x+1)$. These terms do not
exponentiate to $f'(x) \sim (x+1)^p$.
Wallace and Zia\cite{wz} in fact find
the result
\begin{equation}
\left( \frac{\beta m_\sigma^2 }{ M^{\delta -1}} \right) ^{-1}
\rightarrow c_1 + c_2 y^{-\epsilon /2} ~.
\label{ei}
\end{equation}
Both the terms on the right side of (\ref{ei}) must be kept because they
differ in their exponents only by order $\epsilon$.  Also for this
reason,
the constants $c_1$ and $c_2$ calculated by Wallace and
Zia\cite{wz} and given in Ref. [\citelow{us}] are only known
to order $\epsilon$
even though $f(x)$ is known to order $\epsilon ^2$.

Qualitatively, as $H$ is lowered at fixed
$t<0$, at first the $c_1$ term dominates and $m_\sigma^2$ appears to be
decreasing toward a nonzero value at $H=0$ as in the mean field result.
Then, the $c_2$ term takes over and one finds that in fact the sigma
mass goes to zero like $m_\sigma^2 \propto H^{\epsilon/2}$.
Recently, Anishetty {\it et al.}\cite{anishetty}
have done an extensive analysis of the divergence of the
sigma susceptibility produced by the massless pions for $t<0$ and
$H\rightarrow 0$.  In agreement with the
$\epsilon$-expansion result, they find that in $d=3$,
$m_\sigma^2 \propto H^{1/2}$.
The behaviour of the sigma meson mass is illustrated
in Fig. 3 and Fig. 4.
\begin{figure}
\centerline{
\epsfysize=100mm
\epsfbox[72 216 540 576]{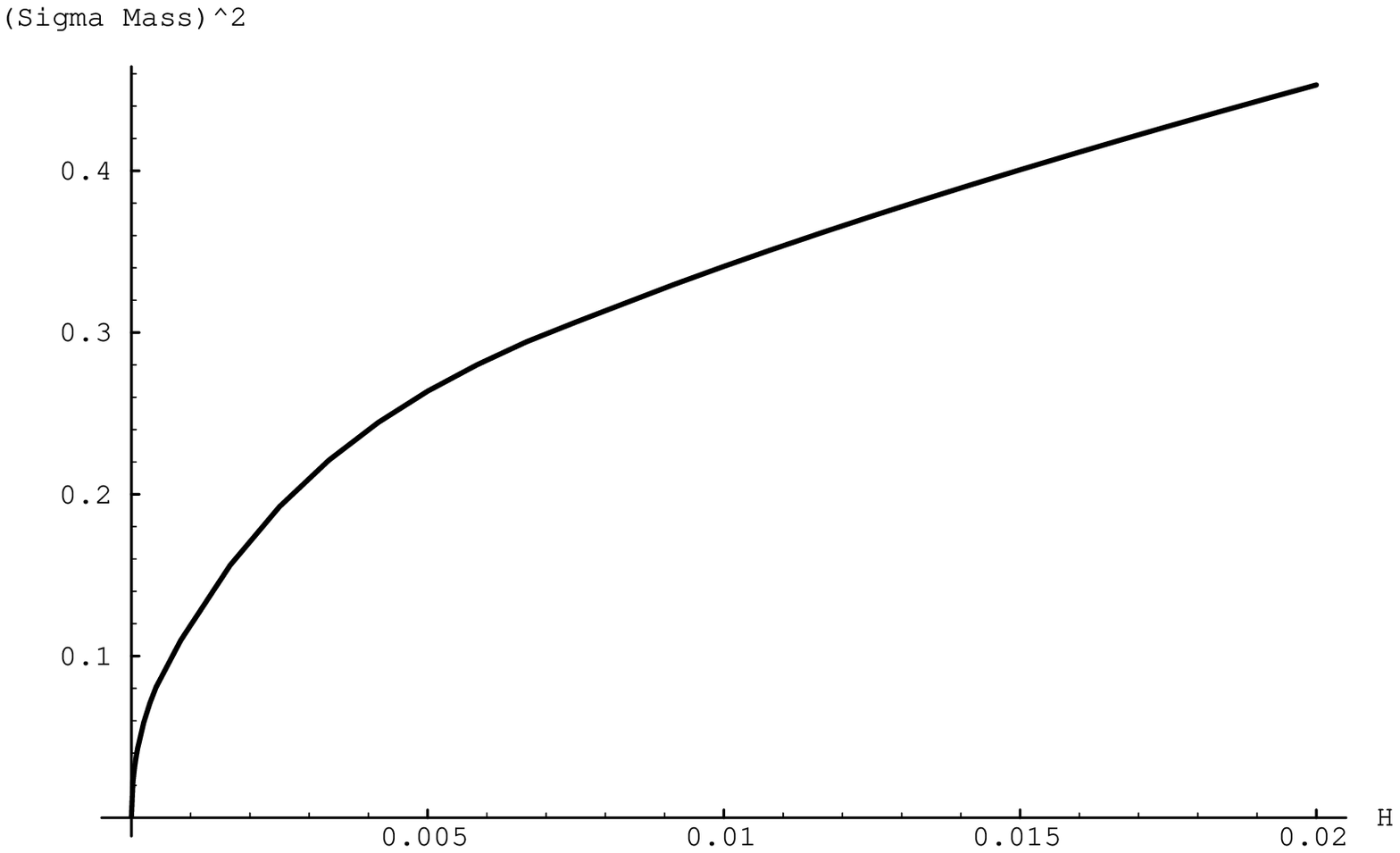}
}
\fcaption{From Ref. [\citelow{us}].
$m_\sigma^2$ as a function of $H$ for $t=-0.2$.
For large $H$ it behaves as if it will be nonzero at $H=0$, but in fact
for $H\rightarrow 0$ it decreases like $m_\sigma^2 \sim H^p$ where to lowest
order in $\epsilon$, $p=\epsilon /2 = 1/2$.}
\label{fig:sigmass2}
\end{figure}
In future lattice simulations, as $m_q$ is lowered toward zero, this behaviour
should be observed.
This result is an example of the power of the renormalization group
techniques in obtaining universal results.  If we had chosen a specific
microscopic model, say that of Gocksch,\cite{gocksch}
or the Nambu--Jona-Lasinio
model of Hatsuda and Kunihiro,\cite{hk} we would have been able to
calculate non-universal quantities far from $T_c$, but would basically
have been limited to using mean field theory, as those authors do.
Then, we would have reached the incorrect conclusion that $m_\sigma\neq 0$
in the chiral limit below $T_c$.  Here, by restricting
ourselves to calculating universal quantities, we are limited to
the region near the critical point, but our
results are model independent and
include the effect of fluctuations.

\subsection{Comparison with Lattice Simulations}

Finite temperature lattice QCD simulations are ideally suited to
testing many of the predictions made in this chapter.
For a pedagogical introduction to and review of finite temperature
lattice QCD simulations, see DeTar's article in this volume.
The static correlation functions of the three dimensional theory are
natural objects to consider in finite temperature (Euclidean)
simulations.  Also, it is much easier to vary parameters like the
temperature and the bare quark mass in a lattice simulation than
in a real experiment.
Hence, it should be
possible to measure the static critical exponents of section 2.1,
and the equation of
state and the scaling behaviour of the pion and sigma masses of section 2.2
on the lattice.

Present
simulations
\cite{oldgott,bernard,columbia,fukugita,gottlieb,zhu,iwasaki,b2,latreviews}
provide evidence that in two flavour
QCD the phase transition is second order, and that the order parameter
is indeed ${\cal M}$.
For example, in the results of Bernard {\it et al.}\cite{bernard} shown in
Fig. 5, there are no signs of any discontinuities
in the expectation value of the order parameter as a function
of temperature.
$\langle \bar q q \rangle$ decreases smoothly as
a function of increasing temperature, as Fig. 1
leads us to expect for simulations done with nonzero
quark masses if the transition is indeed second order for
$m_q=0$.
All these authors have looked unsuccessfully for
signals of two phases coexisting at one temperature, which
would be characteristic of a first order transition.
These simulations are consistent with the transition being second
order, but do not rule out the possibility that
in better simulations with smaller quark masses a small discontinuity
could be found, indicating that the transition is in fact weakly first
order.
Using the zero temperature $\rho$ mass to set the energy scale,
Bernard {\it et al.}\cite{bernard,b2} find that the temperature
at which the decrease in the chiral order parameter
occurs is about $140-160~{\rm MeV}$, significantly lower than
in simulations without dynamical fermions. Over the
same range of temperatures at which the chiral order
parameter decreases, the Polyakov loop expectation value increases
and the specific heat is large.
Also, the calculations slow down --- simulations must be run
for a long time to get reliable
results because large fluctuations occur.
This is
as would be expected near
a second order critical point.
Thus, in QCD with 2 dynamical quarks, it seems that
the crossover associated with deconfinement, for which
there is no order parameter, occurs at the same
temperature as the chiral transition.  The chiral
transition is also a smooth crossover,
but we expect it to approach a second order transition as
$m_q$ is lowered further.
All of this is in marked contrast
to results of lattice gauge theory simulations with
three or four species of light quarks, in which there is a first order
chiral phase transition with latent heat, hysteresis, and
abrupt changes in the values of
observables.\cite{oldgott,columbia,iwasaki,latreviews}
(Our results on the order of the chiral transition ---
second for two massless quarks and first for three or more ---
are also in agreement with results obtained
by treating the QCD vacuum as an
instanton liquid and making a mean-field approximation.\cite{instliq})
\begin{figure}
\vspace{3.truein}
\fcaption{From Ref. [\citelow{bernard}].
The chiral order parameter in the lattice
gauge theory simulations
of Bernard {\it et al.}
with two flavours of staggered quarks.
The chiral order parameter is plotted as a function of $T$ for
$m_q = 0.025$ and $m_q =0.0125$ in lattice units.
The simulations
were done on a $12^3 \times 6$ Euclidian lattice.
On such a lattice, the period in the fourth direction is $1/T$.
Since the number of lattice sites
in this direction (six) is not varied, to vary the temperature one
varies the physical value of the lattice spacing.
The fact that QCD is asymptotically free means that reducing
the lattice coupling constant $g$ reduces the physical value of
the lattice spacing and therefore corresponds to increasing the
temperature.  Hence, in the figure temperature increases to the right.
The curve labeled $m_q = 0$ is a linear extrapolation from
the other two curves.  We predict that if $m_q$ were lowered to
zero, one would find the behaviour of Fig. 1 rather than a linear
extrapolation.}
\label{fig:bernard1}
\end{figure}

Fig. 6. shows the behaviour of the screening mass (inverse correlation
length)
for the $\pi$ and $\sigma$  in the
same simulation as Fig. 5.   If the exponent $\eta$ were zero, then
$m_\pi$ and $m_\sigma$
(which, recall, are susceptibilities$^{-1/2}$) would
have the same scaling behaviour as the screening masses of Fig. 6.
Since $\eta$ is in fact small, the screening masses can
be viewed as crude approximations to $m_\pi$ and $m_\sigma$.
As we expect, the sigma screening mass decreases
and the pion screening mass
increases as the temperature is increased, and above
$T_c$ the two masses each increase and appear to be becoming degenerate.
This is evidence that the order parameter is indeed ${\cal M}$.
\begin{figure}
\vspace{3.truein}
\fcaption{From Ref. [\citelow{bernard}].
The inverse of the pion and sigma correlation
lengths as a function of $T$
for $m_q=0.0125$ in lattice units in the same simulations
as in Fig. 5.  The screening masses
are crude approximations to the inverse susceptibilities
$m_\pi$ and $m_\sigma$ because $\eta$ is small.
The qualitative behaviour should be compared to
Figs. 2 and 3.}
\label{fig:bernard2}
\end{figure}

So far, we have discussed various qualitative lattice results,
that are in
agreement with our expectations.  Before
turning to recent progress on more quantitative
tests, it is worth enumerating the logical possibilities
which could lead to a failure of
the hypothesis that a second order chiral phase transition with
$O(4)$ exponents will be seen in lattice simulations.
Perhaps the most likely alternative is that the transition could
in fact be first order.  As we will see in the next section,
this occurs if the strange quark is too light.  More generally,
we only know that an infrared fixed point with appropriate symmetry
exists.  We do not know that QCD is in fact in the basin
of attraction of this fixed point. A second logical
possibility is that the Ginzburg region may be too small
to see the true critical behaviour.  Fluctuations of the
order parameter are important only close to $T_c$ ---
for $t \equiv (T-T_c)/T_c < t_G$ where $t_G$, like $T_c$,
is not universal.  Outside the Ginzburg region, ({\it i.e.} for $t>t_G$,)
mean field theory is valid.  Thus, even if the chiral phase
transition does have
$O(4)$ exponents, these exponents can only be measured by simulations
done for temperatures satisfying $t<t_G$.  Simulations
with larger $t$ would see the mean field exponents $\eta=0$ and
$\nu=1/2$.  $t_G$, like $T_c$, cannot be predicted by
arguments based on universality, and must be measured
on the lattice.  Can $t_G$ be very small?  This is a logical
possibility.  For example,\cite{landau2} in ordinary, low temperature,
BCS superconductors, $t_G \sim (T_c/\varepsilon_f)^4$ where
$\varepsilon_f$ is the Fermi energy.  In these materials,
$T_c/\varepsilon_f \sim 10^{-3}-10^{-4}$ and $t_G$ is minuscule.
Recent numerical simulations by Kocic and Kogut\cite{KocicKogut}
of the $2+1$ dimensional Gross-Neveu model in which
mean field exponents are seen can be interpreted as
a sign that $t_G$ is small in this theory.
In QCD, there is no analogue of the small parameter $T_c/\varepsilon_f$,
and so there is no reason to expect $t_G$ to be particularly small.
Nevertheless, it remains a logical possibility.  To exclude this
scenario,
it is therefore
crucial to measure the critical exponents in QCD simulations
with two flavours accurately enough to demonstrate that they
deviate from their mean field values.
The third logical possibility is that somewhere in theory space
there is {\it another} infrared fixed point to which QCD
is driven under renormalization.  Were this the case, the
chiral phase transition would be second order but would
not have $O(4)$ exponents.  Of the three logical possibilities,
the third would be the most surprising.

\begin{figure}
\centerline{
\epsfysize=100mm
\epsfbox{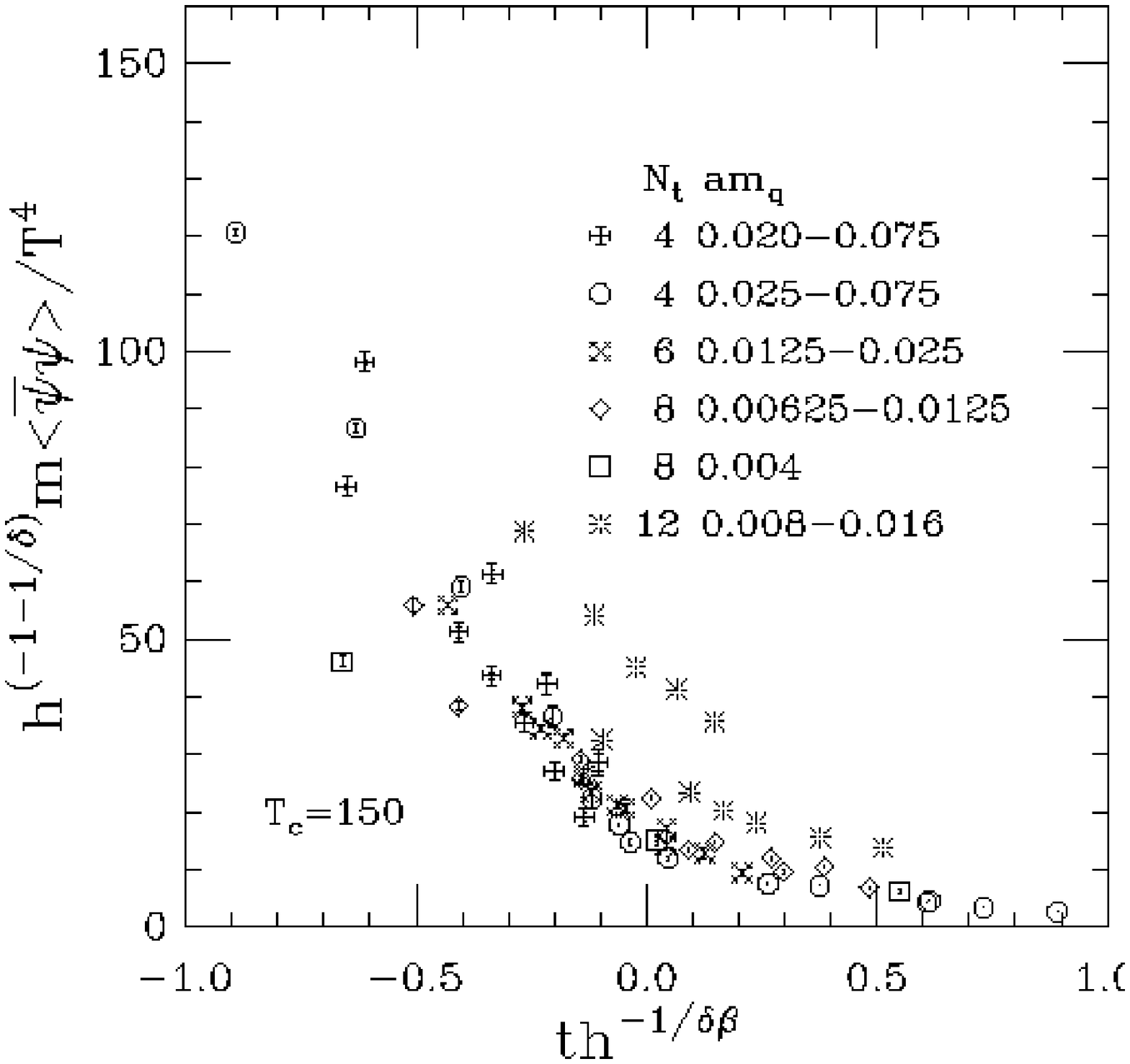}
}
\fcaption{ From Ref. [\citelow{detar}].
DeTar's plot of $\tilde y$ vs. $\tilde x$
using $O(4)$ exponents.}
\label{fig:detarfig}
\end{figure}
Recently, a number of authors have begun to attempt quantitative comparisons
between lattice simulations and expectations based on
the hypothesis that the chiral transition is second
order with $O(4)$ exponents.  DeTar has pointed out\cite{detar} that
it is possible to check that the order parameter $M$, the
temperature $t$,
and the symmetry breaking term $H\sim m_q$ are in fact related
by a single equation of state (\ref{ed}).  Following Binder,\cite{binder}
DeTar writes (\ref{ed}) in a different, but equivalent, fashion.
Define
\begin{equation}
\label{xtilde}
\tilde x \equiv t / H^{1/\beta\delta} ~,~~~~\tilde y \equiv M/H^{1/\delta}~.
\end{equation}
Then, the equation of state (\ref{ed})
is equivalent to
\begin{equation}
\tilde y = g(\tilde x)
\label{neweofs}
\end{equation}
where $g(\tilde x)$ is a universal function related to $f(x)$
of $(\ref{ed})$ by
\begin{equation}
\tilde x = g^{-1}(\tilde y) = x^{-1/\beta\delta} f^{-1}(x)~.
\label{eosrelation}
\end{equation}
Fig. 7 is DeTar's plot of $\tilde y$ versus $\tilde x$
for a number of simulations\cite{bernard,gottlieb,zhu,b2,karschlaer}
with various quark masses,
temperatures, and lattice sizes.  If (\ref{neweofs}) is correct,
all the points should lie on a single curve.  In producing
the figure, DeTar has used the $O(4)$ values for $\beta$ and
$\delta$, and has adjusted $T_c$ to $150~{\rm MeV}$ to obtain
the best agreement, although he notes that $160~{\rm MeV}$
works comparably well.  DeTar's plot is encouraging, but
not yet quantitative.  For one thing, the preliminary
data on the largest lattices with 12 lattice sites in
the $t$ direction are not in good agreement with
the other data, suggesting that finite size effects may
be important.  In the future, with better simulations,
it will be possible to apply this analysis more quantitatively --- the
simulation results will be fit to (\ref{neweofs}) yielding
best fit
values for $T_c$, $\beta$ and $\delta$, along with
error estimates.

\begin{figure}
\centerline{
\epsfysize=80mm
\epsfbox[126 198 504 576]{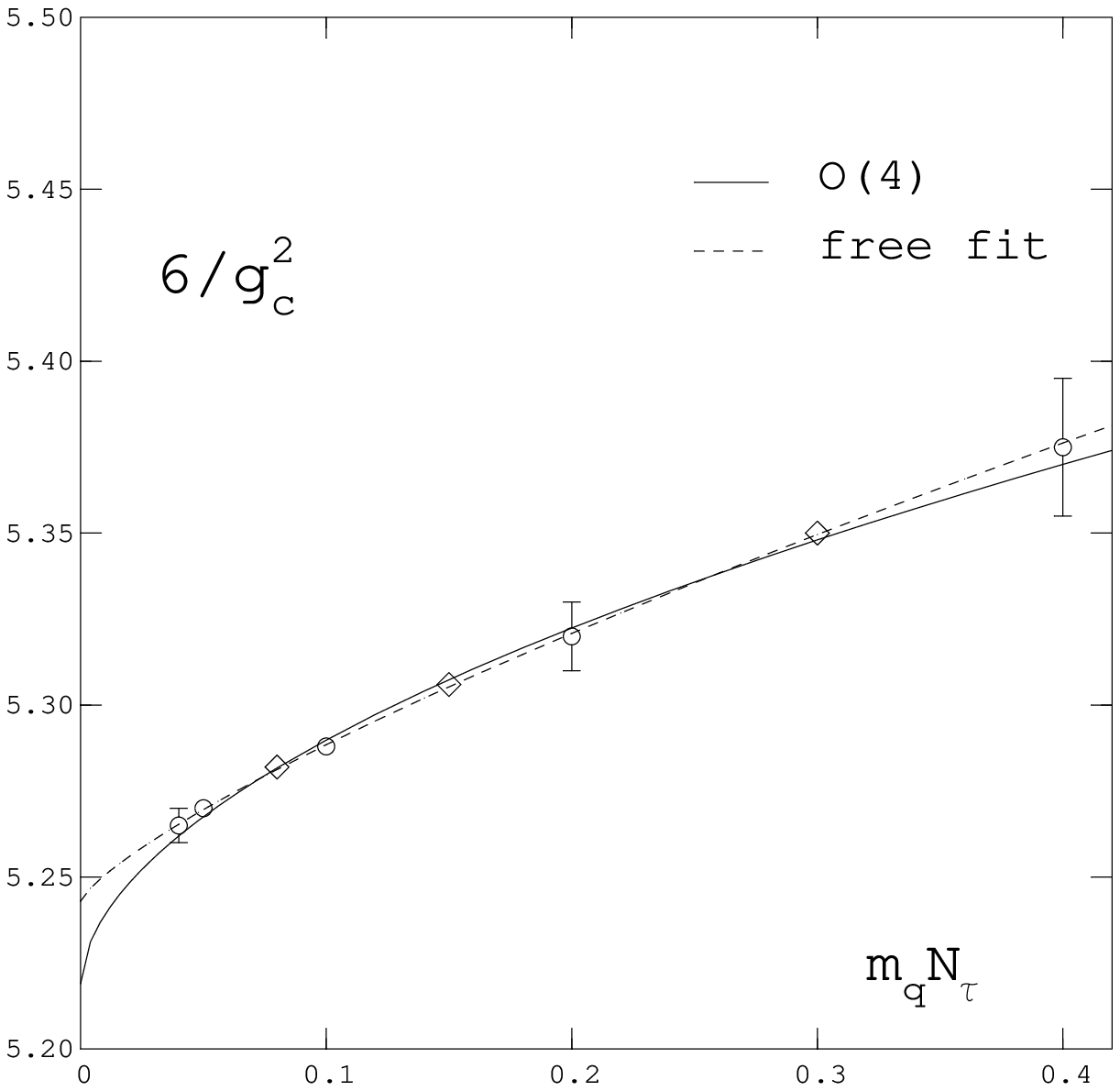}
}
\fcaption{From Ref. [\citelow{karschlaer}].
Pseudocritical coupling as a function of quark
mass for simulations done with $N_\tau =4$
lattice sites in the Euclidean time direction.
The dashed curve is a three parameter fit
to (\ref{karsch2}).  The solid curve is a two
parameter fit with $1/\beta\delta$ set to its
$O(4)$ value.}
\label{fig:karschfig1}
\end{figure}
Karsch\cite{karsch} has pioneered another way of measuring the critical
exponents in a lattice simulation.  From Fig. 3. we see that
for $H\neq 0$, $m_\sigma$ has a minimum at
$t=t_{\rm pc}>0$, which Karsch calls the pseudocritical
temperature.  Because $t$, $H$, and $M$ are related by
(\ref{ed}) or equivalently (\ref{neweofs}),
\begin{equation}
t_{\rm pc}\sim H^{1/\beta\delta}~.
\label{karsch1}
\end{equation}
Karsch therefore notes that in lattice simulations,
the pseudocritical coupling $g_{\rm pc}$ at which
the $\sigma$ susceptibility peaks should depend
on $m_q$ according to
\begin{equation}
\frac{6}{g_{\rm pc}^2(m_q)} = c_0 + c_1 m_q^{1/\beta\delta}\ ,
\label{karsch2}
\end{equation}
where $c_0$ is related to $T_c$ for the second order
transition with massless quarks, and neither $c_0$ nor $c_1$ are
universal.
Karsch and Laermann's plot\cite{karschlaer} of the pseudocritical
coupling in
various
simulations\cite{oldgott,columbia,fukugita,karschlaer}
is shown in Fig. 8.
Fitting the three parameters $c_0$, $c_1$, and $1/\beta\delta$
yields $1/\beta\delta = 0.77\pm 0.14$, which is slightly higher
than the $O(4)$ value $1/\beta\delta = 0.55\pm.02$ and which
is in agreement with the mean field value $1/\beta\delta=2/3$.
However, fitting $c_0$ and $c_1$ with $1/\beta\delta$ fixed
to its $O(4)$ value yields just as good a fit.
Simple inspection of the figure leads to the conclusion
that it is too early to extract the value of
the exponent convincingly --- simulations with lighter quarks are needed.

\begin{figure}
\centerline{
\epsfysize=90mm
\epsfbox[126 198 504 576]{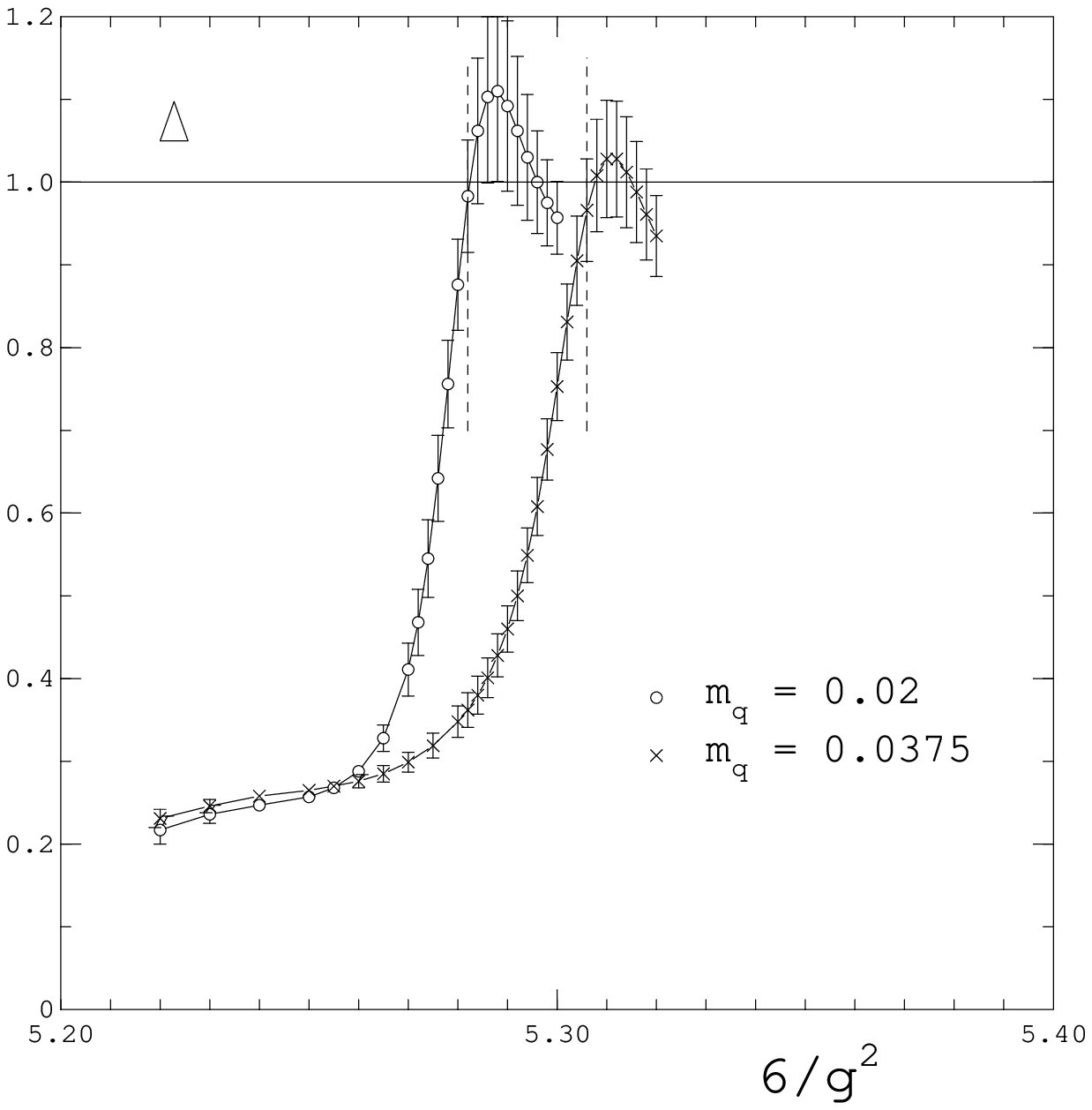}
}
\fcaption{From Ref. [\citelow{karschlaer}].
The ratio of susceptibilities $\Delta \equiv m_\pi^2/m_\sigma^2$
as a function of temperature for two different quark masses.}
\label{fig:karschfig2}
\end{figure}
Karsch and Laermann\cite{karschlaer} have evaluated
both the $\pi$ and the $\sigma$ susceptibilities.  From these,
they extract a direct measurement of the exponent $\delta$.
{}From (\ref{efb}) and (\ref{efc}), we see that the quantity
$m_\pi^2/m_\sigma^2$ which Karsch and Laermann call $\Delta$
is given by
\begin{equation}
\frac{1}{\Delta} = \frac{m_\sigma^2}{m_\pi^2} = \delta - \frac{x f'(x)}
{\beta f(x)}\ .
\label{cumulant}
\end{equation}
For $t>0$, $\Delta\rightarrow 1$ for $m_q\rightarrow 0$
and for $t<0$, $\Delta\rightarrow 0$ for $m_q\rightarrow 0$.
Most important, however, is that
for any quark mass, $\Delta=1/\delta$ at $t=x=0$ ---
{\it i.e.} at $T=T_c$.
Curves of $\Delta$ as a function of $6/g^2$
for
different quark masses should cross at $6/g_c^2$, and at
that coupling they should have $\Delta=1/\delta$.
Karsch and Laermann's results are shown in Fig. 9.
They find
\begin{equation}
0.21<1/\delta <0.26 ~.
\label{deltaresult}
\end{equation}
This is in
agreement with the $O(4)$ value $1/\delta=0.208\pm.002$
and disagrees with the mean field value $1/\delta=1/3$.
To make this result even more compelling, it would be
good to have curves for lighter quarks to see that they
all cross at the same $T_c$ and $\Delta$, and to see
$\Delta$ decreasing toward zero as $m_q\rightarrow 0$
for $T<T_c$.

One critical exponent which has not been mentioned in this section
is $\alpha$, which is associated with the specific heat.
Because $\alpha<0$, there should be a cusp in the specific heat.
However, the specific heat can also have a large analytic
contribution.  In fact, as the temperature is raised
through the chiral transition, deconfinement also occurs
and the number of degrees of freedom in thermal equilibrium
increases substantially, leading to a large specific heat.\cite{karschrev}
This large contribution to the specific heat is an analytic
function of $T$ even for $m_q=0$, while the contribution
from the chiral order parameter is not.  Nevertheless, because the
smooth part of the specific heat is so large, extracting
the non-analytic part and thus $\alpha$ will be difficult.\cite{alpha}

By this point it should be clear that quantitative
tests of the hypothesis that the QCD phase transition
is in the same universality class as the $O(4)$ magnet
are just beginning to be possible.  To date,
lattice simulations of two flavour QCD are in qualitative
agreement with this hypothesis.
As lattice simulations
improve, these tests will become more and more quantitative.
It should soon be possible to exclude conclusively the
possibility that the exponents take on mean field values.
Indeed, the results of Karsch and Laermann for $1/\delta$
are close to doing this already.  This would
demonstrate that the simulations are close enough in
temperature to $T_c$ that the fluctuations of the order
parameter are important --- {\it i.e.} the simulations
are within the Ginzburg region.  This would be reassuring,
as it would imply that as simulations improve further
the true critical
behaviour will be explored, and the hypothesis that
the transition is in the $O(4)$ universality class will
be tested quantitatively.

Now, some {\it caveats}.
Present lattices are small.
Ideally, one wants the correlation length to be large
compared to the lattice spacing and small compared to
the lattice size, and this requires much larger lattices.
(Berera\cite{berera}
has suggested, however, that this problem can be turned
into a virtue --- finite size effects and finite lattice spacing
effects affect the critical behaviour and therefore can be used
to measure critical exponents like $\nu$.)
However, the fundamental reason why present simulations
do not offer more quantitative tests of the predictions
in this chapter is that to date quark masses have been
so large that correlation lengths do not get very long at
$T_c$.  For example, in the work of Bernard {\it et al.},\cite{bernard}
the correlation length in the pion channel at $T_c$ is
only about 2.5 lattice lengths.
To really probe the critical phenomena, lighter
quarks are needed.  Unfortunately, longer correlation
lengths are necessarily accompanied by numerical critical
slowing, and this makes simulations challenging.

Recent work by the Columbia group\cite{shailesh}
offers another indication that lighter quark masses
are necessary in order to quantitatively probe the
critical phenomena.  They vary the ``valence quark mass''
(the quark mass in the $\bar q q$ operator inserted
in order to measure the order parameter) and the
``sea quark mass'' (the mass of the dynamical quarks in
the fermion determinant) independently.  $m_{\rm sea}$ cannot
be made too small because to do so would slow the simulation down;
$m_{\rm val}$ can be made arbitrarily small.
The first result they find is that the order parameter
goes to zero as a temperature dependent power of
$m_{\rm val}$ for fixed $m_{\rm sea}=0.01$.
(All masses are in lattice units.)
They interpret the power law behaviour they see for $m_{\rm val}$
between about $10^{-2}$ and $10^{-4}$ as a sign of some sort of
critical behaviour,
but no interpretation in
terms of the exponents we have discussed is possible, since
these exponents are only relevant for the physical
case where $m_{\rm val}=m_{\rm sea}=m_q$.
They also find indications that for fixed
$m_{\rm val}$, variations of $m_{\rm sea}$ which are
accessible with their present simulations ({\it i.e.} down
to $m_{\rm sea}=0.01$)
seem to renormalize $6/g^2$
and have no other effect.
(It is known\cite{hasenfratz}
that the effect of heavy quarks is to shift $6/g^2$.)
If these preliminary results are confirmed,
they suggest that $m_{\rm sea}$ must be reduced further
than has been possible to date
in order to see the critical phenomena characteristic of
$m_q\rightarrow 0$ discussed in this article.

Another hurdle to be overcome before
lattice simulations can measure the critical properties of the QCD
phase transition is that any lattice implementation of fermion
fields only exhibits the full chiral symmetry in the continuum limit.
Thermodynamic simulations with Wilson fermions are
difficult because Wilson fermions break
chiral symmetry completely.  Larger lattices
or an improved action are required to avoid lattice
artifacts.\cite{detar,wilsonquarks}
If staggered fermions are used,
one starts with four flavours of fermions
which in the continuum have an $SU(4)\times SU(4)$
chiral symmetry. On the lattice, a $U(1)\times U(1)$ subgroup
is all that remains.
In order to study two flavours of fermions, one takes the square root
of the fermion determinant in the lattice action.  It is not at all
clear what this does to the lattice chiral symmetries.
What is known is that all three pions become light
only in the continuum limit.  Finite lattice spacing effects
leave one pion light, while making two of them heavy.
A transition in which a $U(1)\times U(1)$ symmetry
breaks to $U(1)$ would be in the universality class
of the $N=2$ magnet, as
Boyd {\it et al.}\cite{boyd} have discussed.
It is therefore possible that $O(2)$ exponents will
be measured at first, and the $O(4)$ exponents will be seen
only as finer lattices become
possible and the continuum limit is approached.
Therefore, it would be nice
to be able to distinguish $O(4)$ exponents
from $O(2)$ exponents, but this will require much greater
accuracy than is presently possible.  (For example, $\delta=4.808\pm.007$
and $1/\beta\delta=0.60\pm.01$ at an $O(2)$ transition, and whereas
distinguishing $O(4)$ exponents from mean field exponents
seems feasible in the short term,
distinguishing between $O(4)$ and $O(2)$
will be more difficult.)
Therefore, for
staggered as for Wilson fermions, before we are able
to make confident quantitative tests of
our predictions for the critical phenomena we need finer lattices
so that extrapolation to the continuum limit can be done.
Eventually, the results for staggered
and Wilson fermions must agree when so extrapolated.

At present, it is clear, many {\it caveats} remain to be
dealt with before quantitative measurements of
the critical exponents in lattice simulations
become compelling.
The situation will
improve steadily as simulations become possible
on larger lattices with smaller
quark masses.
In the future, simulations
which have longer correlation lengths
and which are closer to the continuum limit
will be able to
measure critical exponents, correlation functions,
susceptibilities, and the
critical equation
of state, and test the hypothesis that the chiral phase
transition in QCD with two flavours of quarks is in
the same universality class as the $O(4)$ magnet.
At present, this hypothesis has passed a number of
qualitative tests.

\subsection{The Influence of the Strange Quark}

To this point in this chapter, we have described a world with two
massless or light quarks, and hence we have implicitly been taking the
strange quark mass to be infinite.  Pisarski and Wilczek
showed\cite{piswil} that if the up, down, and strange quarks
are all massless, then the chiral phase transition
is first order.
With three massless quarks, the order parameter ${\cal M}^i_j$
is a $3\times 3$ matrix and there are two possible quartic
couplings:  $\lambda_1 {\rm tr}({\cal M}^\dagger {\cal M})^2$ and
$\lambda_2({\rm tr}{\cal M}^\dagger {\cal M})^2$. The renormalization group
equations for this model have been studied to lowest order in
$\epsilon$ by Pisarski and Wilczek\cite{piswil} and Paterson.\cite{paterson}
There is indeed a fixed point, but it is not stable.  For three or
more flavours of massless quarks, there are always directions
in $\lambda_1 - \lambda_2$ space
around the fixed point
for which renormalization toward the infrared drives the theory away
from the fixed point.  Assuming the $O(\epsilon)$ result
that there is no infrared stable fixed point is correct, the phase transition
must be first order.  This result has been verified by Gausterer
and Sanielevici.\cite{gausterer}  They did numerical simulations
of the $3 \times 3$ matrix model and found that the transition is indeed
first order.

The chiral phase transition is second order in a world with
two massless quarks and first order in a world with three.
Hence, as the strange quark mass is reduced from
infinity to zero, at some point the phase transition must change from
second order to first order.  This point is called a tricritical point.
In a lattice simulation,
the strange quark mass could be tuned to just the right value to reach
the tricritical point.  We now discuss the critical
exponents that would be observed in such a simulation.

Let us consider the effect of adding a massive but not infinitely
massive strange quark to the two flavour theory.  This will not introduce
any new fields which become massless at $T_c$, and so the arguments
leading to the free energy (\ref{bb}) are still valid.  The only effect of the
strange quark, then, is to renormalize the couplings.  Renormalizing
$\mu^2$ simply shifts $T_c$, as does renormalizing $\lambda$ unless
$\lambda$ becomes negative.  In that case, one can no longer truncate
the Landau-Ginzburg free energy at fourth order.
After adding a sixth order term, the free energy becomes
\begin{equation}
F = \int d^3 x \Biggl\lbrace {1\over 2}(\nabla \phi)^2 + \mu^2\phi^2
+ \lambda(\phi^2)^2 + \kappa (\phi^2)^3 - H\sigma
\Biggr\rbrace ~.
\label{ga}
\end{equation}
While for positive $\lambda$, $\phi^2$ increases continuously from zero
as $\mu^2$ goes through zero, for negative $\lambda$, $\phi^2$
jumps discontinuously from zero to $|\lambda |/(2\kappa)$ when
$\mu^2$ goes through $\lambda^2 /(4\kappa)$.  Hence, the phase transition
has become first order.  Thus, at some $m_s\equiv m_s^*$
at which $\lambda = 0$
at $T=T_c$,
the phase transition changes continuously from second order to first order.

The nonanalytic behaviour
of thermodynamic functions near tricritical points, just as
near ordinary critical points, is universal.  Hence,
it is natural to propose\cite{wilczek} that QCD with two
massless flavours of quarks and with $T$ near $T_c$ and $m_s$ near
its tricritical value $m_s^*$
is in the universality class of the $\phi^6$ Landau-Ginzburg
model (\ref{ga}).  This model has been studied extensively.\cite{lawrie}
Because the $\phi^6$ interaction
is strictly renormalizable in three dimensions,
this model is much simpler to analyze than the $\phi^4$ model of the
ordinary critical point.  No $\epsilon$ expansion is necessary, and
the critical exponents all take their mean field values.  There are calculable
logarithmic corrections to the scaling behaviour of thermodynamic
functions,\cite{lawrie} but we will limit ourselves here to determining
the mean field tricritical exponents.

In mean field theory, the correlation function in momentum space is simply
$G_{\alpha \beta}(k) = \delta_{\alpha \beta} (k^2 + \mu^2)^{-1}$.
Since $\mu^2 \sim t$, this gives the exponents $\eta =0$, $\gamma =1$
and $\nu = 1/2$.  To calculate $\alpha$ and $\beta$, we minimize
$F$ for $H=\lambda=\nabla \phi =0$, and find $\alpha = 1/2$ and
$\beta = 1/4$.
To calculate $\delta$, we minimize $F$ for $t=\lambda =\nabla \phi =0$
and find $\delta = 5$.

The result for the specific heat exponent $\alpha$
is particularly interesting, since it means that
the specific heat diverges at the tricritical point, unlike at the
ordinary critical point.
This means that whereas for $m_s$ large enough that the transition
is second order the specific heat $C(T)$ has a cusp but is finite
at $T=T_c$, as $m_s$ is lowered to the tricritical value $C(T_c)$
should increase since at the tricritical point it diverges.  This behaviour
should be seen in future lattice simulations.

At a tricritical point there is one more
relevant operator than at a critical point, since two physical quantities
($t$ and $m_s$) must be tuned to reach a tricritical point.  Hence,
a new exponent $\phi_t$, the crossover exponent, is required.
For $\lambda \neq 0$, tricritical behaviour
will be seen only for $|t|>t^*$, while for $|t|<t^*$, either ordinary
critical behaviour or first order behaviour (depending on the sign
of $\lambda$) results.  $t^*$ depends on $\lambda$ according to
\begin{equation}
t^* \sim \lambda^{1/\phi_t}
\label{gb}
\end{equation}
The mean field value of $\phi_t$ is obtained by minimizing the free
energy $F$ for $H=\nabla \phi =0$, and is $\phi_t = 1/2$.
These mean field tricritical exponents, $\alpha = 1/2$, $\beta = 1/4$,
$\gamma = 1$, $\delta = 5$, $\eta = 0$, $\nu = 1/2$, and $\phi_t = 1/2$
describe the real
world if $m_s$ is close enough to the tricritical value,
and will describe
future lattice simulations with $m_s$ chosen appropriately.

We now construct a phase diagram showing
the order of the chiral transition as a function
of $m_s$ and the light quark mass $m_q$.
In the $(m_q,m_s)$
plane, there is a region around $(\infty,\infty)$
in which all quarks are sufficiently massive that
a first order deconfinement transition is obtained.
In a region around $(0,0)$, all quarks are light enough
that the first order chiral transition characteristic
of the theory with three massless quarks is obtained.
On the left side of Fig. 10, $m_q=0$ and the chiral
phase transition is first order for $m_s<m_s^*$ and
second order for $m_s>m_s^*$.  For $m_q\neq 0$,
and $m_s$ large enough, the chiral transition is
a smooth crossover.  For $m_s > m_s^*$ and $m_q$ nonzero
but small enough, the crossover is described by $O(4)$
exponents, as discussed in previous sections.
\begin{figure}
\centerline{
\epsfysize=120mm
\epsfbox[72 144 540 648]{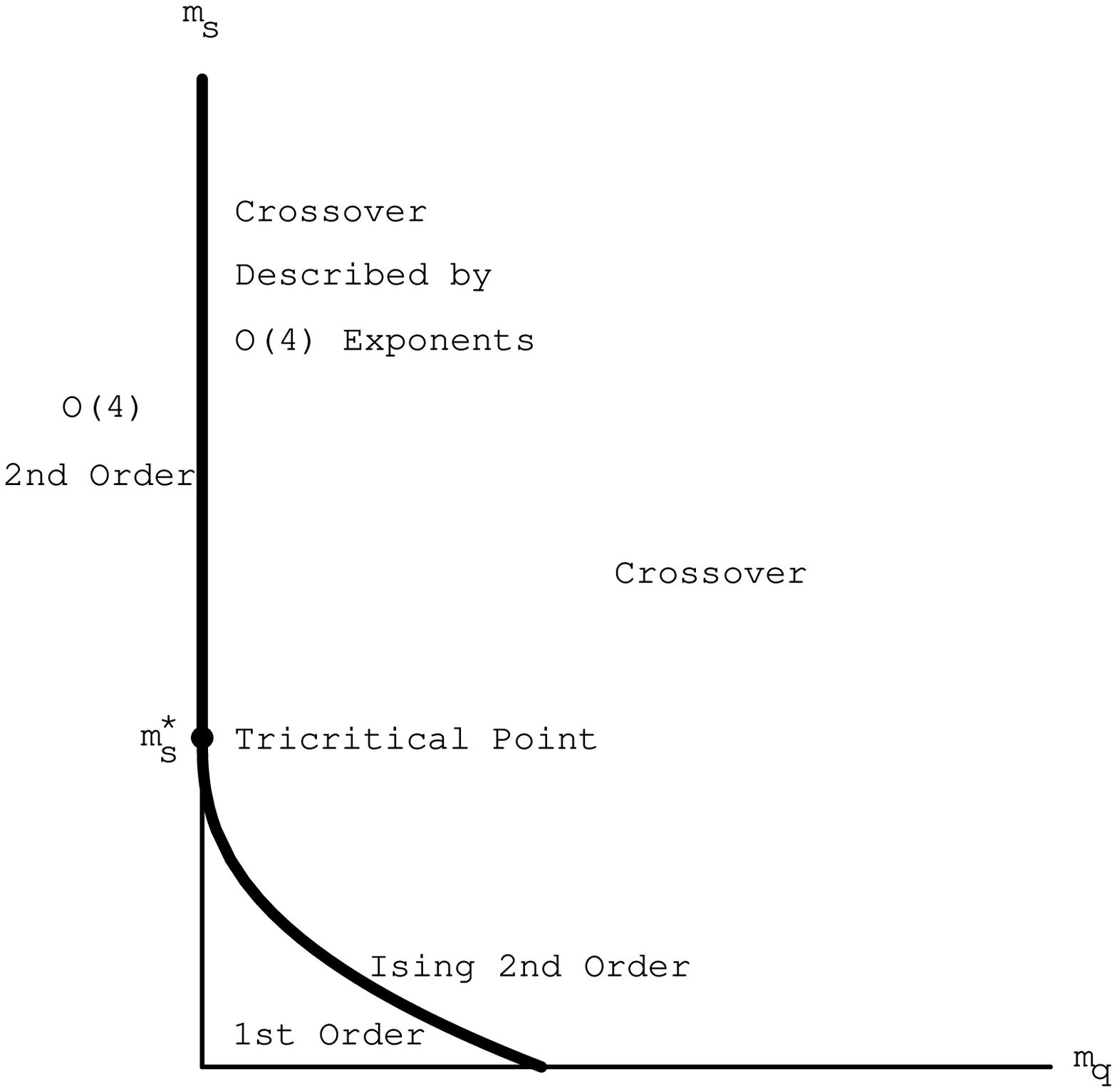}
}
\fcaption{A phase diagram showing how the order
of the chiral transition depends on $m_q$ and $m_s$.
For large $m_s$, there is a second order transition with
$O(4)$ exponents at $m_q=0$.  For $m_q$ nonzero, there
is a smooth crossover which, for small $m_q$, looks like
one of the curves of Fig. 1.  For $m_q=0$, the transition
becomes first order when $m_s$ is reduced below
its tricritical value $m_s^*$.  In a region around $m_s=m_q=0$,
the transition is first order.  The boundary of this region
is a curve of second order transitions with Ising model
exponents.\cite{ggp}  In the figure, this curve is drawn with the shape
$(m_s^*-m_s) \sim m_q^{2/5}$, which is valid near the
tricritical point.}
\label{fig:phasediag}
\end{figure}
At the boundary
between the first order region and the smooth
crossover region, the transition  has been analyzed in mean field
theory by Gavin, Gocksch and Pisarski.\cite{ggp}
They find that along this curve in the $(m_q,m_s)$
plane, a single scalar field is massless and
therefore conjecture that along this curve the transition
is second order with Ising critical exponents.
(The scalar which is massless is pure $s\bar s$ near
$m_s=0$, becomes an $SU(3)$ singlet as one follows the
curve to the point at which $m_s=m_q$ and becomes
the $\sigma$ of previous sections as one
follows the curve further toward the tricritical
point at which $m_q=0$.)
Using (\ref{ga}),
we can deduce the shape of the curve
of second order transitions near the tricritical
point.  For $\lambda < 0$, consider turning on a small
$H$.  $H$ breaks the $O(4)$ symmetry, and makes the
order parameter point in the
$\sigma$ direction.
Working to lowest order in $H$, we find that
$\sigma$ jumps discontinuously between
degenerate minima at $\sigma = 2H\kappa/\lambda^2$
and $\sigma = \sqrt{|\lambda |/(2\kappa)} - (1/2)H\kappa/\lambda^2$
at the temperature at which $\mu^2$ goes through
$\lambda^2/(4\kappa) + H\sqrt{2\kappa /|\lambda |}$.  As $H$ increases, the
size of the jump in $\sigma$ at $T_c$ shrinks, until at
some $H\sim |\lambda |^{5/2}/\kappa^{3/2}$ the two degenerate
minima merge and the transition is second order.  For even
larger $H$, the transition is a smooth crossover.  Since
$H\sim m_q$ and $\lambda\sim (m_s-m_s^*)$, we conclude
that near the tricritical point, the curve of second order
transitions separating the region in which the
transition is first order
from the region in which it is a smooth crossover has the
shape
\begin{equation}
(m_s^* - m_s) \sim m_q^{2/5} \ .
\label{gc}
\end{equation}
Away from the tricritical point, a curve of second
order transitions exists,\cite{ggp} but nothing is known about
its shape.

The real world is one point on Fig. 10.  Lattice simulations,
on the other hand, can be done with varying quark masses.
Thus, on the lattice unlike in the real world it will
be possible to test and explore all the physics of Fig. 10.
It is clearly very important to know where the physical
values of $m_s$ and $m_q$ lie in Fig. 10.  This is not
a universal question, and therefore can only be answered
by experiment or by lattice simulations.

In simulations
done on a $16^3\times 4$ lattice
with staggered fermions,\cite{columbia} the Columbia group
found a first order transition for $m_s=m_q=0.025$ in lattice
units, and found
no sign of first order behaviour in simulations with
only two flavours of fermions and in a simulation
with $m_s=0.1$ and $m_q=0.025$.  By measuring zero
temperature hadron masses, they estimate that in the
latter simulation $m_q$ is larger than its physical value
and, most important, $m_s$ is smaller than its physical
value.  These results suggest that in this simulation,
$m_s$ is greater than $m_s^*$, and furthermore  that the
physical value of $m_s$ is even larger.  Thus, the
Columbia results suggest that the real world is at
a point in Fig. 10 somewhere in the region labelled ``crossover
described by O(4) exponents.''

Recently, Iwasaki {\it et al.}\cite{kanaya}
have announced results from simulations with two light and
one heavy Wilson fermions.  In agreement with the
Columbia results, they find a first order transition when
all three quarks are light, and no sign of a first order
transition in simulations with two light quarks.
However, they report seeing a two state signal characteristic
of a first order transition in simulations
with two light quarks
and a strange quark either of mass $m_s\sim 150 ~{\rm MeV}$
or of mass $m_s\sim 400 ~{\rm MeV}$.
This suggests that the real world lies in the region
of Fig. 10 labelled ``first order'', and is inconsistent
with the Columbia results.  There are a number of
extensions to the work of Iwasaki {\it et al.}
that would make it more compelling.
First, in order to demonstrate consistency with the
result (from their own simulations and those of others)
that the transition is a smooth crossover for
two flavours, it would be good if a simulation was
performed with a heavy enough strange quark
that no signs of a first order transition are seen.
Second, in order to more confidently make
comparisons with the Columbia results,
a measurement of the zero temperature hadron masses
would be helpful.  Finally, as discussed briefly in the previous
section and at length in DeTar's article
in this volume, studying the chiral transition with Wilson
fermions is tricky because of potential lattice artifacts
since the chiral symmetry is completely broken on the
lattice.  At the end of the day, however, when simulations
are done on large enough lattices, it should not matter
whether staggered or Wilson fermions are used.
Although the results of Ref. [\citelow{columbia}]
are at present
more convincing than those of Ref. [\citelow{kanaya}],
the question of where on Fig. 10 the real world lies
remains open.  Time ({\it i.e.} further simulations) will tell.

\subsection{Implications}

We have seen in the previous two sections that the
physics of the chiral phase transition in QCD
is well-suited for study on the lattice.
The hypothesis that the chiral transition in
two flavour QCD is second order with $O(4)$
critical behaviour
has many implications
which are being and will be tested in lattice simulations.
The exploration of the $(m_q,m_s)$ plane has also begun.
Let us now turn to real experiments, as opposed to
those on the lattice. Here, we are not free to dial the
bare quark masses.  We will see that this is unfortunate, particularly
in our discussion of heavy ion collisions.  First, however, let us dispose
of two other possible arenas for testing our results.

The QCD phase transition
certainly occurred in the early universe.  Indeed, much work has been
done on possible observable cosmological
effects of this transition, if it is first
order.
There have been speculations\cite{witten}
that a strongly first order
transition could have affected big bang
nucleosynthesis
and
therefore left a signature in the primordial nuclear
abundances,\footnote{A strong first order
phase transition creates inhomogeneities in the
baryon number density of the universe which
later affect nucleosynthesis.\cite{witten}  Present observations
of nuclear abundances
actually constrain big bang nucleosynthesis
tightly enough that they disallow models in which
the abundances are significantly affected
by inhomogeneities in the baryon number density.\cite{schramm}
These cosmological observations,
therefore, are evidence that the QCD phase transition
is not strongly first order, in agreement with our results.}~~
or could have left an imprint in the form of gravitational waves,
or could even have left relic chunks of new forms of matter.
Unfortunately, we have seen that
for physical values of the strange quark mass,
it is likely but not certain that
the transition is
second order.
In fact, because the up and down quarks are not
massless, the second order transition is smoothed into a crossover.
We can think of no observable
consequences of a second order QCD phase transition in the early universe,
and even fewer consequences of a smooth crossover.

There is one possibly useful cosmological (non)consequence of a QCD transition
which is not first order.
In the early universe, the expansion rate during the transition is very slow
compared to QCD time scales, and thermal equilibrium is
maintained. Therefore, if the transition is not
first order there is no entropy production
at the transition.
Thus, whatever mechanism generates the
baryon number density of the universe at some higher energy scale
need only  generate the present baryon to entropy ratio, and need
not generate a larger baryon number to compensate for dilution
at a first order QCD phase transition.

Now, let us turn from cosmology
to laboratory experiments. Since certain antiferromagnets including
dysprosium have order parameters in the $N=4$ universality class,\cite{bak}
experiments on the phase transition in these materials could
help us understand the QCD transition.\cite{wilczek}
Alas, in these magnets there is
a quartic operator like $(\pi_1^2 + \pi_2^2 )(\pi_3^2 + \sigma^2 )$
which is allowed by the microscopic hamiltonian
of the magnets,\cite{barak} but not by that of QCD.  This operator makes
the symmetric Heisenberg fixed point infrared unstable, and either
makes the phase transition first order or makes it second order but governed
by an anisotropic fixed point.  Hence, dysprosium is not a
suitable arena for learning about the QCD phase transition.

Finally, we turn to
relativistic heavy ion collisions.
As we discussed in the Introduction, in a sufficiently
high energy collision the baryon number of the incident
incident nuclei ends up in that part of the plasma heading
approximately down the beam pipes, and the central rapidity region consists
of a hot plasma with approximately zero baryon number.
In the remainder of this
article, we will
study
the behaviour of the plasma in the central rapidity region as it expands and
cools through $T=T_c$ and eventually hadronizes and becomes pions which
fly off and are detected.

The defining characteristic of a second order phase transition is
the divergence of correlation lengths.  How could this feature be
observed here?  Large volumes of space with the order parameter
correlated and pointing in a direction different than the true
vacuum ({\it i.e.} sigma) direction will become regions in which the
order parameter oscillates coherently about the sigma direction.
After hadronization, correlated volumes will
turn into regions of space where
the ratio of the number of charged pions to neutral pions has some fixed value.
In the standard scenario\cite{bjorken} described in the Introduction,
the longitudinal expansion is boost
invariant and
different positions
in the plasma along the beam direction become
different rapidities as the plasma expands. Therefore, one would
hope that a signal of a second order phase transition would be fluctuations
in the ratio of charged to neutral pions as a function of rapidity.
Such phenomena will be our subject in the next chapter.

A prerequisite for the fluctuations discussed above to be observable is
that the correlation length must get long compared, say, to $T_c$.
To determine whether this does indeed happen, we must leave our
universality safety net behind, since neither $T_c$ nor the magnitudes
of correlation lengths are universal.
{}From Fig 2, it is immediately apparent that we have a problem.
The longest correlation length is in the pion channel, and the
pion mass is increasing with temperature.  This suggests that the
pion mass at $T_c$ is larger than $m_\pi (T=0)\,=\,135\,{\rm MeV}$.
This result can be checked by comparing with (non-universal) results
obtained in a variety of models.
In chiral perturbation theory
using the nonlinear sigma model,\cite{leutwyler} in the linear sigma
model,\cite{loewe} and also in the Nambu--Jona-Lasinio model,\cite{hk}
the pion mass increases from its zero temperature
value as the temperature is increased from zero.
Working in a specific four-dimensional model forces
one to make a mean field\cite{hk} or one-loop\cite{loewe} or
low temperature\cite{leutwyler} approximation which is not valid
near $T=T_c$.  This precludes obtaining good results for
universal quantities, but allows rough estimation of non-universal
quantities like $T_c$ and the low temperature behaviour of
the pion mass.  Non-universal analyses suggest
that $m_\pi$ increases with temperature
as $T$ increases from zero, and our universal
result shows that it is increasing with temperature for $T\sim T_c$.
Hence, it seems clear that the longest correlation length
at $T_c$ will be shorter than $(135 ~{\rm MeV})^{-1}$.  This is to be
compared to $T_c$ itself, which for the case of two massless quarks
is around 140-160 MeV.\cite{bernard,b2}
Hence, even though the quark masses are indeed
small ($\sim 10 ~{\rm MeV}$), the magnetic field $H$ proportional to
$m_q$ is large enough to prevent any correlation lengths from reaching
interesting values.

There is an appropriate quantitative criterion to determine whether
an equilibrium second order phase transition
leads to dramatic effects.  One compares the energy in a correlation
volume just below $T_c$ with the zero temperature pion mass to
determine whether or not the correlated volume can  become a large
number of pions.  Using the lattice simulations
of Ref. [\citelow{bernard}],
we can make a crude attempt
at this comparison.  The
sum of the energy and pressure in a correlation volume
is about $1/4$ the zero temperature $\rho$ mass.  Taken literally, this means
that each correlation volume becomes only one or two pions in the detector.
In these simulations, $m_q$ is somewhat larger than its physical
value, perhaps by a factor of two, and there are many other {\it caveats}
as we have seen,
so this estimate should not be taken literally.  Nevertheless, it seems clear
that the physical value of $m_q$ is large enough that in an equilibrium
phase transition a  correlation volume at $T_c$ does not evolve into a
large number of correlated zero temperature pions.
This is not encouraging.

We pause here for an aside.  The reader may be wondering why, when
the seemingly small {\it equal} quark mass $m_u = m_d = m_q$ has such
deleterious effects, we have completely neglected the difference between
the up and down quark masses.  Unequal
quark masses allow terms of the type\cite{wilczek}
\begin{equation}
\Delta F \propto (\delta m)^2 (\sigma^2 - \pi_3^2 + \pi_1^2 + \pi_2^2 )~.
\label{iz}
\end{equation}
If one is close enough to the critical point that this term matters, one
will discover an anisotropic fixed point rather than the symmetric
Heisenberg fixed point.  However, while the effect of a common quark mass,
namely the mass of the pion, is comparable to $T_c$, the effect of (\ref{iz})
is much smaller.  For example, the QCD contribution to the difference
in mass between the charged and neutral kaons is about 5 MeV.\cite{kaonmass}
Therefore, we need not worry about (\ref{iz}) in real experiments since
it is much less important than the effect of the ``magnetic field''
proportional to the common up and down quark mass.  Of course, (\ref{iz})
could be introduced and studied on the lattice.

It seems likely that if the standard scenario
for heavy ion collisions in which the
QCD plasma cools through $T_c$ while staying close to thermal
equilibrium is correct, then no correlation lengths will get long
enough for there to be any dramatic observable effects of the
phase transition.
The chiral ``transition'', like the confinement/deconfinement
``transition,'' will be a smooth crossover.  If we were able to dial down the
quark masses and hence the pion mass, phenomena associated with
a second order transition would become more prominent.  Alas, in the
real world, unlike on a lattice, we have no such freedom.
Let us hope that we will reach less disappointing conclusions
in the next chapter
when we consider the physics of the QCD phase transition in a
heavy ion collisions in which the chiral order parameter
does not stay in thermal
equilibrium.

The future of the study of the chiral phase transition in QCD
on the lattice looks promising.
As simulations improve,
investigations of the plethora of static critical phenomena we discussed
in the first sections of this chapter
will become better and better.  Critical exponents, the equation
of state, the critical behaviour of the pion and sigma susceptibilities,
and tricritical exponents,  are all
out there waiting to be measured.  The hypothesis that QCD with
two flavours of massless quarks has a second order chiral
phase transition in the same universality class as the
$O(4)$ Heisenberg magnet will be tested more and more quantitatively.

\section{From Disorder to Long Wavelength Oscillations:
The QCD Phase Transition Far From Thermal Equilibrium}

The gloomy penultimate paragraph of the previous chapter began with
a conditional sentence.
In this chapter we will consider the
observable effects in heavy ion collisions if the plasma {\it does not}
stay close to thermal equilibrium through the transition.
There are
tantalizing hints in cosmic ray physics that point in this direction.

Among the zoo of high energy cosmic ray events which have been
detected in emulsion experiments
are a particularly
peculiar class of events called
Centauros\cite{centauro} discovered
in emulsion experiments at the Mount Chacaltaya Observatory.
Centauros are events
with total energy of order 1000 TeV in which many
(of order 100) charged hadrons
each with energies of a several TeV
and very few photons or electrons are seen in a cosmic ray induced shower.
In a normal event, the charged hadrons are accompanied by photons
from the decay of neutral pions.
Centauros have also been seen in emulsions exposed
at the Pamir Observatory and the Mount
Fuji Observatory.\cite{pamir}  The experimental situation is
murky, however, because more recent searches at Mts. Kanbala
and Fuji\cite{kanbala} failed to find any Centauros.\cite{recentcentauro}
Events called mini-Centauros which have fewer
charged hadrons than Centauros but which also have
very few photons or electrons have also been observed.
In the sample of events in described by Lattes {\it et al.},\cite{centauro}
there were 5 Centauros
and 13 mini-Centauros, representing
about 1\% and 2\% of the events seen with energy of the appropriate order of
magnitude.
Cosmic ray shower simulations do predict events which
can be interpreted as mini-Centauros,\cite{ellsworth} but
it is difficult to interpret Centauros normally.\cite{ellsworth}

The JACEE collaboration has flown
balloon borne emulsion detectors to study cosmic rays.\cite{jaceedetails}
They study cosmic ray interactions
in which the primary scattering occurs within
one of the many layers of the detector.  This is a big advantage,
because it means the location of the vertex is known to
within a few microns, the tracks of all the decay products are
well measured,
and therefore the pseudorapidities of the
particles are known.  The disadvantage, however, is that
balloon borne detectors are necessarily exposed for much
shorter periods of time than detectors left on mountain tops.
Mountain top experiments therefore detect higher energy
(rarer) cosmic rays.  In fact, JACEE has seen
very few events with energies in
the 1000 TeV range and no Centauros.\cite{jaceedetails}
However, recently\cite{jacee} they have discovered events
in which there are ``anti-Centauro''
regions in pseudorapidity-azimuth phase space
containing a few tens of photons and almost no
charged tracks.

Centauros are peculiar because so many charged pions are observed without
any of the gammas that would indicate the decay of neutral pions.
This apparent violation of isospin invariance is puzzling, unless one
thinks of it in the language of a second order phase transition in which
these events can be interpreted as the creation of a volume of QCD
plasma in which the $\phi$ field has fluctuated throughout most of the
plasma in some direction in the $\pi_1 - \pi_2$ plane. This implies
correlation
throughout most of a volume
of plasma large enough that it becomes about 100 zero temperature pions.
We convinced ourselves
above that this could not happen if the plasma remains close to
thermal equilibrium.  Hence, these Centauro events provide a tantalizing
hint that it might be wise to consider the effects of going from the
disordered phase above $T_c$ to the ordered phase rapidly without maintaining
thermal equilibrium.
Regardless of what the final word on Centauro
cosmic ray events turns out to be, it is worth considering
qualitative phenomena which can arise if the chiral transition
is far from thermal equilibrium.  In a heavy ion collision,
even if partons reach some initial local thermal equilibrium
at early times, it is far from clear that the long wavelength
modes of the chiral order parameter will be in equilibrium as
chiral symmetry breaking occurs.

In section 3.1, we will
describe the phenomenological consequences
of the existence of a large region of space in which the
order parameter $\phi$ is displaced from the vacuum direction.
Because of the explicit chiral symmetry breaking, $\phi$ oscillates
about the $\sigma$ direction.  In later sections of this chapter,
we will discuss a mechanism by which these
long wavelength pion oscillations could arise.\cite{quenchpaper}
To model
the dynamics of the chiral order parameter in a far
from thermal equilibrium phase transition, we will consider
quenching in the O(4) linear sigma model.  We will argue,
and present numerical evidence, that in the period immediately
following a quench long wavelength modes of the pion field
are amplified.  This could have dramatic
phenomenological consequences in heavy ion collisions.
Simulations of classical evolution following a quench
are a long way from real heavy ion collisions.   A number
of authors have discussed improvements and modifications ---
relaxing the quench approximation; including effects of
expansion; including quantum effects.  Nevertheless,
at this time a quantitative theoretical calculation
is impossible.  The biggest obstacle is that the appropriate initial
conditions (initial meaning just before the transition
occurs) on the long wavelength modes of the order parameter
are not known.  The physics of this chapter rests on much
less solid theoretical foundations than that of the last
chapter.  Away from equilibrium, we cannot use renormalization
group ideas or universality, and our conclusions are
necessarily qualitative.  If seen, long wavelength pion
oscillations and the consequent large fluctuations
in the neutral to charged pion ratio would be a definitive
signature of an out of equilibrium chiral transition.  It
remains an experimental question whether or not these phenomena
occur in heavy ion collisions.

\subsection{Misalignment of the Chiral Condensate}

Among the most interesting speculations regarding
ultra-high energy hadronic or heavy nucleus collisions is the idea
that regions in which the chiral condensate is misaligned may
occur.\cite{anselm,blaizot,recentbjorken,us,kowalski,quenchpaper,bakedalaska}
In such misaligned
regions,
the chiral condensate points in a different direction from that
favoured in the ground state.
If we parametrize the condensate
using the variables of the sigma model, misaligned vacuum regions
are places where the four-component field
$\phi =(\sigma, \vec \pi) $, that in the ground state takes
the value $(v,0)$, is instead
partially aligned in the
$\vec \pi$ directions.  Because chiral symmetry is explicitly broken
by small quark masses, the pion mass is not zero, and therefore
if a misaligned vacuum region were produced, the $\phi$ field
would oscillate about the $\sigma$ direction. Thus, such regions
of misaligned condensate have a nonzero pion field oscillating in time.
They would produce clusters of pions bunched in rapidity
with highly non-Gaussian charge distributions.  A misaligned
vacuum region starting with the field in the $\pi_1 - \pi_2$ plane would
emit only charged pions (equally positive and negative, since the
fields are real), while a misaligned vacuum region starting with the
field pointing
in the $\pi_3$ direction would emit only neutral pions.

More generally if we define
\begin{equation}
f \equiv \frac{ n_{\pi^0}}{n_{\pi^0} + n_{\pi^+ \pi^-} } \ ,
\label{jja}
\end{equation}
then each misaligned region will yield a cluster of pions
with some fixed $f$.  Different regions, of course, will
have different values of $f$.
We can calculate the
probability distribution of the ratio $f$
if we assume that all initial
values on the 3-sphere are equally likely.
We will make
this assumption in order to get a simple analytical result
although it may not be strictly true because the quark
mass term selects a preferred $\sigma$ direction even at high
temperature.
Where $\phi$ starts will determine
in which direction it ends up oscillating about
the sigma direction.   We define angles on the 3-sphere according
to
\begin{equation}
\bigl ( \sigma , \pi_3 , \pi_1 , \pi_2 \bigr ) = \bigl ( \cos \theta ,
\sin \theta \cos \varphi , \sin \theta \sin \varphi \cos \eta ,
\sin \theta \sin \varphi \sin \eta \bigr )~.
\label{jjb}
\end{equation}
Then, the ratio $f$ is given by $f=\cos ^2 \varphi$.
Under the assumption that all initial values on the 3-sphere are
equally probable,
the probability distribution ${\cal P}(f)$ is determined by
\begin{equation}
\int_{f_1}^{f_2}{\cal P}(f)df = {1 \over  \pi ^2}
\int_0^{2\pi} d\eta
\int_0^{\pi} d\theta \sin^2 \theta
\int_{\arccos ( \sqrt{f_2} )}^{\arccos ( \sqrt{f_1} )} d\varphi
\sin \varphi
\label{jjd}
\end{equation}
and turns out to be simply\footnote{This result appears to
have been first known in 1981,\cite{andreev} and has been
rediscovered a number of times since
then.\cite{karmanov,anselm,blaizot,recentbjorken,us,kowalski}}
\begin{equation}
{\cal P}(f) = \frac{1}{2} f^{-1/2} ~.
\label{jje}
\end{equation}
Equivalently, the probability that $f<f_1$ is given by $\sqrt{f_1}$.
If large regions in which $\phi$ is oscillating coherently
develop in a relativistic heavy ion collision, then there should be
clusters of pions with small relative momentum in which
$f$ is constant, and
the values $f$ takes in different such regions should be distributed according
to (\ref{jje}).

As one application of (\ref{jje}),
we note that the probability that the neutral
pion fraction $f$ is less than $.01$ is $0.1$!  This is a graphic illustration
of how different (\ref{jje}) is from what one would expect if individual pions
were independently randomly distributed in isospin space.  It also makes
Centauro events in which less than 1\% of the outgoing particles from a
collision are neutral pions
seem much less surprising than they first appeared.
The analysis of the Centauro data is difficult for several reasons.
Most important of these is the limitation imposed by small statistics.
Also, if a Centauro event occurs too high above
the detector, so many secondary photons will be produced that the
event will not be recognized as a Centauro.
Third, in a Centauro event all of the particles from the collision strike
a small region  of the detector and it is impossible to isolate the
central rapidity region.
The balloon borne JACEE detectors do not have the second and third
problems, but they are only exposed during brief balloon flights
and therefore have very small statistics.
When relativistic heavy ion collisions in a laboratory
colliding beam facility occur
at high enough energies
that there is a low baryon number density central rapidity region,
all of the difficulties
of the cosmic ray experiments will rapidly be overcome.  That will
be the time to look for correlated clusters of pions, and to
look for a distribution like (\ref{jje}).
In the following sections,
we propose and explore a concrete mechanism by which such phenomena
might arise in heavy ion
collisions in which the chiral order parameter
is far from thermal equilibrium.  We then discuss signatures
in heavy ion collision experiments.

\subsection{Emergence of Long Wavelength Pion Oscillations After a
Quench}

In Chapter 2,
we considered the equilibrium phase structure of
QCD.  We argued that QCD with two massless quark flavours probably undergoes
a second-order chiral phase transition.  For many purposes it is a
good approximation to treat the u and d quarks as approximately massless,
and we expect that real QCD has a smooth but perhaps rapid transition as
a function of decreasing
temperature from small intrinsic to large spontaneous chiral
symmetry breaking.  At first sight it might appear that a second-order
phase transition is especially favourable for the development of large
regions of misaligned vacuum.  Indeed the long-lived, long-wavelength
critical fluctuations which provide the classic signature of a second-order
transition {\it are\/} such regions.   Unfortunately
we saw that
light quark masses, even though they are formally much smaller than
intrinsic QCD scales, spoil this possibility.  The pion masses,
or more precisely the inverse correlation length in the pion channel, are
almost certainly {\it not\/} small compared to the transition temperature.
As a result the misaligned regions are modest affairs at best
even near the critical temperature.  They
almost certainly do not contain sufficient energy to
radiate
large numbers of pions.

In this Chapter, we will consider an idealization that is in
some ways opposite to that of thermal equilibrium, that is the
occurence of a sudden
quench from high to low temperatures, in which the
$(\sigma ,\vec \pi)$ fields are suddenly removed from contact with a
high temperature heat bath and subsequently evolve mechanically.
We shall show that long wavelength fluctuations of the
light fields (the pions in QCD, which would be massless if not
for the quark masses) can develop following a quench from
some temperature $T>T_c$ to $T=0$.
The long wavelength modes of the pion fields are unstable and
grow relative to the short wavelength modes.

In a real heavy ion collision, the phase transition must proceed by
some process in between an equilibrium phase transition
in which the temperature decreases arbitrarily slowly
and a quench in which thermal fluctuation ceases instantaneously.
Our goal is to understand the dynamics of
the long wavelength modes of
the chiral order parameter in a quench in the hope
that the qualitative behaviour in this model is representative
of processes occurring in real heavy ion collisions
in which the
long wavelength modes are far from equilibrium, although not quenched.

We shall use the linear sigma model of Gell-Mann and L\'evy\cite{gm}
to describe the
low energy interactions of pions at $T=0$ after a quench:
\begin{equation}
{\cal L}  = \int d^4 x \Biggl\lbrace \frac{1}{2}~\partial^i \phi^\alpha
     \partial_i \phi_\alpha ~-~\frac{\lambda}{4}
     \bigl( \phi^\alpha \phi_\alpha ~-~ v^2 \bigr) ^2 ~+~
      H \sigma ~\Biggr\rbrace ~,
\label{jjg}
\end{equation}
where $\phi$ is a four-component vector in internal space
with entries $(\sigma , \vec \pi )$.
Here, $\lambda$, $v$, and $H\propto m_q$ are to be thought of as
parameters in the low energy effective theory obtained after integrating
out heavy degrees of freedom.
Unlike in the equilibrium case where universality allowed us to
make quantitative predictions, at zero temperature we must make
a non-universal choice of model Lagrangian to obtain equations
of motion for the chiral order parameter.
For now, we
shall treat (\ref{jjg}) as it stands as a classical field theory, since
the phenomenon we are attempting to address is basically classical.
We shall be dealing with energies of order 200 MeV or
less, so that neglect of heavier fields seems reasonable.
Since at zero temperature the $\sigma$ is heavy --- it is a feature
at about 600 MeV which is perhaps too broad to be called a resonance ---
one might consider integrating it out.  This would yield the nonlinear
sigma model.  However, we must implement initial conditions
appropriate to a disordered state in which the $\vec \pi$ and $\sigma$ fields
are equivalent up to small effects due to explicit chiral symmetry
breaking.  Therefore, we must keep the sigma
and use (\ref{jjg}).

To model a quench, we begin
at a temperature well above $T_c$.  The typical
configurations have short correlation lengths
and $\langle \phi \rangle \sim 0$.  ($\langle \phi \rangle
\neq 0$ because $H \neq 0$.)  One then takes the temperature
instantaneously to zero.  The equilibrium configuration
is an ordered state with the $\phi$ field aligned in the $\sigma$
direction throughout space, but this is not the configuration
in which the system finds itself.  The actual, disordered
configuration then
evolves
according to the Lorentz invariant
zero temperature equations of motion obtained
by varying (\ref{jjg}).  Quenching in magnet models has been much studied
in condensed matter physics.\cite{bray}  However in that context it
is usually
appropriate to use diffusive equations of motion, because the magnet is
always in significant contact with other light modes ({\it e.g.}, phonons).
For this reason the condensed matter literature we are aware of
does not directly apply
to our problem.

We will discuss numerical simulations of quenching to zero temperature
in the linear sigma model with an explicit symmetry breaking term $H \sigma$
which makes the pions massive.
Turok and Spergel\cite{turgel} have considered
this scenario with no explicit symmetry breaking term as a cosmological
model for large scale structure formation in the early universe.
They find a scaling solution in which the size of correlated
domains grows without bound at the speed of light.
This is easily understood qualitatively, and is peculiar to the
case where $m_\pi=0$.
The field $\phi$ is at different points on the vacuum manifold in
regions of the universe which have not been in causal contact
after the quench.
Because the vacuum manifold is degenerate in potential energy,
the system evolves to reduce gradient energy.
As time continues, larger and larger
regions come into causal contact, align, and the size of correlated
domains grow at the speed of light.
When the
$O(N)$ symmetry
is explicitly broken, however, we do not expect a scaling solution.
The $H \sigma$ term tilts the potential,
the vacuum manifold is not
degenerate,
and in a time of order $m_\pi^{-1}$ the scalar field
in all regions (whether in causal contact or not) will be oscillating
about the sigma direction, and the physics cannot be described
in terms of differently aligned domains coming into alignment.

\subsubsection{A Numerical Simulation}

We now describe numerical simulations of quenching in the linear
sigma model which were reported in Ref. [\citelow{quenchpaper}].
We choose $\phi$ and $\dot \phi$
randomly independently on each site of a cubic lattice.  This means
that the lattice spacing $a$ represents the correlation length
in the initial configuration.  In the initial conditions,
$\phi$ is disordered,
the $O(N)$ symmetry is not spontaneously broken, and the lattice
spacing $a$ represents
the $\pi$ and
$\sigma$ correlation lengths which are approximately
degenerate.
In the simulation whose results are shown
in Fig. 11,
we chose $\phi$ and $\dot \phi$ randomly from Gaussian
distributions centred around $\phi = \dot \phi =0$ and with
$\langle \phi^2 \rangle^{1/2} = v/2$ and
$\langle \dot\phi^2 \rangle ^{1/2} = v$.
The three parameters $v$, $H$, and $\lambda$ in (\ref{jjg})
determine
$m_\pi$, $m_\sigma$, and $f_\pi=\langle 0|\sigma |0\rangle$
according to
\begin{equation}
\lambda \langle 0 | \sigma |0 \rangle \Bigr(
\langle 0 | \sigma |0 \rangle^2 ~-~v^2 \Bigr) ~-~ H ~=~0~~,
\label{jjh}
\end{equation}
\begin{equation}
m_\pi^2={H \over  \langle 0 | \sigma |0 \rangle }~~,~~~~{\rm and}~~~~
m_\sigma^2=3\lambda  \langle 0 | \sigma |0 \rangle^2 - \lambda v^2~~.
\label{jji}
\end{equation}
Note that $f_\pi = \langle 0| \sigma |0 \rangle > v$ for $H\neq 0$.
In interpreting our results we must remember that
while in the code we are free to choose the energy scale
by setting the lattice spacing $a=1$,
$a$ actually represents the initial correlation length.
In choosing
the parameters for the simulation shown in Fig. 11, we
assumed that $a=(200~{\rm MeV})^{-1}$ and then chose
$v=87.4 ~{\rm MeV} = 0.4372~a^{-1}$,
$H=(119 ~{\rm MeV})^3 = 0.2107~a^{-3}$,
and $\lambda = 20.0$ so that $f_\pi=92.5~ {\rm MeV}$,
$m_\pi=135~ {\rm MeV}$,
and $m_\sigma = 600 ~{\rm MeV}$.

\begin{figure}
\centerline{
\epsfysize=120mm
\epsfxsize=155mm
\epsfbox[73 45 738 549]{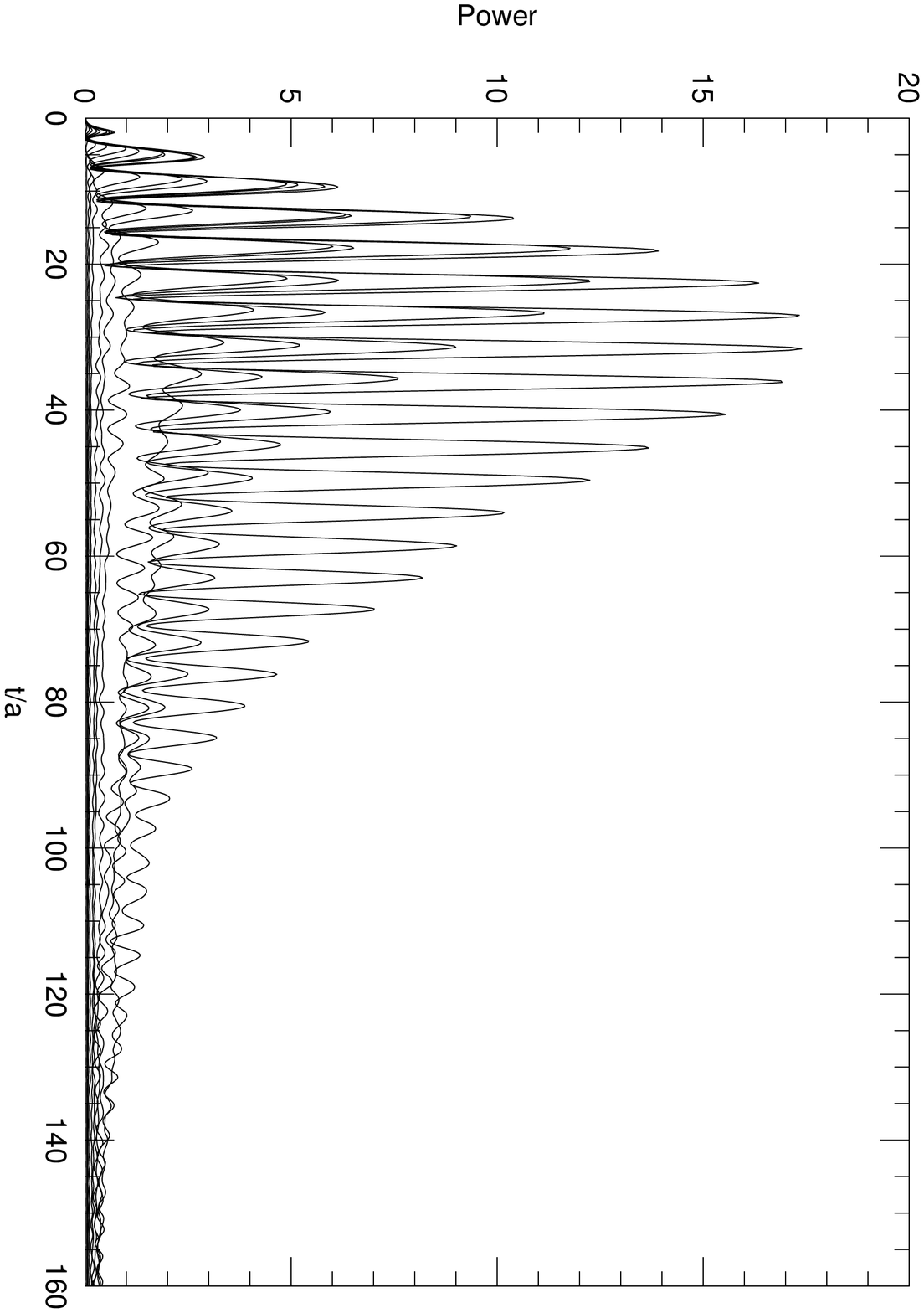}
}
\fcaption{
The time evolution of the power spectrum of
the pion field at several spatial
wavelengths after a quench at $t=0$.
The parameters of the potential
and the initial configurations
for $\phi$  and $\dot \phi$ are described in the text.
Time is measured in lattice units, and since $a=(200{\rm ~MeV})^{-1}$
the unit of time is approximately $1 {\rm ~fm}$.
The figure shows the time evolution of the angular
averaged power spectrum (defined in the text)
of one of the three components of the pion field.
The curves plotted are (top to bottom in the figure)
the average power in the modes in the momentum bins centred at
$ka=$~0.20, 0.26, 0.31, 0.37, 0.48, 0.60, 0.71, 0.94, 1.16, 1.39, and 1.84.
The initial power spectrum is white and all the curves start
at $t=0$ at approximately 0.01 in lattice units.
The power in the
longest wavelength pion modes is amplified by a factor of order
1000 relative to that in the shortest wavelength modes, which
is not amplified at all.
{}From Ref. [\citelow{quenchpaper}]; extended to later times
here than in Ref. [\citelow{quenchpaper}].}
\label{fig:pionosc}
\end{figure}
With parameters and initial conditions chosen, we evolved
the initial configuration according to the equations of motion
using a standard finite difference,
staggered leapfrog scheme.
The simulation
was performed on a $64^3$ lattice
and the time step
was $a/10$.
After every two time steps,
we computed the spatial
fourier transform of $\phi$.  For each component of
$\phi$, we computed the power
\begin{equation}
P(\vec k,t) = \frac{1}{L^3}
\Biggl| \int {\rm d}^3x \exp (i \vec k \cdot \vec x)
\phi(\vec x,t) \Biggr|^2 =
\frac{1}{L^3}\int {\rm d}^3y \int {\rm d}^3x
\exp (i \vec k \cdot \vec x ) \phi (\vec x + \vec y,t) \phi (\vec y,t)
\label{powerdef}
\end{equation}
and averaged over all
modes with $k\equiv |\vec k|$ in bins of width $0.057a^{-1}$.
$L^3=(64a)^3$ is the lattice volume.
Fig. 11 shows
the power in modes of one component of the pion field
with spatial wave vectors of
several different magnitudes $k$ as a function of time.
The curves all start at approximately the same value at
$t=0$ because the initial power spectrum is white
since we chose $\phi$ independently at each lattice site.
For a lattice with an infinite number of sites,
with the normalization used in (\ref{powerdef})
the initial value of the power spectrum is related
to the mean square of the initial $\phi$ distribution by
$P(\vec k,0) = (1/N) \langle \phi^2 \rangle$, with $N=4$.
This means that
all the curves in Fig. 11 begin at $t=0$ at about 0.01 in lattice units.
The behaviour of the low momentum pion modes is striking.
While the initial power spectrum is white and, as
ergodicity arguments would predict, the system at late times
is approaching an equilibrium configuration in which the
equipartition theorem holds, at intermediate times of order
several times $m_\pi^{-1}$ the low momentum pion modes are oscillating
with large amplitudes.

We verified that
finite size effects are not important even for the longest
wavelength ($ka=0.20$) mode shown in Fig. 11 by checking that the
behaviour of the $ka=0.31$ mode is the same in a $32^3$ and
a $64^3$ box.  The longer the wavelength of a mode, the
more that mode is amplified.  This holds up to the longest
wavelength mode for which we can trust the results of the
simulation, namely the $ka=0.20$ mode which corresponds
to a wavelength $\lambda=2\pi /k$ of about half the lattice size.

The behaviour of only one component of the
pion field is shown in Fig. 11.  The other two look qualitatively
the same.  It should be noted that the exact height of the peaks
in the curves depend on the specific initial conditions.
If the simulation is run with the same initial distributions for
$\phi$ and $\dot \phi$ but with a different seed for the random
number generator, the heights of the peaks
change, and the relative sizes of the peaks in the three different
pion directions change.  The qualitative features of Fig. 7 ---
the growth of long wavelength modes of the pion field --- do not depend
on the specific realization of the initial conditions.
Amplification occurs for a range of choices of the parameters
in the potential including physically motivated values.
We will discuss the effects of varying the choice of initial
distributions below.

These numerical simulations have been repeated by
Gavin, Gocksch and Pisarski,\cite{ggp2} who report
that they have replicated the power spectra of Fig. 11.
There has been some confusion in the literature, however.
A number of authors have tried to describe the
result seen in Fig. 11 in terms of a correlation length.
In thermal equilibrium, this would be reasonable,
as the power spectra would be completely specified
by a single length --- the correlation length.
The essence of the phenomena of Fig. 11
is, however, that the spectra are nonthermal.  They
cannot be described by a single length scale, and
computing a correlation length is meaningless.
The classical simulation of Fig. 11 ceases to be physically
relevant at late times, when instead of classical
fields one has noninteracting pions.  Fig. 11 has
been extended formally to late times, however, to show how
at late times in
the classical simulation the energy is equipartitioned
and thermal equilibrium is reached.  Hence, comparing
intermediate and late times in the figure is instructive.
At intermediate times, short wavelength modes have
about the same amplitude as they do at late times
in thermal equilibrium, while  long wavelength modes have
been  amplified.  It is impossible to describe this
behaviour in terms of a single length scale.
For a visual image, think of long wavelength ocean swells
with short wavelength
chop in addition; perhaps it is easier just to look at Fig. 11.
The essential qualitative phenomenon
whose cause and consequences we now discuss
is that long wavelength pion oscillations
have been excited --- the short wavelength modes have
not been damped out.  This picture is more complicated
than the idealized ``smooth'' region of disoriented
chiral condensate described in the previous section.

Although no meaningful correlation length
or domain size can be defined,
it is possible to describe the physics using
the time dependent correlation functions
\begin{equation}
C(\vec x,t) = \int {\rm d}^3y \phi(\vec x + \vec y,t)\phi(\vec y)
\label{corrfneq}
\end{equation}
for each component of $\phi$.  For each component of $\phi$,
the power $P(\vec k,t)$ is the spatial Fourier
transform of this correlation function.  Since the essence of the
physics we are describing is the amplification of long
wavelength power, a description using the power spectrum
as in Fig. 11 is the natural one, but the same information is
of course encoded in the correlation function.
Another reason why a description of the physics
using $P(\vec k,t)$ is more natural than one using
$C(\vec x,t)$ is that it is closer to what experimentalists
measure --- they measure the momenta of particles not
positions in space. Long wavelength oscillations
seen as amplification of $P(\vec k,t)$ at some
small magnitude $\vec k$
become pions with similar momenta
(in the $\hat k$ direction) which are
close to each other in an experimentalist's detector.

\subsubsection{Understanding the Results}

The emergence of long wavelength pion oscillations in
the numerical simulations is a striking qualitative
phenomenon.  We now discuss a simple qualitative
explanation for the result.
Whenever one has spontaneous breaking of a
continuous global symmetry, massless
Nambu-Goldstone bosons occur.  The masslessness of these modes occurs
because of a cancellation:
\begin{equation}
m^2~=~ -\mu^2 ~+~ \lambda \phi^2~,
\label{jjf}
\end{equation}
where the second term arises from interaction with a condensate whose
expectation value $\langle \phi \rangle^2$
satisfies $\langle \phi \rangle^2=\mu^2/\lambda~\equiv~v^2$
in the ground
state.  However following
a quench the condensate starts with its average
at $0$ rather than $v$,
and generally oscillates before settling to its
final value.  Whenever $\langle \phi \rangle ^2 < v^2$, $m^2$ will be
negative, and {\it sufficiently long wavelengths fluctuations in the
Nambu-Goldstone
boson field will grow exponentially}.
The same mechanism will work, though
less efficiently, if one has only an approximate symmetry and approximate
Nambu-Goldstone modes, or for that matter
other modes whose effective $m^2$ is
`accidentally' pumped negative by their interactions with the condensate.

Let us compare the growth of long wavelength modes found numerically
with expectations based on the mechanism just discussed.
Suppose the potential $V$ in (\ref{jjg})
were simply $V(\phi) = (m^2/2)\phi^\alpha
\phi_\alpha$.  Then, the equations of motion would be linear, and
modes with different spatial wave vector $\vec k$ would be uncoupled.
Each curve in Fig. 11 would be a sinusoid with period $\pi / \sqrt{m^2 +
\vec k ^2}$ and constant amplitude.  (The power spectrum, being quadratic
in the fields, oscillates with one half the period of the fields.)
The period of the oscillations in Fig. 11 is indeed given by
$\pi / \sqrt{m_\pi^2 + \vec k ^2}$, but the amplitudes are far from
constant.  This behaviour can be understood qualitatively by approximating
$\phi^2$ in the nonlinear term in the equation of motion
by its spatial average:
\begin{equation}
\phi^\alpha \phi_\alpha (\vec x, t) \sim \langle \phi_\alpha \phi^\alpha
\rangle (t) ~~.
\label{jjj}
\end{equation}
Note that the Hartree or mean field
approximation (\ref{jjj}) is not
controlled.  We make it only to obtain some
understanding of the qualitative behaviour of the system.
Using (\ref{jjj}) and doing the spatial fourier transform, the equation
of motion for the pion field becomes
\begin{equation}
{d^2 \over dt^2} \vec \pi (\vec k,t) =
- m_{eff}^2 (k,t) \vec \pi (\vec k,t)
\label{jjk}
\end{equation}
where the time dependent ``mass'' is given by
\begin{equation}
m_{eff}^2 (k,t) \equiv - \lambda v^2 + k^2
+ \lambda \langle \phi^2 \rangle (t)
\label{jjl}
\end{equation}
and where $k=|\vec k|$.

\begin{figure}
\centerline{
\epsfysize=90mm
\epsfbox[36 54 756 576]{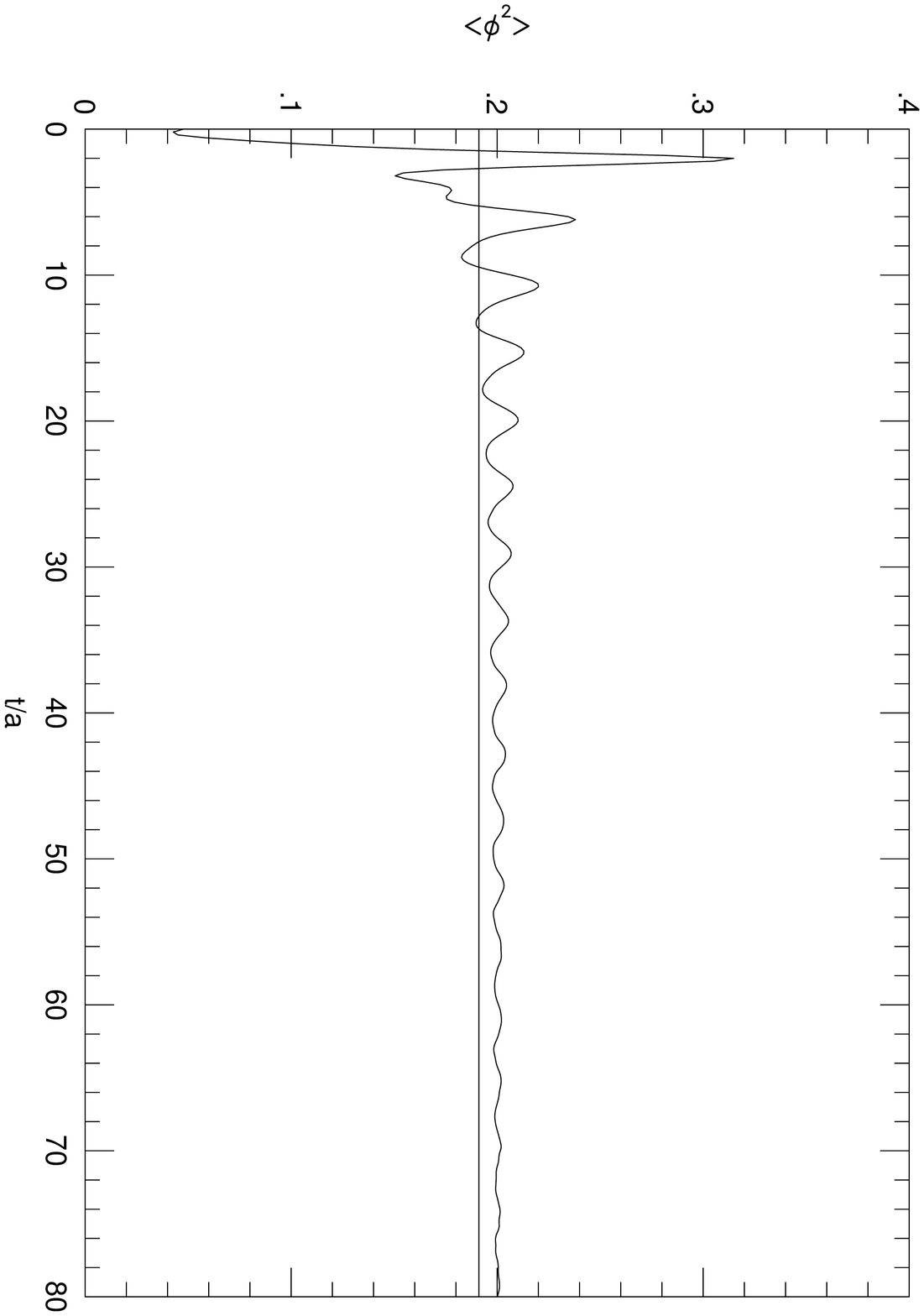}
}
\fcaption{Time evolution of the spatially averaged
$\langle \phi^2 \rangle$ for the same simulation whose results
are shown in Figure~1.  The horizontal line is at
$\langle \phi^2 \rangle = v^2$.  From Ref. [\citelow{quenchpaper}].}
\label{fig:vevosc}
\end{figure}
Fig. 12 shows the time evolution of $\langle \phi^2 \rangle$ in the
same simulation whose results are shown in Fig. 11.
In the
initial conditions, $\phi$ is Gaussian distributed with
$\langle \phi^2 \rangle < v^2$.
Therefore, for a range of wavevectors with $k$ less than
some critical value, $m_{eff}^2 < 0$ and the long wavelength
modes of the pion field start growing exponentially.
$\langle \phi^2\rangle$ grows and then executes damped oscillations
about its ground state value
$\langle 0|\sigma |0 \rangle ^2$.
A wave vector $k$ mode of the pion field is
unstable and grows exponentially whenever $\langle \phi^2 \rangle <
v^2 - k^2/\lambda$.  Modes with $k^2 > \lambda v^2$ can never be unstable.
The $k=0$ mode is unstable during
the periods of time when the $\langle \phi^2
\rangle$ curve in Fig. 12 is below $v^2$.
Since $\langle \phi^2 \rangle$ is
oscillating about
$\langle 0|\sigma |0 \rangle^2 > v^2$, after some time the
oscillations have damped enough that $\langle \phi^2 \rangle$ never drops
below $v^2$, and from that time on all modes
are always stable and oscillatory.
In general, longer wavelength modes are unstable
for more and for longer
intervals of time than shorter wavelength modes
as $\langle \phi^2 \rangle$ oscillates.
Also, $m_{eff}^2$ is more negative and more growth occurs for
modes with smaller $k$.
Making the approximation (\ref{jjj}) and thus using (\ref{jjk})
cannot be expected to completely reproduce the effects
of the nonlinear term in (\ref{jjg}) which is local in position space.
Nevertheless, it predicts that the long wavelength modes of
the pion field go through periods of
exponential growth and predicts that the longer the wavelength
the greater the amplification,
and therefore
gives us some understanding of the behaviour of these modes
in Fig. 11.
In an equilibrium phase transition, explicit symmetry breaking
keeps the correlation length at $T_c$ finite and,
in the case of QCD, too short to be of interest.
After a quench, on the other hand,
arbitrarily long
wavelength modes of the pion field are amplified even though
the pion mass is nonzero.

What about the sigma modes?
In Ref. [\citelow{quenchpaper}] the angular averaged power
spectrum for the sigma field is plotted, as was
done in Fig. 11 for one component of the pion field.
For $\sigma^2 < v^2 /3$, the effective
mass squared for the $k=0$ mode is negative.  This condition
is far less likely to be satisfied than $\langle \phi^2 \rangle < v^2$.
Hence, while some growth of the
low momentum $\sigma$ modes is
possible, they do not grow nearly as much as the
low momentum pion modes do.\cite{quenchpaper}

At late times $\langle \phi^2 \rangle$
oscillates with small enough amplitude that it never
goes below $v^2$ and consequently no modes are ever unstable.
Hence, if we make the approximation (\ref{jjj}) we would expect that at
late times each mode in Fig. 11 would continue to oscillate with
approximately constant amplitude, with the longer wavelength modes
maintaining the large amplitudes acquired during their exponential
growth spurts.
That this is not what is seen in Fig. 11
reflects the effects we neglected in making the approximation (\ref{jjj}).
Because
the modes are in fact coupled and the equations of motion are
actually nonlinear, ergodicity arguments suggest that eventually
equipartition should apply.
If the energy is equally divided among
modes, then $({\rm amplitude})^2 (m_\pi^2 + k^2)$ should be approximately
constant.  Hence, the power in a mode at late times should be
approximately constant in time and inversely proportional to
$(m_\pi^2 + k^2)$, and this is indeed the behaviour at
late times in Fig. 11.
It is reasonable that longer wavelength modes take
longer to decrease in amplitude than shorter wavelength modes.
As the system heads toward equipartition, power has to flow from
low momentum modes to higher momentum modes.
This plausibly
happens more quickly at higher momentum because higher momentum
modes are coupled to a larger number of modes, or, in the continuum,
a larger volume of momentum space.
Hence, we end up with the striking phenomena of Fig. 11:
while the system begins and ends  with
small $\vec \pi (\vec k,t)$, in between the long wavelength pion modes
are dramatically amplified.

The late time behaviour of the
simulation in Fig. 11 is not physical.
When the expansion and cooling of the plasma is taken
into account, there will be a time after which the
energy density is low enough that
the description in terms of classical fields
no longer makes sense.  After this time, one has individual
pions flying off towards the detector.
{}From Fig. 11, we see that if this occurs sometime
in a broad range of times of about several tens of fm,
it is plausible
that at this time long wavelength modes of the
pion field are greatly
amplified compared to modes with higher $|k|$.

It is important to note that we have only used the
Hartree approximation (\ref{jjj}) as a qualitative guide.
The emergence of long wavelength pion oscillations apparent
in Fig. 11 occurs in a simulation of the
strongly coupled nonlinear
equations of motion.  We have already seen that the Hartree
approximation leads to incorrect conclusions about the
late time behaviour in Fig. 11.
In fact, the Hartree approximation
can lead one astray in another way if it is taken too literally.
A number of authors have noted\cite{ggp2,boyanovsky}
that $\langle \phi^2 \rangle$ of Fig. 12 is less
than $v^2$ only at early times.  Indeed, in the Hartree
approximation, after a time of order $m_\sigma^{-1}\sim 1/3 ~{\rm fm}$,
$\langle \phi^2 \rangle$ is oscillating about its vacuum
value.\cite{ggp2,boyanovsky}  This suggests that
the growth of long wavelength modes should cease after
a time of this order.  However, from Fig. 11
we see that such growth occurs for much longer times --- until
$t\sim 30~{\rm fm}$.
Thus, in this respect, the Hartree approximation fails
to give a qualitative explanation of the phenomena
observed in the full nonlinear equations.
The possibility that the Hartree approximation can
fail in this way was noted by Boyanovsky {\it et al.};\cite{boyanovsky}
Fig. 11 is evidence that it does fail in this way.

In section 3.1, we discussed the phenomenological
consequences of long wavelength oscillations of the pion field.
Regions in which the oscillations are in the $\pi_3$ direction
become clusters of neutral pions, while regions in which
the oscillations are in the $\pi_1-\pi_2$ plane become clusters
of charged pions.  Implicit in this description was the
assumption that the latter clusters include pions with
both positive and negative electric charge, and are approximately
neutral.
We can now justify this assumption.\cite{borlange}
While the total electric charge must be conserved, there is
no conservation law forbidding the emergence of regions
of net positive and negative charge.
We must determine
whether long wavelength oscillations of the electric
charge density $\rho^e$ occur.  The electromagnetic current
due to the charged pions is given by
\begin{equation}
j_\mu^e = \pi_1 \partial_\mu \pi_2 - \pi_2 \partial_\mu \pi_1
\label{jjla}
\end{equation}
so the charge density is
\begin{equation}
\rho^e = \pi_1 \dot \pi_2 - \pi_2 \dot \pi_1~.
\label{jjlb}
\end{equation}
It is easy to verify that the equations of motion obtained
by varying (\ref{jjg}) do indeed conserve $j_\mu^e$.
Because $j_\mu^e$ is conserved, we expect it to
be more difficult to excite long wavelength charge
oscillations than to excite long wavelength oscillations
of $\pi_1$, $\pi_2$, or $\pi_3$.
If we write $(\pi_1,\pi_2)$ as a complex field $r \exp (i \theta)$
then $\rho^e = r^2 \dot \theta$.  We have seen that there are periods
of time when the pion field is exponentially unstable.  This instability
which manifests itself dramatically in Fig. 11 results in growth of
the magnitude of $\vec \pi$.  Thus, it leads to growth of $r$ but does
not tend to change $\theta$.  Equivalently, the instability
tends to kick $\vec \pi$ radially away from $\vec \pi = \vec 0$,
and results in oscillations with
$\vec \pi$ oscillating approximately through the origin
with $\dot \theta \sim 0$ rather than with $\vec \pi$ ``orbiting''
the origin with $\dot \theta \neq 0$.
We have verified that the dramatic phenomena of Fig. 11 are not
manifest in the electric charge density $\rho_e$ by
computing the spatial fourier transform of $\rho^e$ in the
simulation of Fig. 11.  First, we have verified that the total charge
in the $64^3$ box is indeed conserved.  Second, we found that the
low $k$ modes of $\rho^e$ do not grow dramatically.
Long wavelength oscillations of $\pi_1$, $\pi_2$, and $\pi_3$
are dramatically amplified
after a quench, while long wavelength oscillations of the
conserved electric
charge density $\rho_e$ are relatively unaffected.

\subsubsection{Other Scenarios}

Bjorken, Kowalski, and Taylor have proposed
that regions of disoriented chiral condensate
may arise in proton-proton
collisions.\cite{recentbjorken,kowalski,bakedalaska}
They suggest that after a collision, a ``cold'' region
throughout which the condensate is
misaligned in a similar direction in isospin space
is surrounded
by a ``hot'' region of outgoing particles --- hence
the name Baked Alaska for the scenario.
The MiniMax experiment\cite{minimax} at Fermilab
(J. D. Bjorken and C. C. Taylor, co-spokespersons)
is currently taking data to search for such phenomena.
The scenario we have described
is somewhat different, in that while long wavelength
oscillations of the chiral condensate arise, we do
not envision ``smooth'' regions of disoriented condensate.
The long wavelength sloshing, although dramatic because
it is completely different than what is obtained in thermal
equilibrium, nevertheless has short wavelength oscillations
superimposed on top of it.
The emergence of long wavelength pion oscillations as
described in this article requires that a large volume of space
(certainly several fm across) is first in a high
temperature phase in which the condensate is
disordered ({\it n.b} disordered not disoriented).
The volume has to be large enough that it makes sense to
discuss long wavelength modes of the chiral order parameter.
In this volume, the plasma must then experience a
transition to the low temperature phase which
occurs in such a fashion that the long wavelength
modes of the order parameter do not stay in thermal
equilibrium.
We suggest that this physics
may be relevant in the central
rapidity region of a sufficiently energetic heavy ion collision,
but it is not
clear that it is possible to use this sort of language
to describe a proton-proton collision.  Although
the Baked Alaska scenario is somewhat different
than the one we are discussing, it has similarities.
It is certainly far from thermal equilibrium.  Furthermore,
Bjorken\cite{latestbjorken} has suggested that the processes
at the boundary between the ``hot'' and ``cold'' regions
of the Baked Alaska can be described as a quench.
Finally, and most important, although the settings are
different --- hadron-hadron collisions vs. heavy ion collisions ---
at the end of the day long wavelength pion oscillations
lead to large fluctuations in the neutral to charged pion
ratio in either case.

The idea that coherent pion emission from a classical
pion field occurs in collision experiments has
a long history,\cite{heisenberg,hornsilver,botke,andreev,karmanov}
and solutions to the classical equations of motion
have been studied in other contexts also.\cite{enikova}
In addition to the ideas we have described to this point,
another possibility which has been discussed in the literature
is that when there are a lot of pions in a small volume
the fact that they are bosons can play a role --- bose
condensation can occur,\cite{lamlo} and the wave function
must be symmetrized.\cite{pratt}  This physics can have
the effect of broadening the probability distribution
for $f$ from a narrow Gaussian, although it does not lead
to a $1/\sqrt{f}$ distribution.  Pehaps this
could be associated with Centauro events.
Greiner, Gong and M\"uller\cite{ggm}
have argued that for these effects to be significant,
pion densities larger than those likely to occur at RHIC
are necessary, and that in addition if such large densities
were to occur, a description in terms of a bose gas of pions would
not be appropriate.
There have also been recent
proposals that classical pion fields could arise in
hadron-hadron collisions\cite{amado,martinis} and in the
leading (high-rapidity) regions of heavy ion collisions.\cite{martinis}

It may turn out that the amplification of long wavelength
modes described in this article may not be the appropriate
explanation for the Centauro events, which may turn out to
be hadron-hadron or hadron-nucleus collisions in which
a large volume in which the chiral order parameter in
its high temperature phase is never formed.  Related
ideas like Bjorken, Kowalski, and Taylor's
Baked Alaska scenario for the formation of a
region of disoriented chiral condensate,
or some of the other ideas just mentioned
may turn out to be more appropriate descriptions of Centauro events,
and the interesting phenomena which we hope Bjorken, Taylor, {\it et al.}
discover experimentally\cite{minimax} in proton-proton collisions.

In studying
the chiral phase transition, perhaps the role of the Centauro
events will prove to be merely inspirational, and not of direct
relevance.  Whatever the inspiration, it is important
to consider the qualitative phenomena which are possible
when the plasma in the central rapidity region of a
heavy ion collision goes from the high temperature phase
in which the chiral order parameter is disordered to
the low temperature phase in such a way that long wavelength
modes of the order parameter do not stay in equilibrium.
We have used the quench as an idealization which may
capture the relevant non-equilibrium physics.
We have seen numerical evidence (supported by qualitative understanding)
that long wavelength oscillations
of the order parameter in the pion directions may be
greatly amplified, whereas this does not happen
if thermal equilibrium is maintained.  In the next sections,
we describe  various improvements to the treatment,
and discuss experimental consequences.

\subsection{Initial Conditions, Expansion, and Quantum Effects}

The results shown in Fig. 11 do depend on the initial
distributions for $\phi$ and $\dot \phi$.  If the initial
conditions are such that the energy per unit volume
in the simulation is much  more than $\frac{\lambda}{4} v^4$,
no striking growth of long wavelength modes occurs.
The system quickly achieves a configuration in which
the energy is equipartitioned among modes corresponding
to thermal equilibrium at a temperature {\it above} $T_c$.
Thus, in order for long wavelength modes to grow as in
Fig. 11, the energy of the fluctuations in $\phi$ in the
initial conditions must be less than it would be in thermal
equilibrium.  Another way of saying this is that whereas
in the initial condition for Fig. 11 $\langle \phi \rangle$
is in a configuration appropriate to a high temperature,
$\langle \phi^2 \rangle$ and $\langle \dot\phi^2 \rangle$
have magnitudes characteristic of a lower temperature.
These admittedly {\it ad hoc} quench initial conditions
were chosen in the hope of capturing qualitative behaviour
typical of more general nonequilibrium physics.  We now discuss
efforts to improve upon this treatment.

Gavin and M\"uller\cite{gavmul} note that if upon starting with
quench initial conditions, the condensate were to evolve
from $\langle \phi \rangle \sim 0$
to $\langle \phi \rangle \sim v$
much
more slowly than it does in the classical linear sigma model
considered above, then the amplification of
long wavelength pion oscillations would be much greater than
in Fig. 11.
They call this scenario annealing, although
this is a somewhat different usage than in condensed matter
physics.  Annealing usually means slow cooling in which
both condensate and fluctuations stay close to their thermal
equilibrium configurations as the temperature is slowly
reduced.  In this standard annealing, since approximate
thermal equilibrium is maintained, no growth of long
wavelength modes would occur.
Gavin and M\"uller's scenario, in which the
condensate evolves slowly (as in standard annealing) but
in which nonequilibrium conditions which allow amplification
of long wavelength modes are maintained nevertheless,
is in a sense more out of equilibrium than a quench.
It would be marvelous if this turned out to be the appropriate
idealization for treating heavy ion collisions, but that
may be too much to hope for.

A much more ambitious project than the classical simulations
described so far is to include the effects of quantum
fluctuations (in addition to classical
thermal fluctuations) on the dynamics of the order parameter.
A number of groups have made progress in this
direction.\cite{das,boyanovsky,kluger}
In order to treat the quantum fluctuations, however, all
groups have had to make linearizing approximations
either by working in the Nambu--Jona-Lasinio model,\cite{das}
or by using the Hartree approximation\cite{boyanovsky}
or the  large-N approximation.\cite{kluger}
Such approximations cannot be avoided at present,
but we have already seen that they are fraught with risk.
Using the Hartree approximation to study the classical
simulations of Fig. 11 leads to the prediction that growth
of long wavelength modes occurs for a much shorter time
than is in fact the case.  Thus, it is perhaps not surprising
that in treatments of a quench
in which quantum effects have been included,
short correlation lengths have been found.\cite{das,boyanovsky}
Also, as we have noted before, the essence of the phenomenon
discovered in the classical simulations cannot be described
simply as the development of a long correlation length.
Long wavelength modes are amplified, while short wavelength
oscillations remain.  Further work is needed in order to
understand quantum effects on the dynamics.

A less ambitious improvement on the simulations of the
previous section is to include the effects of expansion.
Following Bjorken,\cite{bjorken} we wish to implement
the boost invariant longitudinal expansion characteristic of the
central rapidity region of a sufficiently
energetic heavy ion collision.  The natural
coordinates to use are the rapidity $y$ of equation (\ref{posrap})
and the proper time $\tau$ of equation (\ref{taudef}), instead
of the time $t$ and the longitudinal position $z$.
This can be implemented by taking the linear sigma model
equations of motion and performing the replacement
\begin{equation}
\Bigl[ \frac{\partial^2}{\partial t^2} - \frac{\partial^2}{\partial z^2}
\Bigr] \ \rightarrow \ \Bigl[ \frac{1}{\tau}\frac{\partial}{\partial \tau}
\Bigl( \tau \frac{\partial}{\partial \tau} \Bigr) - \frac{1}{\tau^2}
\frac{\partial^2}{\partial y^2} \Bigr] \ .
\label{replacement}
\end{equation}
The first treatments to include expansion were $1+1$ dimensional ---
they assumed the configuration was uniform in the transverse
directions.\cite{blaizot,krz1,blaizotkrz,khlebnikov,kogan1,huangwang,krz2}
By making the further simplifying assumption
that $\phi$ is independent
of rapidity, Blaizot and Krzywicki\cite{blaizot,blaizotkrz}
reduce the problem to a $0+1$ dimensional nonlinear dynamical
system.  Following the evolution of these equations is instructive
for analyzing the physics of a large smooth disoriented chiral
condensate as envisioned in the Baked Alaska scenario of
Bjorken {\it et al.},
and is a good toy model
in which calculations can be done without
resorting to numerical work.\footnote{Some
authors\cite{anselm,blaizot,latestbjorken} have
used the further simplification gained by
working in the nonlinear sigma model
in which there is no $\sigma$ degree of freedom.
In this article, we are interested
in the amplification of long wavelength pion oscillations
from an initial state in which $\phi$ is disordered
and the $\vec \pi$ and $\sigma$ fields are
equivalent up to small effects due to explicit
chiral symmetry breaking. We
have therefore
had to use the linear sigma model.  If, however, one is interested
in studying the evolution of long wavelength pion oscillations
at later times
without regard to how they may have arisen
at earlier times, then the nonlinear sigma
model suffices.}~~
Krzywicki\cite{krz1} has considered
how the dynamics of this simple system can be affected by
coupling to a bath representing all the other degrees of
freedom in the system.  Huang and Wang\cite{huangwang}
have done a $1+1$ dimensional simulation in which $\phi$
and $\dot\phi$ are random functions of rapidity with
the same quench initial distributions that were used in the simulation of
Fig. 11.  They find that oscillations in the $\pi$ fields as
a function of rapidity
develop over distances in rapidity of up to 2 or 3.

One reason why it is important to include expansion is
that one can then ask whether it is possible to begin with
initial conditions in thermal equilibrium at a high temperature
and have something analogous to a quench occur as the system
cools by expanding.  This question has been addressed in
numerical simulations by Asakawa, Huang, and Wang\cite{asakawa}
and by Cooper {\it et al}.\cite{kluger} We will describe
the work of both groups below, but let us
first consider the qualitative answer to the question just posed.
Both groups find that starting with thermal equilibrium,
one stays close enough to thermal equilibrium that
no significant growth of long wavelength modes occurs.
Both groups also find, however, that if the long wavelength
modes of the condensate are not in thermal equilibrium at
the start of their simulations, then long wavelength pion
oscillations can, in fact, be amplified.  It would have been
wonderful if simply adding expansion to the linear sigma model
had been enough to turn thermal equilibrium into a quench,
but perhaps this was too much to ask.
Real heavy ion collisions are much more complicated than
the linear sigma model, however.  Indeed, using the sigma model
to describe the dynamics cannot be valid for $T\gsim T_c$.
The unanswered question
can be phrased as follows:  In a heavy ion collision, at the
time when it becomes appropriate to use the linear sigma model
to describe the dynamics, are the long wavelength modes of the
order parameter in thermal equilibrium with the short wavelength
modes and the other degrees of freedom, or are they out of
equilibrium?  Nobody knows the answer.  If they are out of
equilibrium, then perhaps the quench discussed above or the
quench with expansion to be discussed below yield a
qualitative description of the
physics.  If that is the case, then long wavelength
pion oscillations arise.

Asakawa, Huang, and Wang\cite{asakawa} have simulated the
classical
equations of motion of the linear sigma model
in $2+1$ dimensions, where the two spatial dimensions
are the transverse ones.  They
include the effect of
longitudinal expansion by using $(1/\tau)\partial/\partial \tau
(\tau \partial/\partial \tau )$ instead of $\partial^2 /\partial t^2$
in the equations of motion, but they do not do a $3+1$ dimensional
simulation.  They are therefore able to use a large 2-dimensional
lattice, and can implement transverse expansion.  They do so by
choosing initial conditions which are vacuum outside a cylindrical region
of transverse radius 5 fm, and with $\phi$ disordered with
a correlation length of 0.5 fm and with
$\langle \phi \rangle=0$ inside
the cylindrical region.  They do two simulations.  The first
has $\langle \phi^2 \rangle$ large enough to approximate
thermal initial conditions, and not much happens in this simulation.
\begin{figure}
\centerline{
\epsfysize=140mm
\epsfbox[144 108 504 657]{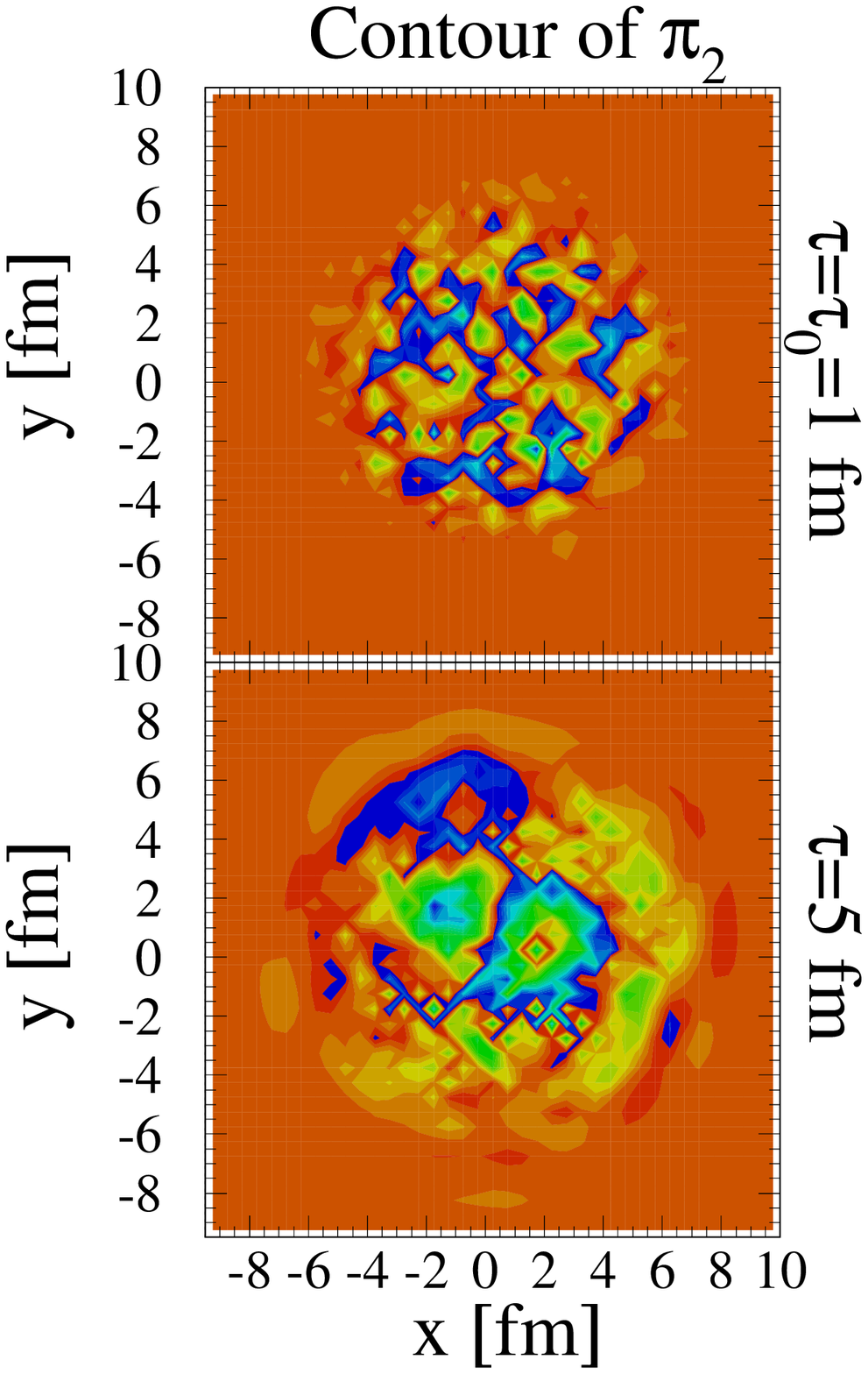}
}
\fcaption{From Ref. [\citelow{asakawa}].
Contour plot of a component of the $\pi$ field ($\pi_2$)
at the initial proper time $\tau = 1 ~{\rm fm}$ and later
at $\tau=5 ~{\rm fm}$ in a $2+1$
dimensional simulation of the linear sigma model including expansion
with quench initial conditions done by Asakawa {\it et al}.
The uniform grey around the outside of both figures represents
$\pi_2 =0$.  In the lower panel, there are prominent features
at longer wavelengths than in the initial conditions shown
in the upper panel.  For example, the structure centred at
about $(2,0)$ is a region of positive $\pi_2$ peaking
at about $85~{\rm MeV}$, and the structure centred at
about $(-1.5,1.5)$ is a region of negative $\pi_2$ with
a minimum at about $\pi_2=-65~{\rm MeV}$.  These
structures are not smooth, however.
(The grey-scale in this black and white representation
of the figure does not make all this apparent --- the original
in Ref. [\citelow{asakawa}] is in colour, and the authors also have
a  three dimensional colour version which is the clearest
way of presenting the results, but which is
not clear in black and
white.)
Thus, prominent long wavelength
oscillations with wavelengths of about 8 fm
have been
excited. Because of the cylindrical boundary conditions
used to implement the transverse expansion, these results
cannot simply be Fourier transformed, as was done in
producing Fig. 11.
However, the phenomena are clearly analogous.  Long wavelength
modes have been markedly amplified, but they cannot be described
as smooth domains as
they have short wavelength oscillations superimposed on them.}
\label{fig:asakawafig}
\end{figure}
The second simulation, shown in Fig. 13, has $\langle \phi^2 \rangle$
smaller, as in a quench.  The transverse expansion is apparent
in the figure, as is the growth of long wavelength pion
oscillations.  Just as in Fig. 11, one obtains a configuration
with prominent long wavelength oscillations (about 4 fm from peak to
trough; hence about 8 fm in wavelength) superimposed with short
wavelength oscillations which have not been damped out.
As in the simulations of Ref. [\citelow{quenchpaper}],
the longest wavelengths which are amplified are comparable to
the system size.

\begin{figure}
\centerline{
\epsfysize=80mm
\epsfxsize=120mm
\epsfbox{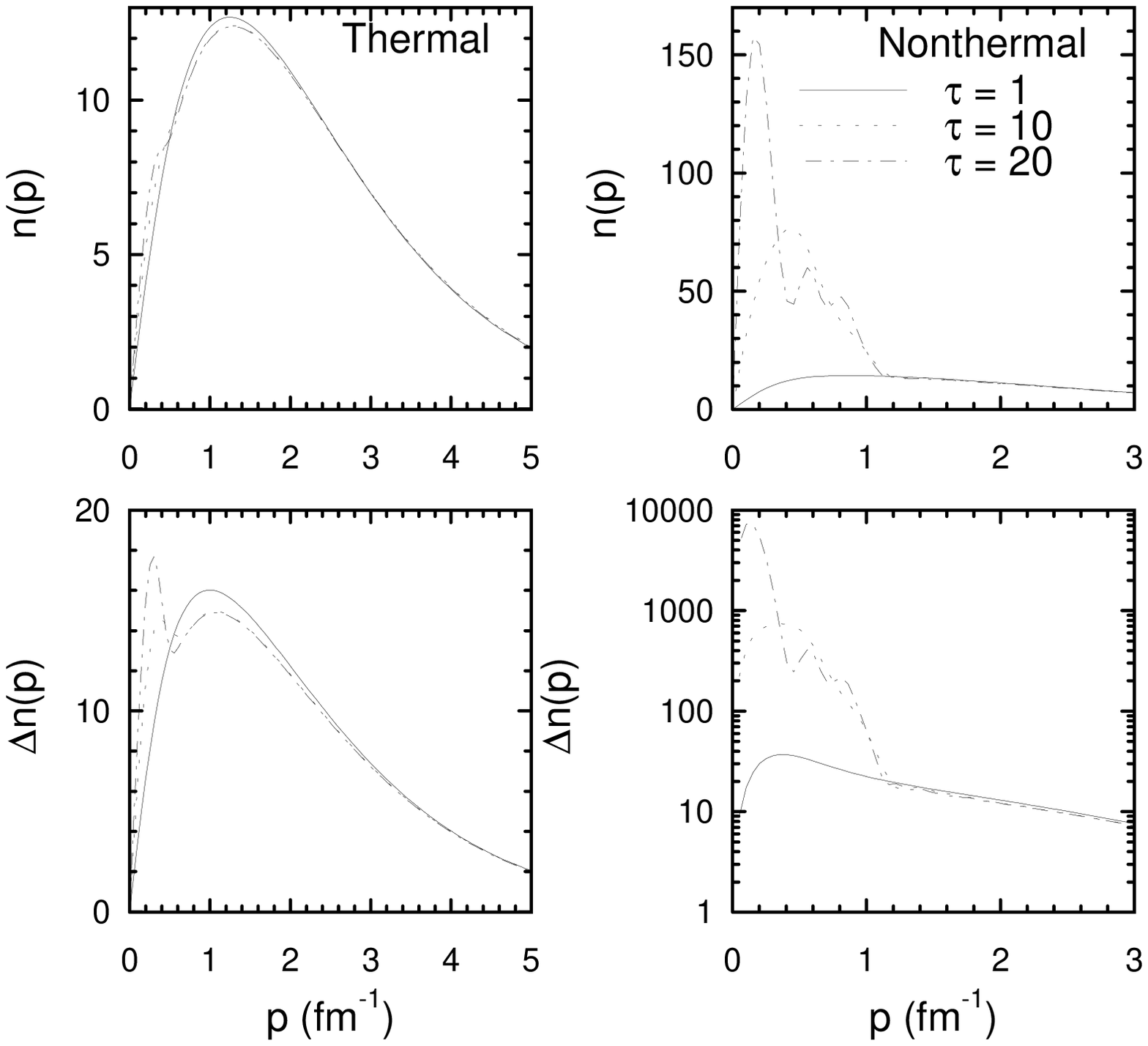}
}
\fcaption{From the final paper of Ref. [\citelow{kluger}].
Results from simulations
by Cooper {\it et al.}
of the linear sigma
model in a mean field approximation including quantum
and thermal fluctuations and boost invariant longitudinal expansion.
For nonthermal initial conditions, long wavelength oscillations
are amplified, while this does not occur for thermal initial conditions.
See text for further discussion.}
\label{fig:klugerfig}
\end{figure}
The linear sigma model simulations of
Cooper {\it et al.}\cite{kluger} are $3+1$ dimensional,
include the effects of quantum fluctuations, include the
time dependent
renormalization of the couplings in the potential due to
the changing quantum and thermal fluctuations,
include boost invariant
longitudinal expansion, and do not include transverse
expansion.
In order to include quantum effects, they work in a mean field
approximation, and so risk missing some of the physics as
we discussed in section 3.2.2.
Because longitudinal expansion is included, the
amplitudes of all the fields becomes small at very late times,
the equations of motion linearize, and it becomes possible
to define and measure particle numbers.
Results from their simulations are shown in Fig. 14.
They plot the mean number density $n(p)$ and its
fluctuations as a function of transverse momentum $p$
at several proper times.  (See
Ref. [\citelow{kluger}] for definitions,
but growth of $n(p)$ for low $p$ corresponds to growth of
long wavelength oscillations.)
The left column is for a simulation with thermal initial conditions
with an initial temperature of $T=200~{\rm MeV}>T_c$.
Because of the longitudinal expansion, the system cools through
$T_c$, but long wavelength modes are not amplified.
The right column is for a simulation with nonthermal initial
conditions in which $\langle \sigma \rangle$ has been given
an initial nonzero time derivative.  Cooper {\it et al.}
note that using quench-like initial conditions
(long wavelength modes in a thermal distribution at one
temperature with short wavelength modes in a thermal distribution
at a lower temperature)
produces
similar results to those shown in the right column.
Long wavelength oscillations are dramatically
amplified.  It is impossible
to define a domain size at late times in the right column,
because multiple length scales have emerged.
For nonthermal initial conditions, amplification of long wavelength
modes is manifest in the particle number distribution, while
for thermal initial conditions, this does not happen.
This result is in qualitative agreement with that of
Asakawa {\it et al.},\cite{asakawa} and confirms
the observation first made in Ref. [\citelow{quenchpaper}]:
if the long wavelength modes of the
order parameter are not in thermal
equilibrium as the chiral transition occurs, there is a
simple mechanism which can lead to dramatic amplification
of long wavelength pion oscillations.

To this point, the only quantum effects we have discussed
are the quantum corrections to the dynamics.
There are several other inherently quantum mechanical
questions, however.  First, once a long wavelength oscillation
has been created, at late times
as the classical treatment is becoming inappropriate
one might worry that there may be enough quantum scattering
of pions ({\it e.g} $\pi^0\pi^0\rightarrow\pi^+\pi^-$) to
wash out the effect of interest before the pions are far enough
apart on their way to the detector that they no longer interact.
This question is being studied by Huang and Suzuki,\cite{huangsuz}
and their preliminary result is that such effects are small
because pions, as Goldstone bosons, have only derivative couplings
and so have small scattering rates at low momentum.

Another
very interesting question which has received much attention
is: To what quantum state does a classical configuration
in which the chiral condensate is misaligned and
oscillating correspond?
In other words, once the oscillating
classical field becomes noninteracting
quantum mechanical pions, what is the quantum state of those pions?
Following Horn and Silver,\cite{hornsilver} a number
of authors\cite{karmanov,kowalski,bakedalaska,ggm}
have considered the unique quantum state with $N$
pions with zero total isospin.
Computing the distribution
of the number ratio $f$ of neutral pions obtained in this
state does indeed yield the $(1/2)f^{-1/2}$ distribution.
Greiner, Gong, and M\"uller note that this state also
has unusual correlations.  Defining
\begin{equation}
{\cal C}_{\alpha\beta}\equiv\frac{\langle n_{\pi^\alpha} n_{\pi^\beta} \rangle}
{\langle n_{\pi^\alpha}\rangle \, \langle  n_{\pi^\beta} \rangle} - 1
\label{corr}
\end{equation}
where $\alpha$ and $\beta$ can be $0$, $+$, or $-$, they find
that for states with many pions, $C_{00}=4/5$,
$C_{0+}=C_{0-}=-2/5$ and $C_{++}=C_{--}=C_{+-}=1/5$.
Note that in ordinary intensity interference experiments
for bosons coherent sources give $C=0$,\cite{gyulassy}
and $C\neq 0$ is only possible for identical bosons.
Here, in contrast, neutral and charged pions are each
positively correlated, and anti-correlated with the other.
It seems likely that
these charge dependent correlations will be qualitatively
similar for other quantum states which may correspond
to classical regions in which the condensate is misaligned.
While it is reasonable to consider states with small
isospin, there is no reason to require $I\equiv 0$.
It is therefore reassuring that Cohen {\it et al.}\cite{cohen}
have shown that the $(1/2)f^{-1/2}$ distribution is significantly
affected only when $I$ becomes larger than about $30\%$
of the number of pions.
Once one allows consideration of quantum states which
do not have a definite particle number, there are
many possibilities.  Amado and Kogan\cite{kogan2,amadokogan}
have enumerated several, including the isospin projected
coherent states of Botke, Scalapino, and Sugar\cite{botke}
and squeezed states.\cite{yuen}  Other authors\cite{boyanovsky,kluger}
have noted that mixed states described by density matrices are
a possibility.
The argument for the
relevance of squeezed states\cite{kogan2} is an interesting one.
In the Hartree approximation, each mode of the
pion field satisfies the equations of motion (\ref{jjk})
for a harmonic oscillator with a time dependent mass.
Replacing the classical modes by quantum harmonic oscillators,
one obtains parametric quantum oscillators, and this is just
the setting in which squeezed states naturally arise.\cite{yuen}
The argument does rely on using the Hartree approximation, which
we already know to be problematic, and so it is not conclusive.
The question of what quantum state corresponds to a
region in which long wavelength amplifications of a classical
field remains open.

\subsection{Signatures and Experiments}

The definitive signature of long wavelength pion oscillations
is large fluctuations in $f$, the neutral pion fraction,
and we discuss this below.
However, a number of other signatures have also been proposed.
Huang, Suzuki, and Wang\cite{suzuki,latesthuang} have noted that in a region
in which the chiral condensate is misaligned, the isovector
part of the electromagnetic current is no longer dominated by
the $\rho^0$ meson as in the normal vacuum --- it couples both to
$\rho$ mesons and to their  chiral partners the $a_1$ mesons.
Huang {\it et al.} consider a large smooth region of disoriented
vacuum, and average over all possible directions of disorientation,
and find that the $\rho$ peak at 770 MeV in the dilepton spectrum will
be reduced by a factor of 2 and a broad bump in the dilepton
spectrum will emerge at the $a_1$ energy (1260 MeV).  The $\omega$ meson,
on the other hand, is a chiral singlet and so will not be affected.
They therefore propose that seeing a reduction in the $\rho$
peak in the dilepton spectrum while the $\omega$ peak remains
unchanged would be a signal for the formation of disoriented
chiral condensate.  It is not clear, unfortunately, that this
signal would be prominent in the more realistic situation
in which there are long wavelength oscillations
of the condensate, but not large smooth regions.

A direct signal of the amplification of long wavelength
pion oscillations would be simply
to look for an enhancement of the number of pions produced
at low momentum, without regard to their
charge.\cite{latestgavin,latesthuang}  It is not clear that such
a signal would be unambiguous in and of itself.  However,
if this signal were detected in coincidence with the detection
of large fluctuations in $f$, then it might provide a estimate
of the wavelengths of importance --- excess production of
pions below some transverse momentum $p_T$ would show that
modes with wavelengths longer than $1/p_T$ had been amplified.

It is also worth looking for the correlations
${\cal C}_{\alpha\beta}$ defined in (\ref{corr}).
In particular,\cite{ggm} the unusual charge dependence of the
correlations (neutral and charged pions correlated with
themselves, and anti-correlated with the other) may be
detectable.  Gavin\cite{latestgavin} has also considered
distinguishing the effect of regions of disoriented
chiral condensate on the correlation between like
sign charged pions from that of standard interference effects,
but concludes that this will be difficult.

We now turn to the definitive signature of
long wavelength oscillations of the chiral order parameter.
If a particular $\vec k$ mode ends up oscillating in
the $\pi_0$ direction, then it will become neutral pions
moving in the direction of $\vec k$.  Similarly, oscillations
in the $\pi_1 - \pi_2$ plane will become charged pions.
Thus, in some regions of the detector, $f$ would be large,
and in others it would be small.  If this were the only
physics occurring, $f$ would be distributed like $(1/2)f^{-1/2}$
as we have discussed, but of course there will be a large
background due to all the other pions in the event.
(These correspond very roughly to the short wavelength
oscillations which, as we have seen, remain even as
long wavelength oscillations are amplified.
This correspondence is only rough, however,
because at short wavelengths the
linear sigma model cannot capture all of the physics.)

What is needed is a detector which covers
as much of $4\pi$ solid angle as possible, which counts
both charged pions and photons from the decay of neutral pions,
and which is segmented in pseudorapidity $\eta$ and azimuth
$\varphi$ into fine enough segments that the number of photons
and charged pions in each segment can be accurately measured.
(Ideally, a theorist's detector would be segmented enough
that the typical segment had either zero or one photon or charged
pion in it.)  One would use this detector to look at
the central rapidity region of the collision, where almost all
the charged particles are pions and where almost all the photons
come from neutral pion decay, so that particle identification
is not crucial.  It would be very helpful to have enough
information about the transverse momenta $p_T$ so that
high momentum charged pions and photons can be cut from
the data.  (In looking for a signal, one of the things to play
with would be where to put the $p_T$ cut, but somewhere around
200 MeV would seem reasonable.)
The data, then, would consist of an event by event catalogue
of the positions in $(\eta,\varphi)$ of the photons and
charged pions.
There are many conceivable ways of analyzing the data
to look for large fluctuations in $f$.  Several are described
below.

Let us first make a very crude attempt to estimate the number
of pions that might arise from the excitation of a long
wavelength pion oscillation.  Inspired by Fig. 13,
consider a wave with wavelength
$\lambda \sim 8~{\rm fm}$ and with $k=2\pi/\lambda \sim 150~{\rm MeV}$.
The energy density of such an oscillation with amplitude $A$
is
\begin{equation}
\varepsilon\sim(m_\pi^2 + k^2)A^2
\sim 25\ \Biggl(\frac{A}{f_\pi}\Biggr)^2~{\rm MeV}/{\rm fm}^3 \ .
\label{endens}
\end{equation}
The number of pions, then, is
\begin{equation}
N\sim\frac{V\varepsilon}{\sqrt{m_\pi^2 + k^2}} \sim 125 \
\Biggl( \frac{V}{(8~{\rm fm})^3}\Biggr) \Biggl(\frac{A}{f_\pi}\Biggr)^2
\label{estimate}
\end{equation}
where $V$ is the volume of space in which this mode is oscillating
with this amplitude.
The uncertainties in this estimate are huge.
There is no information in the simulations of Fig. 13 about
the longitudinal extent of the region which is oscillating.
However, if it extends in rapidity from -1 to 1, then
at a time $t=5~{\rm fm}$ this corresponds to an extent in
$z$ of $8~{\rm fm}$.  So, in (\ref{estimate}) we have
taken $V\sim (8~{\rm fm})\lambda^2$.  Note, however, that
there is no reason why $V$ cannot be larger than $\lambda^3$ ---
$V$ is the volume over which the mode is excited, and is not
in principle related to $\lambda$.
We do not really know what amplitudes
can arise --- $A\sim 75~{\rm MeV}$ in Fig. 13, and in general
it seems reasonable that $A\sim f_\pi$ is possible.
Gavin has also
tried to make an estimate of the number of pions from
an event like that in Fig. 13;\cite{latestgavin}
he comes up with 30-300 pions, which seems reasonable.
Following Horn and Silver,\cite{hornsilver} Gavin\cite{latestgavin}
notes that the momentum distribution of pions emitted from
a smooth region of disoriented condensate of radius $R$
goes like $\exp(-p^2 R^2)$.  Taking
$R=\lambda/2$,  this suggests
typical transverse momenta of about $50~{\rm MeV}$.

A few words of warning about the estimates just given
seem in order.  A reader at this point may be crying out
that we should do a better job of analyzing the simulations
done by various authors, and do better simulations, to make the estimates of
the typical transverse momenta and pion multiplicity
more accurate.  This would be misleading.
The theoretical foundations of the work described in this chapter are very
speculative.  Using the linear sigma model
to describe the dynamics
seems reasonable for the long wavelength modes,
but  cannot be justified in a controlled way.
Either neglecting quantum effects or including them in
a Hartree approximation risks missing some of the physics.
Most important of all, we do not know what
initial conditions to impose on the long wavelength modes
of the chiral order parameter at the time when partonic language
ceases to be appropriate and a description of the dynamics
in terms of the linear sigma model becomes appropriate.
Thus, estimates of the kind given above
are necessarily very crude.
Indeed, the best advice
for an experimentalist is
to ignore all attempts at quantitative
estimates, and just try
to look at what is out there.  Furthermore, although
there are fewer complications at RHIC than at current
lower energy experiments, given the lack of theoretical
certainty it would be crazy not to look for fluctuations
in $f$
in current experiments at CERN and at the AGS. This is being
or will soon be
done,\cite{bolekcraig} but no results have yet been reported.

Let us now consider how future data might be analyzed.
The first thing to do will be simply to plot the data
on a lego plot ({\it i.e.} in the $(\eta,\varphi)$ plane)
using one colour for charged particles and another for
photons, and then simply look by eye to see what is there.
One of the consequences of the amplification of long
wavelength pion oscillations is that rare dramatic events
in which regions of the lego plot are almost all neutral
or almost all charged will be much less rare than in
thermal equilibrium.  As Krzywicki\cite{krz2} has emphasized,
even if such dramatic events remain rare,
({\it i.e.} less rare than in equilibrium but rare
nevertheless) they are worth
waiting for and looking for.

Another simple way of analyzing the data will be to
count all the charged particles and all the photons in each event,
compute $f\sim \frac{n_\gamma/2}{n_{\rm charged} + n_\gamma/2}$
for each event, and plot a histogram of the $f$'s for different
events.\cite{bolek}
As an example, the WA98 experiment\cite{WA98} at CERN detects
both photons and charged particles in a window of rapidity
of width 1.1 near central rapidity,
over all $2\pi$ of azimuth.  In a typical event,
of order $600-800$ charged particles are accepted.
The task will be to compare the $f$ histogram from the
data with the $f$ histogram for  ``background'' events ---
presumably Monte Carlo generated --- which do not
contain any unusual fluctuations in $f$ and so have
a Gaussian $f$ distribution with width proportional
to the mean value of $(n_\gamma/2+n_{\rm charged})^{-1/2}$.
Long wavelength
oscillations of the chiral order parameter would lead
to a broadening of this histogram.
Clearly, understanding
the $f$ distribution of the background events is crucial.
There is reason for optimism, however.
It seems likely that if a Monte Carlo event generator
predicts, say,  more charged pions than
actually occur, then it will also predict
more neutral pions than actually occur, while
the $f$ distribution will be unaffected.  This suggests that $f$ distributions
for background events obtained from event generators
will be much more robust and accurate than, say, distributions
for the total multiplicity.  This must of course be studied and tested
further.
Preliminary work suggests that
a population of events given by taking
Monte Carlo generated background
events, superimposing
150 pions with momenta distributed like
$\exp(-(p/50\,{\rm MeV})^2)$
in the centre of mass frame,
uniformly distributed in azimuth,
and with $f$ distributed
like $(1/2)f^{-1/2}$,
and then imposing WA98 acceptances,
would yield an $f$ histogram which
is clearly detectably broader than that of a population of background
events.  It might also be helpful to look
for broadening of the $f$ histogram as a control parameter is
varied which changes the energy of the collision.
The beam energy is the obvious choice, but if that is
not possible, looking at events with varying impact parameter
may prove useful.
At RHIC, the analysis analogous to the one sketched above
would be to average over all charged particles and photons in a rapidity bin
of some width, say 1, rather than to average over
all particles in an entire event.

It is important to realize, however, that
averaging over azimuth is pessimistic.
It is entirely possible that in a given event (or in a given
region of rapidity in an event) long wavelength modes
corresponding to one direction in azimuth will end up oscillating
in one direction in $\vec \pi$ space, while modes corresponding
to a different direction in azimuth oscillate in a different
direction in $\vec \pi$ space.  Stated in another way,
it is clear that the pions from the long wavelength
$\pi_2$ oscillation in Fig. 13 will not be uniformly distributed
in azimuth.
Thus, the generic signature of
long wavelength pion oscillations is not large fluctuations of
$f$ which are uniform in $\varphi$, but rather large
fluctuations in $f$ which depend both on $\varphi$ and $\eta$.
Thus, the data should be binned in bins of varying extents
$\Delta \eta$ and $\Delta \varphi$, and $f$ histograms should
be done for varying bin sizes.  Again, for each bin size,
the $f$ histogram
from the data must be compared with
that from background events.

Another way of organizing the data analysis goes as follows.\cite{paulo}
Bin the data in such small bins that there are, say, about
10 particles per bin.  Then for a bin centred at $(\eta,\varphi)$
define
$\tilde f(\eta,\varphi) = f - \langle f \rangle$.
(In a real experiment, once experimental
acceptances are taken into account, $\langle f\rangle$ will not be precisely
$1/3$.)
Because $\tilde f$ is defined using very small bins,
a histogram of $\tilde f$ would be broad due to
fluctuations even in the absence of interesting
effects, and so would not be useful.
However, compute the correlation function
\begin{equation}
C(\Delta\eta,\Delta\varphi) = \langle \tilde f(\eta_1,\varphi_1)
\tilde f(\eta_2,\varphi_2) \rangle
\label{corrfn}
\end{equation}
where the averaging $\langle \ldots \rangle$ is done over
all bins with $|\eta_1-\eta_2|=\Delta\eta$ and
$|\varphi_1-\varphi_2|=\Delta\varphi$.
Seeing structure in this correlation function would suggest
the appropriate scales in rapidity and azimuth on which to
bin the data to see unusual $f$ histograms.
One could also do this analysis averaging over an
ensemble of events.  If all the events in the ensemble
are selected from a single statistical distribution
which is the sum of some signal distribution and some
noise distribution, then the signal seen in
$C(\Delta\eta,\Delta\varphi)$ would grow like
the square root of the number of events in the ensemble.
In this case, computing $C(\Delta\eta,\Delta\varphi)$
using an ensemble of events would be very valuable.
However, if events are not selected from the
same distribution, then averaging over an ensemble
could wash out interesting effects.  In particular,
using an ensemble of events could miss
seeing structures in which the interesting scales in
rapidity and azimuth are different in different events,
and could miss detecting effects which only occur rarely.
Thus, the correlation function
$C(\Delta\eta,\Delta\varphi)$ should be
computed both event by event and for ensembles of events.

There are many ways to look for the fluctuations
in the number ratio of neutral to charged pions
which characterize long wavelength oscillations of the chiral
condensate.  We look forward to the day when the choice
among analyses becomes data driven.

\subsection{Concluding Remarks}

Several different groups\cite{quenchpaper,asakawa,kluger}
have now done complementary simulations
in which different physics has been included and
different assumptions have been made.
All have found that
long wavelength pion oscillations
can be dramatically amplified if the chiral transition
is out of  thermal equilibrium.
This behaviour can be understood at least qualitatively
in terms of a simple mechanism:
as the system evolves from a disordered
configuration toward a low temperature
ordered configuration, nonequilibrium dynamics of the
condensate can give long wavelength pion modes
a negative effective mass squared, making them unstable
to growth.
The abrupt transition, or quench, considered in section 3.2.1
is an idealization.  In a heavy ion collision, the
transition presumably proceeds by a mechanism which
is in some sense between the quench idealization and
the thermal equilibrium idealization.
There remains room for improved simulations, complementary
to all that have been done so far.  A $3+1$ dimensional
classical
simulation with boost invariant longitudinal expansion
done on a large enough lattice that transverse expansion
can be included as was done in $2+1$ dimensions by Asakawa
{\it et al.} would be very interesting.  Furthermore,
if the lattice is large enough that the simulation can
be run to a late enough time that the classical field
amplitudes linearize, then a direct computation
of the particle number would be possible.
Such a simulation would complement
that of Cooper {\it et al.} because although it would
not include quantum fluctuations,
it would not require
a mean field approximation and would include transverse expansion.
However, none of
these simulations can come close to
describing a real heavy ion collision.
First of all, the linear sigma model cannot
capture all the physics, although it is likely a reasonable
approximation for the long wavelength modes of the chiral order
parameter.
Perhaps the largest uncertainty arises from
the fact that we do not know what initial
conditions to impose on the $\sigma$ and $\pi$ fields at the time
at which a linear sigma model description of the dynamics
becomes appropriate.
This is crucial, because simulations
have shown that expansion beginning with thermal initial
conditions is not sufficient
to obtain the growth of long wavelength modes.  It is
necessary that at the time when the linear sigma model
description becomes appropriate, the long wavelength
modes of the order parameter are not in equilibrium with
the short wavelength modes.

The conclusions to take away from this chapter
are essentially
qualitative.  Consider
a heavy ion collision which is
energetic enough that there is a central rapidity region
of high energy density and low baryon number.
If, as this region cools through the chiral
transition, long wavelength modes do not stay in equilibrium,
then there is a robust mechanism which can lead to amplification of
long wavelength pion oscillations.
This cannot happen in thermal equilibrium.
Long wavelength pion oscillations
would be observed by detecting regions in rapidity-azimuth
space in which the fraction of neutral pions is unusual.
The pions from different long wavelength oscillations
will have neutral pion fractions $f$ with an
inverse square root distribution.  This will of course be masked
by all the other pions in the events, but the signal to noise
ratio can be increased by focussing on low $p_T$ pions.
In contrast to the equilibrium physics of chapter 2 of
this article in which we were able to use the notion
of universality,
the nonequilibrium physics
which may lead to long wavelength pion oscillations
is speculative.
Therefore,
this subject will only become mature when it becomes experimentally
driven.  A dedicated experiment to look for related phenomena
is underway at Fermilab.\cite{minimax}  Experimentalists
are looking for the phenomena described in this article in current
generation heavy ion experiments and will look for them in
colliding beam experiments at RHIC and the LHC, where
sufficient energy should be available.  Detection of
unusual fluctuations of $f$ in the central rapidity region
of a heavy ion collision would be a dramatic and definitive
signature of an out of equilibrium chiral transition.
The ball is in the experimentalists' court, and we wish them well.

\section{Acknowledgements}

I first want to thank Frank Wilczek.  Learning from him and working
with him
has been fruitful and fun.  I also want to thank him for
his careful reading of this article.
I have had very helpful
conversations with many people
about different aspects
of the physics I have presented.
Some of these
conversations happened recently, in response to
an earlier draft of this article, and I am grateful for all the
comments I got; others happened longer ago.
My thanks to
P. Bedaque, Bj, J.-P. Blaizot, W. Busza, S. Chandrasekharan,
N. Christ, C. DeTar, S. Gavin, S. Hegyi,
S. Hsu, Z. Huang, K. Kanaya, F. Karsch,
I. Kogan, K. Kowalski, Y. Kluger, E. Laermann, R. Mawhinney, E. Mottola,
B. M\"uller, C. Ogilvie, B. Petersson, R. Pisarski, D. Schramm,
S. Selipsky, D. Spergel, M. Suzuki, C. Taylor, N. Turok, X. Wang,
and B. Wyslouch.

This work was supported by the Harvard University
Society of Fellows, the Milton Fund of Harvard University,
and by the National Science Foundation under grant PHY-92-18167.


\section{References}

\begin{thebibliography}{9}

\bibitem{asympfree} D. J. Gross
and F. Wilczek, {\it Phys. Rev. Lett.} {\bf 30}
(1973) 1343; \hfill \break
H. Politzer, {\it Phys. Rev. Lett.} {\bf 30} (1973) 1346.

\bibitem{collinsperry} J. C. Collins and M. J. Perry {\it Phys. Rev. Lett.}
{\bf 34} (1975) 1353.

\bibitem{us} K. Rajagopal and F. Wilczek, {\it Nucl. Phys.} {\bf B399}
(1993) 395.

\bibitem{quenchpaper} K. Rajagopal and F. Wilczek,
{\it Nucl. Phys.} {\bf B404} (1993) 577.

\bibitem{gpy}D. J. Gross, R. D. Pisarski, and L. G. Yaffe, {\it Rev. Mod.
Phys.} {\bf 53} (1981) 43.

\bibitem{mclerran} L. McLerran, {\it Rev. Mod. Phys.} {\bf 58} (1986) 1021.

\bibitem{shuryak} E. V. Shuryak, {\it The QCD Vacuum, Hadrons, and the
Superdense Matter}, (World Scientific, Singapore, 1988).

\bibitem{hwa}R. Hwa, ed., {\it Quark-Gluon Plasma}, (World Scientific,
Singapore, 1990).

\bibitem{muller} B. M\"uller, {\it Physics and Signatures of
the Quark-Gluon Plasma.}, to appear in {\it Rept. Prog. Phys.},
DUKE-TH-94-76 (1994) nucl-th/9410005; \hfill\break
B. M\"uller, {\it Physics of the Quark-Gluon Plasma},
Duke preprint DUKE-TH-92-36, 1992.

\bibitem{qmi} Proceedings of Quark Matter '88, {\it Nucl. Phys.} {\bf A498},
(1989).

\bibitem{qmii}Proceedings of Quark Matter '90, {\it Nucl. Phys.} {\bf A525},
(1991).

\bibitem{qmiii}Proceedings of Quark Matter '91, {\it Nucl. Phys.} {\bf A544},
(1992).

\bibitem{qmiv}Proceedings of Quark Matter '93, {\it Nucl. Phys.} {\bf A566},
(1994).

\bibitem{svetitsky} B. Svetitsky and L. G. Yaffe, {\it Nucl. Phys.} {\bf B210}
(1982) 423;  \hfill \break
B. Svetitsky, {\it Phys. Reports} {\bf 132} (1986) 1.

\bibitem{landau} L. D. Landau and E. M. Lifshitz, {\it Statistical Physics,
2nd ed.} (Pergamon, Oxford, 1969).

\bibitem{wilson} K. Wilson,
{\it Phys. Rev.} {\bf D10} (1974) 2445. \hfill \break
Reviewed in J. Kogut, {\it Rev. Mod. Phys.} {\bf 51} (1979) 659.

\bibitem{montecarlo} K. Wilson, in {\it Recent Developments in Gauge
Theories}, eds. G. 't Hooft {\it et al.} (Plenum, New York, 1980);
\hfill \break
M. Creutz, L. Jacobs, and C. Rebbi, {\it Phys. Rev. Lett.} {\bf 42}
(1979) 1390; \hfill \break
M. Creutz, {\it Phys. Rev. Lett.} {\bf 43} (1979) 553.

\bibitem{kogut} J. B. Kogut {\it et al.}, {\it Phys. Rev. Lett.}
{\bf 50} (1983) 393, {\bf 51} (1983) 869;\hfill \break
S. A. Gottlieb {\it et al.}, {\it Phys. Rev. Lett.} {\bf 55} (1985) 1958;
\hfill \break
N. H. Christ and A. E. Terrano, {\it Phys. Rev. Lett.} {\bf 56} (1986) 111.

\bibitem{oldgott} S. Gottlieb {\it et al., Phys. Rev.} {\bf D35} (1987) 3972;
{\it Phys. Rev. Lett.} {\bf 59} (1987) 1513; {\it Phys. Rev.} {\bf D41}
(1990) 622.

\bibitem{bernard} C. Bernard {\it et al.},
{\it Phys. Rev.} {\bf D45} (1992) 3854.

\bibitem{columbia} F. R. Brown
{\it et al.}, {\it Phys. Rev. Lett.} {\bf 65}
(1990) 2491.

\bibitem{fukugita} M. Fukugita, H. Mino, M. Okawa, and A. Ukawa, {\it Phys.
Rev. Lett.} {\bf 65} (1990) 816; and {\it Phys. Rev.} {\bf D42} (1990) 2936.

\bibitem{gottlieb} S. Gottlieb {\it et al., Phys. Rev.} {\bf D47}
(1993) 3619;\hfill\break
S. Gottlieb {\it et al., Nucl. Phys.} {\bf B} {\it Proc.
Suppl.} {\bf 30} (1993) 315.

\bibitem{zhu} D. Zhu, {\it Nucl. Phys.} {\bf B} {\it Proc. Suppl.}
{\bf 34} (1994) 286; R. Mawhinney, {\it ibid.} {\bf 30} (1993) 331.

\bibitem{iwasaki} Y. Iwasaki, K. Kanaya, S. Sakai, and
T. Yoshi\'e, {\it Nucl. Phys.} {\bf B} {\it Proc. Suppl.} {\bf 34}
(1994) 314; {\it ibid.} {\bf 30} (1992) 311; and {\it Chiral
Phase Transition in Lattice QCD with Wilson Quarks}, UTHEP-300 (1995)
hep-lat/9504019.

\bibitem{b2} C. Bernard {\it et al.}, {\it QCD Thermodynamics
at $N_t=8$ and 12}, (1994) hep-lat/9412112.

\bibitem{latreviews} Reviews of
finite temperature lattice gauge theory include\hfill \break
F. Karsch, p. 61 in Ref. [8];  N. Christ, p. 81 in Ref. [12];
B. Petersson, p. 237 in Ref. [11]; \hfill \break
C. DeTar, {\it Finite Temperature QCD: Progress and Outstanding
Problems}, Talk given at Lattice '94, hep-lat/9412010;\hfill\break
F. Karsch, {\it Nucl. Phys.} {\bf B} {\it Proc. Suppl.} {\bf 34} (1994) 63;
\hfill \break
B. Petersson, {\it Nucl. Phys.} {\bf B} {\it Proc. Suppl.} {\bf 30} (1993) 66;
\hfill \break
D. Toussaint, {\it Nucl. Phys.} {\bf B} {\it Proc. Suppl.} {\bf 26} (1992) 3;
\hfill \break
S. Gottlieb, {\it Nucl. Phys.} {\bf B} {\it Proc. Suppl.} {\bf 20} (1991) 247;
\hfill\break
C. DeTar, this volume.


\bibitem{piswil} R. Pisarski and F. Wilczek, {\it Phys. Rev.}
{\bf D29} (1984) 338.

\bibitem{feynman}
R. P. Feynman, {\it Photon-Hadron Interactions}, (Benjamin, Reading,
MA, 1972).

\bibitem{perkins}D. H. Perkins, {\it Introduction to High Energy Physics},
(Addison-Wesley, Menlo Park, 1987).


\bibitem{bjorken}J. D. Bjorken, {\it Phys. Rev.} {\bf D27} (1983) 140.

\bibitem{salmeron}
R. A. Salmeron, p. 1 in {\it Quark-Gluon Plasma}, B. Sinha {\it et al.}, eds.,
(Springer-Verlag, Berlin, 1990).


\bibitem{satz}H. Satz, p. 371 in Ref. [12].

\bibitem{bialas}A. Bia\l as and W. Czy\'z, p. 271 in Ref. [8].

\bibitem{geiger}K. Geiger and B. M\"uller, {\it Nucl. Phys.} {\bf B369},
(1992) 600; \hfill \break
K. Geiger, {\it Phys. Rev.} {\bf D46} (1992) 4965 and {\it ibid.} p. 4986.

\bibitem{biro}T. S. Bir\'o {\it et al.},
{\it Phys. Rev.} {\bf C48} (1993) 1275.

\bibitem{twostage}E. Shuryak, {\it Phys. Rev. Lett.} {\bf 68} (1992)
3270;\hfill\break
E. Shuryak and L. Xiong, {\it Phys. Rev. Lett.} {\bf 70} (1993)
2241;\hfill\break
K. Geiger and J. I. Kapusta, {\it Phys. Rev.} {\bf D47} (1993)
4905;\hfill\break
T. Altherr and D. Seibert, {\it Phys. Lett.} {\bf B313} (1993) 149.



\bibitem{hydroreview}J.-P. Blaizot and J.-Y. Ollitrault, p. 393 in
Ref. [8].

\bibitem{hardtojustify} Ref. [38], p. 399.

\bibitem{wilczek}F. Wilczek, {\it Int. J. Mod. Phys.} {\bf A7} (1992) 3911.
This elaborates earlier work of Ref. [28].

\bibitem{thooft} G. 't Hooft, {\it Phys. Reports}  {\bf 142} (1986) 357.

\bibitem{amitma} See, for example, S.-K. Ma, {\it Modern Theory of Critical
Phenomena},
(Benjamin/Cummings, Reading, MA, 1976) and \hfill \break
D. J. Amit, {\it Field Theory, the Renormalization
Group, and Critical Phenomena}, (World Scientific,
Singapore, 1984).

\bibitem{gm}M. Gell-Mann and M. L\'evy, {\it Nuovo Cimento}
{\bf 16} (1960) 705.

\bibitem{ginsparg}P. Ginsparg, {\it Nucl. Phys.} {\bf B170} (1980) 388,
 and references therein.

\bibitem{bmn}G. Baker, B. Nickel and D. Meiron, {\it Phys. Rev.} {\bf B17}
(1978) 1365, and {\it Compilation of 2-pt. and 4-pt. graphs for continuous
spin models}, unpublished, University of Guelph report (1977).

\bibitem{wilsonfisher}K. G. Wilson and M. E. Fisher,
{\it Phys. Rev. Lett.} {\bf 28} (1972) 240; \hfill \break
K. G. Wilson and J. Kogut, {\it Phys. Reports} {\bf 12} (1977) 77.

\bibitem{kanayakaya}K. Kanaya and S. Kaya, {\it Critical exponents
of a three dimensional O(4) spin model}, UTHEP-284 (1994) hep-lat/9409001.

\bibitem{bww}E. Br\'ezin, D. J. Wallace and K. G. Wilson,
{\it Phys. Rev.} {\bf B7} (1973) 232.

\bibitem{bw}E. Br\'ezin and D. J. Wallace, {\it Phys. Rev.} {\bf B7}
(1973) 1967.

\bibitem{wz}D. J. Wallace and R. K. P. Zia, {\it Phys. Rev.} {\bf B12}
(1975) 5340.

\bibitem{anishetty}R. Anishetty {\it et al.}, {\it Infrared
Behaviour of Systems with Goldstone Bosons}, IMSc-94/52 (1995) hep-th/9502003.


\bibitem{gocksch}A. Gocksch, {\it Phys. Rev. Lett.} {\bf 67} (1991) 1701.

\bibitem{hk}T. Hatsuda and T. Kunihiro, {\it Phys. Rev. Lett.}
{\bf 55} (1985) 158 and references therein.


\bibitem{instliq}M. A. Nowak, J. J. M. Verbaarschot and I. Zahed, {\it Nucl.
Phys.} {\bf B325} (1989) 581; \hfill \break
E. M. Ilgenfritz and E. V. Shuryak, {\it Nucl. Phys.} {\bf B319}
(1989) 511;\hfill \break
E. V. Shuryak {\it Nucl. Phys.} {\bf A522} (1991) 377c.

\bibitem{landau2} E. M. Lifshitz and L. P. Pitaevskii
{\bf Statistical Physics Part 2}, (Pergamon, Oxford, 1980).

\bibitem{KocicKogut} A. Kocic and J. Kogut, {\it Can Sigma Models
Describe Finite Temperature Chiral Transitions?}, ILL-TH-94-19 (1994)
hep-lat/9407021.

\bibitem{detar} Articles by C. DeTar in Ref. [27].

\bibitem{binder} K. Binder, in {\it Phase Transitions and Critical
Phenomena, Vol. 5B}, ed. C. Domb and M. S. Green, (Academic Press,
London, 1976), 2.

\bibitem{karsch} F. Karsch, {\it Phys. Rev.} {\bf D49} (1994) 3791.

\bibitem{karschlaer} F. Karsch and E. Laermann, {\it Phys. Rev.} {\bf D50}
(1994) 6954.

\bibitem{karschrev} For results for the energy density as
a function of temperature in lattice simulations, see F. Karsch {\it The Phase
Transition to the Quark Gluon Plasma: Recent Results from Numerical
Simulations}, BI-TP-95-11 (1995) hep-lat/9503010, and references
therein.

\bibitem{alpha} This point has been stressed by
E. Shuryak, {\it Comm. Nucl. Part. Phys.} {\bf 21} (1994) 235.

\bibitem{berera} A. Berera, {\it Phys. Rev.} {\bf D50} (1994) 6949;
A. Berera, {\it A Renormalization Group Approach to the
Chiral Phase Transition}, AZPH-TH-93-07 (1994) hep-ph/9407212.

\bibitem{shailesh} S. Chandrasekharan, {\it Critical Behavior of
the Chiral Condensate at the QCD Phase Transition}, CUTP-663,
(1994) hep-lat/9412070; and N. Christ and S. Chandrasekharan, private
communication.

\bibitem{hasenfratz}A. Hasenfratz and T. DeGrand, {\it Phys. Rev.}
{\bf D49} (1994) 466.


\bibitem{wilsonquarks} Y. Iwasaki {\it et al.}, {\it Nucl. Phys.}
{\bf B} {\it Proc. Suppl.} {\bf 26} (1992) 311; {\it ibid.} {\bf 30} (1993)
327; {\it ibid.} {\bf 34} (1994) 314; {\it Phys. Rev. Lett.} {\bf 69}
(1992) 21; Y. Iwasaki {\it et al.} UTHEP-291 (1994)
hep-lat/9412054;\hfill\break
C. Bernard {\it et al.}, {\it Phys. Rev.} {\bf D46}
(1992) 4741; {\bf D49} (1994) 3574; T. Blum {\it et al.}, {\it Phys. Rev.}
{\bf D50} (1994) 3377.


\bibitem{boyd}G. Boyd {\it et al.}, {\it Nucl. Phys.} {\bf B376}, (1992) 199.

\bibitem{paterson}A. J. Paterson, {\it Nucl. Phys.} {\bf B190} (1981) 188.

\bibitem{gausterer}H. Gausterer and S. Sanielevici,
{\it Phys. Lett.} {\bf B209} (1988) 533.

\bibitem{lawrie}I. Lawrie and S. Sarbach in {\it Phase Transitions and Critical
Phenomena}
{\bf 9} (1984) 1, ed. C. Domb and J. Lebowitz (Academic
Press).

\bibitem{ggp} S. Gavin, A. Gocksch, and R. D. Pisarski,
{\it Phys. Rev.} {\bf D49} (1994) 3079.\hfill\break
See also H. Meyer-Ortmanns and B.J. Schaefer, {\it How sharp
is the chiral crossover phenomenon for realistic meson masses?},
HD-TVP-94-16 (1994) hep-ph/9409430.

\bibitem{kanaya} Y. Iwasaki, K. Kanaya, S. Kaya, S. Sakai, and
T. Yoshi\'e, {\it Nature of the finite temperature transition
in QCD with strange quark}, UTHEP-290 (1994) hep-lat/9412012;\hfill\break
Y. Iwasaki, {\it Phase Diagram of QCD at Finite Temperatures
with Wilson Fermions}, UTHEP-293 (1994) hep-lat/9412103;\hfill\break
K. Kanaya, {\it Deconfining Chiral Transition in QCD on the
Lattice}, UTHEP-296 (1995) hep-lat/9502020.

\bibitem{witten}E. Witten, {\it Phys. Rev.} {\bf D30}
(1984) 272; \hfill \break
G. Fuller, C. Alcock, and G. Mathews, {\it Phys. Rev.} {\bf D37}
(1988) 1380;\hfill\break
J. Applegate and C. Hogan, {\it Phys. Rev.} {\bf D31} (1985) 3037.

\bibitem{schramm} D. Thomas, D. N. Schramm, K. A. Olive, G. J. Mathews,
B. S. Meyer, B. D. Fields, {\it Astrophysical J.} {\bf 430} (1994) 291.

\bibitem{bak}P. Bak and  D. Mukamel, {\it Phys. Rev.} {\bf B13} (1976) 5086.

\bibitem{barak}Z. Barak and
M. B. Walker, {\it Phys. Rev.} {\bf B25} (1982) 1969.


\bibitem{leutwyler}J. Gasser and H. Leutwyler, {\it Phys. Lett.} {\bf B184},
(1987) 83; \hfill \break
P. Gerber and H. Leutwyler, {\it Nucl. Phys.} {\bf B321} (1989) 387.
\hfill \break
For a review of chiral perturbation theory at finite temperature
and finite volume, see H. Leutwyler {\it Nucl. Phys.} {\bf B} {\it
Proc. Suppl.} {\bf 4} (1988) 248.


\bibitem{loewe}C. Contreras and M. Loewe, {\it Int. J. Mod. Phys.} {\bf A5},
(1990) 2297.

\bibitem{kaonmass}H. Leutwyler, {\it Nucl. Phys.} {\bf B337} (1990) 108.

\bibitem{centauro}C. M. G. Lattes,
Y. Fujimoto, and S. Hasegawa, {\it Phys. Rept.}
{\bf 65} (1980) 151.

\bibitem{pamir}S. G. Bayburina {\it et al.}, {\it Nucl. Phys.} {\bf B191}
(1981) 1.

\bibitem{kanbala}J. R. Ren
{\it et al.}, {\it Phys. Rev.} {\bf D38} (1988) 1417.

\bibitem{recentcentauro}For recent discussions of Centauro events
see: Chacaltaya and Pamir Collaboration (L. T. Baradzei {\it et al.}),
{\it Nucl. Phys.} {\bf B370} (1992) 365; and articles by
S. A. Slavatinsky, D. Linkai, S. Hasegawa and M. Tamada, K. Goulianos,
and F. Halzen
in the proceedings of the 7th Int. Symp. on Very High Energy Cosmic
Ray Interactions, 1992, Ann Arbor, ed. L. Jones, (American Institute
of Physics, New York, 1993).

\bibitem{ellsworth}R. W. Ellsworth {\it et al.}, {\it Phys. Rev.} {\bf D23}
(1981) 764; {\it ibid.} 771.

\bibitem{jaceedetails}T. H. Burnett {\it et al.}, {Phys. Rev. Lett.} {\bf 50}
(1983) 2062; {\it Nucl. Phys.} {\bf A447} (1985) 197;
{\it Phys. Rev. Lett.} {\bf 57} (1986) 3249;
{\it Phys. Rev.} {\bf D35} (1987) 824;\hfill \break
Y. Takahashi and S. Dake, {\it Nucl. Phys.} {\bf A461} (1987) 263.

\bibitem{jacee}J. J. Lord and J. Iwai, Univ. of Washington preprint
(paper 515 submitted to the International Conference on High Energy
Physics, Dallas, August 1992); \hfill \break
J. Iwai (JACEE collaboration), UWSEA
92-06.

\bibitem{anselm}A. Anselm, {\it Phys. Lett.} {\bf 217B} (1988)
169;\hfill\break
A. A. Anselm and M. G. Ryskin, {\it Phys. Lett.} {\bf B266}
(1991) 482.

\bibitem{blaizot}J.-P. Blaizot and A. Krzywicki, {\it Phys. Rev.} {\bf D46}
(1992) 246.

\bibitem{recentbjorken}J. D Bjorken, {\it Int. J. Mod. Phys.} {\bf A7}
(1992 4189; \hfill \break
J. D. Bjorken, {\it Acta Physica Polonica\/} {\bf B23} (1992) 561.


\bibitem{kowalski}K. L. Kowalski and C. C. Taylor, {\it Disoriented Chiral
Condensate:  A White Paper for the Full Acceptance Detector}, CWRUTH-92-6
(1992) hep-ph/9211282.

\bibitem{bakedalaska} J. D. Bjorken,
K. L. Kowalski and C. C. Taylor, {\it Baked
Alaska}, Proceedings of Les Rencontres de Physique de la Vall\'ee
D'Aoste, La Thuile,
SLAC-PUB-6109 (1993); and {\it Observing Disoriented
Chiral Condensates}, Proceedings of the workshop on Physics
at Current Accelerators and the Supercollider, Argonne (1993)
hep-ph/9309235.

\bibitem{andreev} I. V. Andreev, {\it JETP Lett.} {\bf 33} (1981) 367.

\bibitem{karmanov} V. Karmanov and A. Kudrjavtsev, ITEP-88 (1983).


\bibitem{bray}A. J. Bray,
{\it Phys. Rev.} {\bf B41} (1990) 6724, and references
therein; \hfill \break
T. J. Newman,
A. J. Bray, and M. A. Moore, {\it Phys. Rev.} {\bf B42} (1990) 4514, and
references therein;\hfill\break
J. A. N. Filipe, A. J. Bray, and S. Puri, {\it Phase Ordering
Kinetics with External Fields and Biased Initial Conditions},
(1995) cond-mat/9504079.

\bibitem{turgel}N. Turok
and D. N. Spergel, {\it Phys. Rev. Lett.} {\bf 66} (1991) 3093.


\bibitem{ggp2} S. Gavin, A. Gocksch, and R. Pisarski, {\it Phys. Rev.
Lett.} {\bf 72} (1994) 2143.

\bibitem{hornsilver} D. Horn and R. Silver {\it Ann. Phys. (N.Y.)}
{\bf 66} (1971) 509.

\bibitem{boyanovsky} D. Boyanovsky, H. J. de Vega, and R. Holman,
{\it Phys. Rev.} {\bf D51} (1995) 734; and work presented at
Journ\'ee Cosmologique, Paris (1994) hep-th/9412052.

\bibitem{borlange} K. Rajagopal, {\it Nucl. Phys.} {\bf A566} (1994) 567c.

\bibitem{minimax} J. D. Bjorken, C. C. Taylor, {\it et al.}, {\it MiniMax:
A Revised Proposal for T-864}, (1993);\hfill\break
J. D. Bjorken,
{\it MiniMax: Multiparticle Physics at the Tevatron},
SLAC-PUB-6430 (1994).

\bibitem{latestbjorken}J. D. Bjorken, {\it Disoriented Chiral Condensate},
SLAC-PUB-6488, (1984); and {\it Geometry of Multi-Hadron Production},
SLAC-PUB-6707, (1994).

\bibitem{heisenberg}W. Heisenberg, {\it Z. Phys.} {\bf 133} (1952) 65.

\bibitem{botke} J. C. Botke, D. J. Scalapino, and R. L. Sugar,
{\it Phys. Rev.} {\bf D9} (1974) 813.

\bibitem{enikova} M. M. Enikova, V. I. Karloukovski and C. I. Velchev,
{\it Nucl. Phys.} {\bf B151} (1979) 172.

\bibitem{lamlo} C. Lam and S. Lo, {\it Phys. Rev. Lett.} {\bf 52}
(1984) 1184; {\it Phys. Rev.} {\bf D33} (1986) 1336; {\it Int.
J. Mod. Phys.} {\bf A7} (1992) 4189.

\bibitem{pratt} S. Pratt, {\it Phys. Lett.} {\bf B301} (1993) 159;\hfill\break
S. Pratt and V. Zelevinsky, {\it Phys. Rev. Lett.} {\bf 72} (1994) 816.

\bibitem{ggm} C. Greiner, C. Gong and B. M\"uller, {\it Phys. Lett.}
{\bf B316} (1993) 226.

\bibitem{amado} R. D. Amado {\it et al.}, {\it Phys. Rev. Lett.} {\bf 72}
(1994) 970.

\bibitem{martinis}M. Martinis {\it et al.}, {\it Phys. Rev.} {\bf D51}
(1995) 2482; and hep-ph/9411329 and hep-ph/9501210.

\bibitem{gavmul} S. Gavin and B. M\"uller, {\it Phys. Lett.} {\bf B329}
(1994) 486.

\bibitem{das} P. F. Bedaque and A. Das, {\it Mod. Phys. Lett.}
{\bf A8} (1993) 3151.

\bibitem{kluger} F. Cooper, Y. Kluger, E. Mottola, and J. P. Paz,
{\it Phys. Rev.} {\bf D51} (1995) 2377; \hfill\break
F. Cooper, S. Habib, Y. Kluger, E. Mottola, J. P. Paz, P. R. Anderson,
{\it Phys. Rev.} {\bf D50} (1994) 2848;\hfill\break
Y. Kluger, LA-UR-94-1566 (1994) hep-ph/9405279; and
LA-UR-94-2754 (1994) hep-ph/9408286;\hfill\break
Y. Kluger {\it et al.}, LBL-36904 (1995) hep-ph/9503205.

\bibitem{krz1} A. Krzywicki, {\it Phys. Rev.} {\bf D48} (1993) 5190.

\bibitem{blaizotkrz} J.-P. Blaizot and A. Krzywicki, {\it Phys. Rev.}
{\bf D50} (1994) 442.

\bibitem{khlebnikov} S. Yu. Khlebnikov, {\it Mod. Phys. Lett.} {\bf A8}
(1993) 1901.

\bibitem{kogan1} I. I. Kogan, {\it Phys. Rev.} {\bf D48} (1993) 3971.

\bibitem{huangwang} Z. Huang and X. Wang {\it Phys. Rev.} {\bf D49} (1994)
R4335.

\bibitem{krz2} A. Krzywicki, {\it Disoriented Chiral Condensates},
talk at 29th Rencontres de Moriond, LPTHE-ORSAY-94-43 (1994) hep-ph/9405244.

\bibitem{asakawa} M. Asakawa, Z. Huang, and X. Wang, LBL-35981 (1994)
hep-ph/9408299.

\bibitem{huangsuz} Z. Huang and M. Suzuki, private communication.

\bibitem{gyulassy} M. Gyulassy, S. K. Kauffmann, and L. W. Wilson,
{\it Phys. Rev.} {\bf C20} (1979) 2267.

\bibitem{cohen} T. D. Cohen, M. K. Banerjee, M. Nielsen, X. Jin,
{\it Phys. Lett.} {\bf B333} (1994) 166.

\bibitem{kogan2}I. I. Kogan, {\it JETP Lett.} {\bf 59} (1994) 307.

\bibitem{amadokogan} R. D. Amado and I. I. Kogan, {\it Phys. Rev.} {\bf D51}
(1995) 190.

\bibitem{yuen} H. P. Yuen, {\it Phys. Rev.} {\bf A13} (1976) 2226;
\hfill\break B. L. Schumacher, {\it Phys. Rep.} {\bf 135} (1986) 317.

\bibitem{suzuki}Z. Huang, M. Suzuki, and X. Wang, {\it Phys. Rev.}
{\bf D50} (1994) 2277.

\bibitem{latesthuang}Z. Huang, {\it Disoriented Chiral Condensate},
LBL-36740 (1995) hep-ph/9501366.

\bibitem{latestgavin}S. Gavin, {\it Dynamics of Chiral
Symmetry Breaking in Nuclear Collisions}, (1994) hep-ph/9407368;\hfill\break
S. Gavin, contribution to proceedings of Quark Matter '95.

\bibitem{bolekcraig} For example, by the WA98 experiment
at CERN (B. Wyslouch, private communication); by
the NA49 experiment at CERN (S. Hegyi, private communication); and
by the E866 experiment at Brookhaven (C. Ogilvie,
private communication).

\bibitem{bolek} This paragraph is based on preliminary studies
done by B. Wyslouch and collaborators using the acceptances
for the WA98 detector, discussed by him
at an informal meeting at M.I.T. and in further conversations.

\bibitem{WA98} A. L. S. Angelis {\it et al.},
{\it Proposal for a Large Acceptance Hadron and Photon
Spectrometer}, CERN/SPSLC/91-17, May 1991;\hfill\break
Status Report for WA98, CERN/SPSLC/94-32, October 1994.

\bibitem{paulo} P. Bedaque, private communication.



\end{thebibliography}
\end{document}